\documentclass[prd,
superscriptaddress,preprintnumbers,tightenlines,nofootinbib, eqsecnum]{revtex4-2}

\usepackage{amsmath}
\usepackage{amsfonts}
\usepackage{amssymb}
\usepackage{bm}
\usepackage{hyperref}
\usepackage{mathrsfs}
\usepackage{graphicx}
\usepackage{tabularx}
\usepackage{empheq}
\usepackage{ulem}
\usepackage[usenames]{color}
\usepackage{ytableau}
\usepackage{comment}
\usepackage{tikz}
\usetikzlibrary{arrows.meta, positioning, calc, fit, backgrounds}
        
\allowdisplaybreaks

\DeclareSymbolFontAlphabet{\mathrsfs}{rsfs}
\DeclareMathAlphabet{\mathcal}{OMS}{cmsy}{m}{n}

\newcommand{\FP}{\mathop{\mathrm{FP}}_{B=0}}

\newcommand{\nn}{\nonumber}
\newcommand\calO{\mathcal{O}}
\newcommand{\dd}{\mathrm{d}}
 
\newcommand{\de}{\mathrm{e}} 
\newcommand{\dI}{\mathrm{I}}
\newcommand{\dJ}{\mathrm{J}}
\newcommand{\dK}{\mathrm{K}}
\newcommand{\dM}{\mathrm{M}}

\newcommand{\dS}{\mathrm{S}}

\newcommand{\dW}{\mathrm{W}}
\newcommand{\dX}{\mathrm{X}}
\newcommand{\dY}{\mathrm{Y}}
\newcommand{\dZ}{\mathrm{Z}}

\newcommand{\calF}{\mathcal{F}}

\newcommand{\ab}{{}^{\alpha\beta}}
\newcommand{\RR}{_{\mathrm{RR}}}

\newcommand{\It}{\widetilde{\mathrm{I}}}
\newcommand{\Jt}{\widetilde{\mathrm{J}}}
\newcommand{\Kt}{\widetilde{\mathrm{K}}}
\newcommand{\Wt}{\widetilde{\mathrm{W}}}
\newcommand{\Xt}{\widetilde{\mathrm{X}}}
\newcommand{\Yt}{\widetilde{\mathrm{Y}}}
\newcommand{\Zt}{\widetilde{\mathrm{Z}}}

\newcommand{\Pf}{\mathop{\mathrm{Pf}}_{s_1, s_2}}

\newcommand{\be}{\begin{equation}}
\newcommand{\ee}{\end{equation}}
\newcommand{\bse}{\begin{subequations}}
\newcommand{\ese}{\end{subequations}}

\definecolor{darkgreen}{rgb}{0,0.5,0}

\newcommand{\red}{\textcolor{red}}

\hypersetup{
unicode=false,
 pdftoolbar=true, 
 pdfmenubar=true,
 pdffitwindow=false,
 pdfstartview={FitH},
 pdftitle={My title},
 pdfauthor={Author},
 pdfsubject={Subject},
 pdfcreator={Creator},
 pdfproducer={Producer},
 pdfkeywords={keyword1} {key2} {key3},
 pdfnewwindow=true,
 colorlinks=true,
 linkcolor=red,
 citecolor=cyan,
 filecolor=magenta,
 urlcolor=darkgreen,
 linktocpage=true
}

\makeatletter
\g@addto@macro\bfseries{\boldmath}
\makeatother

\begin{document}
	
\title{Gravitational radiation reaction for compact binary systems \\at the fourth-and-a-half post-Newtonian order in harmonic coordinates}

\author{Luc \textsc{Blanchet}}\email{luc.blanchet@iap.fr}
\affiliation{$\mathcal{G}\mathbb{R}\varepsilon{\mathbb{C}}\calO$, 
	Institut d'Astrophysique de Paris, \\UMR 7095, CNRS, Sorbonne Universit{\'e},
	98\textsuperscript{bis} boulevard Arago, 75014 Paris, France}

\author{Guillaume \textsc{Faye}}\email{faye@iap.fr}
\affiliation{$\mathcal{G}\mathbb{R}\varepsilon{\mathbb{C}}\calO$, 
Institut d'Astrophysique de Paris, \\UMR 7095, CNRS, Sorbonne Universit{\'e},
98\textsuperscript{bis} boulevard Arago, 75014 Paris, France}

\author{Emeric \textsc{Seraille}}\email[Corresponding author: ]{emeric.seraille@phys.ens.fr}
\affiliation{Laboratoire de Physique de l’Ecole Normale Sup{\'e}rieure, ENS, CNRS, Universit{\'e} \\ PSL, Sorbonne Universit{\'e}, Universit{\'e} Paris Cit{\'e}, F-75005 Paris, France
}
\affiliation{$\mathcal{G}\mathbb{R}\varepsilon{\mathbb{C}}\calO$, 
Institut d'Astrophysique de Paris, \\UMR 7095, CNRS, Sorbonne Universit{\'e},
98\textsuperscript{bis} boulevard Arago, 75014 Paris, France}

\author{David \textsc{Trestini}}\email{david.trestini@southampton.ac.uk}
\affiliation{School of Mathematical Sciences and STAG Research Centre, University of Southampton, Southampton SO17 1BJ, United Kingdom}

\date{\today}

\begin{abstract}
We derive the gravitational radiation-reaction (RR) force in the harmonic coordinate system  for general orbits in a general frame at the fourth-and-a-half post-Newtonian (4.5PN) order in the case of compact binary systems. Dimensional regularization is used to treat the ultra-violet divergences which appear at that order. We prove that the RR acceleration implies the known radiation fluxes at infinity associated with energy, angular momentum, linear momentum, and center-of-mass position.
As a consistency check, we verify the manifest Lorentz invariance of the RR acceleration in harmonic coordinates. Our result should be useful for comparisons with other approaches such as the gravitational self force (GSF) and the post-Minkowskian (PM) 
expansion. In particular, it agrees with the recent 2PM derivation of the RR force using effective field theory.
\end{abstract}


\maketitle

\section{Introduction}\label{sec:introduction}

It is well known~\cite{CE70, BuTh70, Bu71, MTW, Miller74, WalkW80, Ehl80, K80a, K80b, PapaL81, Schafer82, BD84, Dcargese} that the leading effect in the dynamics of an isolated source due to the reaction to the emission of gravitational radiation  appears at the so-called second-and-a-half post-Newtonian (2.5PN) order beyond the Newtonian acceleration (\textit{i.e.}, corrections  $\propto 1/c^5$). The radiation-reaction (RR) terms are inherently coordinate dependent, and several physically equivalent expressions for the RR in different gauges are known (see \textit{e.g.},~\cite{S83}). Burke \& Thorne (BT)~\cite{Bu71,BuTh70} derived a simple expression for the RR acceleration at 2.5PN order. In the case of a compact binary system, the RR acceleration of the first particle (labelled ``1'') is given by 
\begin{equation}\label{eq:aBT}
	a_{\text{RR}1}^i\Big|_\text{BT} = - \frac{2G}{5 c^5}\,y_1^j \frac{\dd^5\dI_{ij}}{\dd t^5} + \calO\left(\frac{1}{c^7}\right)\,,
\end{equation} 
where $\dI_{ij}(t) = m_1 y_1^{\langle i}y_1^{j\rangle} + (1\leftrightarrow 2)$ denotes the symmetric and trace-free (STF) quadrupole moment of the binary system, with $y_1^i(t)$ the trajectory of the particle~1, $m_1$ its mass, and the angle brackets denote the STF projection. This expression is valid in a particular coordinate system nowadays called the BT coordinates. However, the BT coordinates have a drawback: the expression~\eqref{eq:aBT} is not invariant against shifting the origin of the coordinate system, and more generally, it is not Lorentz invariant (in fact, it is not even Galileo invariant). Around the same time, Chandrasekhar and Esposito obtained an alternative expression valid in a different coordinate system~\cite{CE70}. The expression in ADM coordinates was derived in~\cite{Schafer82}. Yet a different expression for the RR acceleration at 2.5PN order was obtained by Damour~\cite{D82,D83} in harmonic coordinates for the case of compact binaries, motivated by the comparison to the observations of the Hulse-Taylor binary pulsar. In the ``modern'' notation that we shall adopt in the present paper, the RR acceleration of the particle~1 at 2.5PN order in harmonic-coordinates reads 
\begin{equation}\label{eq:aharm}
	a_{\text{RR}1}^i\Big|_\text{harm} = \frac{G}{c^5}\left[ \frac{3}{5} y_1^j \dI_{ij}^{(5)} + 2\frac{\dd}{\dd t}\left(v_1^j \,\dI_{ij}^{(3)}\right)+ 4\frac{\dd^2}{\dd t^2}\left(v_1^i \dW^{(1)} - \dY_i^{(1)}\right) - (\partial_{ijk}\chi)_1 \dI_{jk}^{(3)} - 4 v_{12}^j (\partial_{ij}U)_1 \dW^{(1)}\right] + \calO\left(\frac{1}{c^7}\right)\,,
\end{equation} 
where $v_1^i=\dd y_1^i/\dd t$ is the velocity of the particle~1, $v_{12}^i=v_{1}^i-v_{2}^i$, and where the superscript $(n)$ denotes the $n$-th derivative with respect to time $t$. As seen from Eq.~\eqref{eq:aharm}, the multipolar structure is more complicated than in the BT expression~\eqref{eq:aBT}, with two supplementary moments $\dW$ (a monopole) and $\dY_i$ (a dipole) in addition to the quadrupole $\dI_{ij}$. We refer below to these supplementary moments as the ``gauge'' moments which play a crucial role in harmonic coordinates. Besides, we denote
$U=G m_1/r_1+ (1 \leftrightarrow 2)$
the Newtonian potential and $\chi=G m_1 r_1 + (1\leftrightarrow 2)$
the superpotential such that $\Delta\chi=2U$, where $(\partial_{ijk}\chi)_1$ and $(\partial_{ij}U)_1$ denote the regularized values at the position of the particle~1. The great advantage of the harmonic-coordinate expression~\eqref{eq:aharm}, in contrast to the BT expression, is that it is Lorentz invariant; this follows from the harmonic-coordinate condition, which preserves Lorentz invariance.
Consequently, even at higher PN orders, we shall find that the RR acceleration in harmonic coordinates is made up of drastically fewer terms than in BT coordinates, since many terms combine together to form the \textit{relative} position and velocity of the bodies.
The expression for the RR acceleration of compact binaries has been worked out by many authors at the next 3.5PN order (\textit{i.e.}, at 1PN relative order). It was derived in the frame of the center-of-mass (CM) and in a general (parametrized) gauge by Iyer and Will (IW)~\cite{IW93, IW95}, relying on the balance equations between the energy and angular momentum, which were parametrized, and the associated fluxes at future null infinity, whose expressions were known at 1PN order.
We refer to this procedure as the ``flux-balance'' method. In the balance method, because of the arbitrariness of the choice of gauge, the result for the RR acceleration is expressed in terms of a set of arbitrary (IW) gauge parameters~\cite{IW93, IW95} which uniquely determine the gauge. At 2.5PN order, the choice of gauge is parametrized by exactly two parameters called $\alpha_3$ and $\beta_2$, adopting the latest notation in~\cite{GII97}. For instance, the BT gauge corresponds to $(\alpha_3,\beta_2)=(5,4)$, the harmonic gauge to $(\alpha_3,\beta_2)=(0,-1)$, and the ADM gauge to $(\alpha_3,\beta_2)=(3,5/3)$. 
In the extension of this method to 4.5PN order by Gopakumar, Iyer and Iyer (GII)~\cite{GII97}, twelve additional parameters are needed, which amounts to a total of twenty. 

Besides the flux-balance method, there are also ``first-principles'' calculations in specific gauges, which do not rely on balance equations. For instance, the BT gauge has been extended to 3.5PN order for general systems in~\cite{B93, B97}, and after specialization to compact binary sources and reduction to the CM frame, it was shown to correspond to a unique set of IW gauge parameters. Other first-principles derivations at 3.5PN order were performed in various gauges~\cite{JaraS97,PW02,KFS03,NB05,itoh3}: in particular, the result in harmonic coordinates has been obtained by Pati and Will~\cite{PW02} and Nissanke and Blanchet~\cite{NB05}, again corresponding to a unique set of gauge parameters. It is important to keep in mind that the balance method~\cite{IW93, IW95} is limited to the frame of the CM, where the only invariants are the energy and the angular momentum. By contrast, the first-principles approach computes the RR acceleration in a general frame, and only after that, specializes the result to the CM where the gauge parameters are computed.\footnote{In principle, nothing prevents the generalization of the balance approach to a general frame (taking also into account the fluxes of linear momentum and CM position), except for the likely proliferation of the gauge parameters.}

At the 4.5PN order, the first-principles calculation of the RR was performed in Ref.~\cite{BFT24} in a general frame (where there is the full set of invariants: energy, angular momentum, linear momentum and position of the CM) and in the (extended) BT coordinate system.
The BT gauge was chosen because it significantly simplifies intermediate calculations and allows a complete derivation in terms of multipole moments; it thus provides a natural extension of Eq.~\eqref{eq:aBT} to 4.5PN, without expanding the moments in terms of the binary's positions and velocities. 
In the CM frame, Ref.~\cite{BFT24} found a discrepancy when comparing with the flux-balance approach developed at 4.5PN order in~\cite{GII97}. It was indeed proven that, at 4.5PN order, the definition of the CM frame needs to be corrected  by the contribution from the radiation field and, in particular, must include the physical effect of the binary's recoil by gravitational radiation~\cite{BFT24}. This implies that the CM acceleration at 4.5PN order is no longer local in time as it involves the integrated flux of linear momentum. This contradicts~\cite{GII97}, which assumed as a basic hypothesis that the RR acceleration in the CM is local at this order (note that Ref.~\cite{LPY23}, which followed the first-principles approach, also omitted this contribution). Nevertheless, Ref.~\cite{BFT24} was able to correct the end result of~\cite{GII97} by incorporating the non-local recoil effect (as well as some specific local corrections) and to uniquely determine the twenty GII gauge parameters
that correspond to the extended BT gauge.

In the present paper, we shall extend~Ref.~\cite{BFT24}  to obtain the RR acceleration at 4.5PN (2PN relative) order for compact binaries in \textit{harmonic} coordinates.
First, we shall determine the transformation of the particles' trajectories (or ``shift'') going from the extended BT gauge to harmonic coordinates. The harmonic-coordinate condition specifies a unique prescription for a global coordinate system covering a general extended isolated and smooth source, and satisfying the appropriate boundary condition at infinity (no incoming radiation). In the multipolar post-Minkowskian (MPM) formalism~\cite{BD86,B87,BD88,BD92}, this uniqueness results from the fact that the metric in harmonic coordinates is uniquely determined as the solution of the matching between the MPM metric outside the source and the metric in the near zone, the latter being in the form of a PN expansion~\cite{B98mult,PB02,BFN05}. After matching, the MPM metric in harmonic coordinates is parametrized by two families of ``source'' multipole moments $\dI_L$ (mass type) and $\dJ_L$ (current), such as the mass quadrupole moment $\dI_{ij}$ entering the lowest order result~\eqref{eq:aharm}, and by four supplementary families of ``gauge'' moments $\dW_L$, $\dX_L$, $\dY_L$ and $\dZ_L$, such as the monopole $\dW$ and dipole $\dY_i$ moments in~\eqref{eq:aharm}. All the source and gauge moments are determined as explicit integrals over the matter distribution in the source. For a compact binary source they are given by specific (in general non-local) functionals of the particles' trajectories and velocities.

We shall obtain the 4.5PN RR acceleration in harmonic coordinates by directly performing the required shift upon the end results of Ref.~\cite{BFT24}. We shall explicitly check that all the flux-balance equations (for energy, angular and linear momenta, and CM position) are preserved, with the same expressions for the fluxes as in BT coordinates; only the ``Schott'' terms~\cite{Schott}, which we compute explicitly, are modified with respect to the extended BT coordinate system. Once the 4.5PN RR acceleration is known in an arbitrary frame, we restrict to the CM frame (carefully accounting for the hereditary contributions due to recoil) and compare with the parametrization of Sec.~VII in~\cite{BFT24}, which corrects the parametrization of~\cite{GII97}. In this way, we obtain the full set of gauge parameters corresponding to harmonic coordinates at 4.5PN order. The high PN harmonic-coordinate RR acceleration should be compared to recent calculations of the gravitational scattering of two body systems in the post-Minkowskian approximation (see~\cite{BDG24} and references therein).\footnote{Donato Bini (private communication) has checked, and we have independently confirmed, that our results for the RR at the 4.5PN order perfectly agree with the recent PM RR results (truncated at 4.5PN, and up to order $G^2$) given in the Tables~IV and~V of Ref.~\cite{BDG24}.} 

This work completes the full (conservative and dissipative) equations of motion of compact binaries (without spins) at 4.5PN order in harmonic coordinates.
The conservative part of the dynamics was previously determined to 4PN order from the Hamiltonian formalism in ADM coordinates~\cite{JaraS12, JaraS13, BiniD13, DJS14, JaraS15, DJS15eob}, the Fokker action in harmonic coordinates~\cite{BBBFMa, BBBFMb, BBBFMc, MBBF17, BBFM17} and the Effective Field Theory in harmonic coordinates~\cite{FS4PN, FMSS17, PR17, FS19, FPRS19, BlumMMS20a}. The non-local 4PN tail term therein, including both conservative and RR contributions, is known from~\cite{BD88,BD92,FStail,GLPR16}. From the 4.5PN RR terms in the equations of motion (this work and the previous one~\cite{BFT24}) we conclude that the flux-balance equations are proven with 2PN relative precision.\footnote{The RR part of the non-local tail term at 4PN order enters the balance laws at 1.5PN relative order, and has recently been shown to yield an important non-local Schott contribution in the energy balance equation~\cite{T25}.} However, as a caveat, we recall that such precision is insufficient for state-of-the-art calculations of the orbital phase evolution of circular compact binaries; the recent phasing obtained at 4.5PN order beyond quadrupolar radiation~\cite{BFHLT23a,BFHLT23b,T25} implicitly assumed that the energy balance law holds up to the same (relative) 4.5PN level. 

The plan of this paper is as follows. In Sec.~\ref{sec:metricharmonic}, we recall the construction with the MPM-PN formalism (and matching) of the unique metric outside and inside a general matter source in harmonic coordinates. All the multipole source and gauge moments are given in arbitrary $d$ dimensions. In Sec.~\ref{sec:BTtoharmonic} we perform a coordinate transformation from BT to harmonic coordinates and determine the corresponding shift (or ``contact'' transformation) of the particles' acceleration. In Sec.~\ref{sec:MultipoleMoments} we compute the multipole moments of the compact binary and find that one of the gauge moments involves a pole treated with dimensional regularization. Sec.~\ref{sec:acceleration_general} presents our results in a general frame, including the general expression for the RR acceleration and the Schott terms entering the flux balance laws. Sec.~\ref{sec:CM} deals with the results in the CM frame, namely the passage to the CM frame, the CM RR acceleration, the verification of the complete set of balance equations, and the computation of the IW parameters. The Lorentz invariance of the RR acceleration is checked in Sec.~\ref{sec:Lorentz} and the paper ends with a short conclusion in Sec.~\ref{sec:conclusions}. In Appendix~\ref{appendix:innerzone} we present an alternative derivation of our main results, based on the coordinate transformation to harmonic coordinates implemented in the near zone, in contrast with the ``exterior-zone'' implementation in Sec.~\ref{sec:BTtoharmonic}. Other Appendices contain useful but lengthy expressions.

\section{Metric in harmonic coordinates in the MPM-PN formalism}
\label{sec:metricharmonic}

As reviewed in the Introduction, the solution of the Einstein field equations in harmonic coordinates for a general isolated matter source has been determined by matching the MPM expansion in the exterior of the source to the PN expansion in the source's interior and near zone. The solution is valid formally up to any PN order, including all conservative and RR effects, with the ``no incoming radiation'' condition imposed at past null infinity. In the exterior, it admits the MPM expansion
\begin{equation}\label{eq:MPM}
\mathcal{M}(h^{\alpha\beta})\equiv h^{\alpha\beta}_\text{MPM}=\sum_{n=1}^{+\infty}G^n\,h^{\alpha\beta}_{(n)}\,,
\end{equation}
where $\mathcal{M}(h^{\alpha\beta})$ stands for the multipole expansion of the fully-fledged metric $h^{\alpha\beta}$ in harmonic coordinates. The formal $G$-expansion (or PM expansion) in~\eqref{eq:MPM} is constructed algorithmically starting from and iterating the linearized approximation with $n=1$, which represents the most general solution of the linearized vacuum Einstein field equations in harmonic coordinates~\cite{SB58,Pi64,Th80,BD86}. 

The construction of the MPM metric and definition of the moments are usually done in 3 spatial dimensions, but in the present work, we generalize it to an arbitrary number of dimensions $d$ because we shall apply dimensional regularization (DR) to the multipole moments for the case of point-like binary sources. In $d$ dimensions, the MPM algorithm proceeds essentially in the same way, but we must use the $d$-dimensional flat retarded integral operator to define the iteration. The multipole moments are defined from the linearized approximation $n=1$, which can be split into a ``canonical'' contribution and a linear gauge transformation, namely
\begin{equation}\label{eq:h1}
	h^{\alpha\beta}_{(1)} = h^{\alpha\beta}_{\text{can}\,(1)} + \partial^\alpha\zeta_{(1)}^{\beta}+\partial^\beta\zeta_{(1)}^{\alpha}-\eta^{\alpha\beta}\partial_\gamma\zeta_{(1)}^{\gamma}\,.
\end{equation}
The canonical form defines the ``source'' moments, whereas the gauge vector defines the ``gauge'' moments. They are determined from an irreducible decomposition of the linearized metric~\eqref{eq:h1} and their 3-dimensional expressions are known in terms of the PN expansion of the source's pseudo stress-energy tensor~\cite{B98mult}. In $d$ dimensions, the problem is more complicated because the irreducible decomposition is not straightforward to generalize. The problem was first solved for the easier case of the mass-type moments in~\cite{BDEI05dr}, then for the more subtle case of the current-type moments in~\cite{HFB21}; here we report (and complete) the end results.

The canonical part of the metric~\eqref{eq:h1} is parametrized by three irreducible moments $\dI_L$, $\dJ_{i\vert L}$ and $\dK_{ij\vert L}$, 
\begin{subequations}\label{h1can}
	\begin{align}
		h^{00}_{\text{can}\,(1)} &= - \frac{4}{c^2} \sum_{\ell=0}^{+\infty}
		\frac{(-)^\ell}{\ell!} \partial_L \It_{\text{ret}\,L} \, ,\\
		h^{0i}_{\text{can}\,(1)} &= \frac{4}{c^3} \sum_{\ell=0}^{+\infty}
		\frac{(-)^\ell}{\ell!} \bigg[ \partial_{L-1} \dot{\It}_{\text{ret}\,iL-1}
		+ \frac{\ell}{\ell+1} \partial_L \Jt_{\text{ret}\,i|L} \bigg] \, ,\\
		h^{ij}_{\text{can}\,(1)} &= - \frac{4}{c^4} \sum_{\ell=0}^{+\infty}
		\frac{(-)^\ell}{\ell!} \bigg[ \partial_{L-2} \ddot{\It}_{\text{ret}\,ijL-2}
		+ \frac{2\ell}{\ell+1} \partial_{L-1} \dot{\Jt}_{\text{ret}\,(i|\underline{L-1}j)}
		+ \frac{\ell-1}{\ell+1}\partial_L \Kt_{\text{ret}\,ij|L}\bigg] \,,
\end{align}\end{subequations}
where $\partial_L\equiv\partial_{i_1}\cdots\partial_{i_\ell}$ are multi-spatial derivatives (with the multi-index $L=i_1\cdots i_\ell$). The overdots denote differentiation with respect to time. In the last line, the underlined indices are excluded from the symmetrization operation, indicated by parenthesis $(\dots)$.  Note that we have here opted for the convention of Ref.~\cite{HFB21} for the ordering of the indices in $\dJ_{i|L}$; this corresponds to the ``HFB'' convention discussed in Footnote 3 of~\cite{BFHLT23b}.  The retarded multipole-moment functions, defined for any multipole-moment and denoted $\dI_{L}(r,t)$ without loss of generality, are given by \footnote{The retarded and advanced Green's functions of the flat d'Alembertian operator read (following the notation in~\cite{BBBFMa}): 
\begin{align*}
	G_\text{ret}(\mathbf{x},t) = -
	\frac{\tilde{k}}{4\pi}\frac{\theta(c t-r)}{r^{d-1}}
	\,\gamma_{\frac{1-d}{2}}\left(\frac{c t}{r}\right)\,, \qquad
    G_\text{adv}(\mathbf{x},t) = -
	\frac{\tilde{k}}{4\pi}\frac{\theta(-c t-r)}{r^{d-1}}
	\,\gamma_{\frac{1-d}{2}}\left(\frac{c t}{r}\right)\,,
\end{align*}
with $\tilde{k}=\Gamma(\frac{d}{2}-1)/\pi^{\frac{d}{2}-1}$
and the useful definition
\begin{align*}
	\gamma_{\frac{1-d}{2}}(z) \equiv \frac{2\sqrt{\pi}}{\Gamma(\frac{3-d}{2})\Gamma(\frac{d}{2}-1)}
	\,\big(z^2-1\bigr)^{\frac{1-d}{2}}\,,\qquad\int_1^{+\infty} \dd z \,\gamma_{\frac{1-d}{2}}(z) = 1\,,\qquad\lim_{d\to 3}\gamma_{\frac{1-d}{2}}(z)=\delta(z-1)\,.
\end{align*}
}
\begin{subequations}\label{notationtilde}
\begin{align}\label{tilderet}
	\It_{\text{ret}\,L}(r,t) \equiv
	\frac{\tilde{k} \ell_{0}^{d-3}}{r^{d-2}}\int_1^{+\infty} \dd
	z\,\gamma_{\frac{1-d}{2}}(z)\,\dI_L(t-z r/c)\,,\quad\text{such that}\quad\Box\,\It_{\text{ret}\,L}(r,t)=0\,.
\end{align}
We have also introduced $\ell_0$, an arbitrary length scale associated to dimensional regularization. We can also define the advanced multipole-moment functions  
\begin{align}\label{tildeadv}
	\It_{\text{adv}\,L}(r,t) \equiv
	\frac{\tilde{k}\ell_{0}^{d-3}}{r^{d-2}}\int_1^{+\infty} \dd
	z\,\gamma_{\frac{1-d}{2}}(z)\,\dI_L(t+z r/c)\,,\quad\text{such that}\quad\Box\,\It_{\text{adv}\,L}(r,t)=0\,.
\end{align}
\end{subequations}
In the absence of a $1/\varepsilon$ pole in the expression for the moment as $\varepsilon\equiv d-3 \rightarrow 0$, note that $\It_{\text{ret}\,L}$ and $\It_{\text{adv}\,L}$ reduce, respectively, to $r^{-1}\,\dI_{L}(t-r/c)$ and $r^{-1}\,\dI_{L}(t+r/c)$. Similarly, the gauge vector in~\eqref{eq:h1} defines the four irreducible gauge moments $\dW_L$, $\dX_L$, $\dY_{L}$ and $\dZ_{i|L}$ through
\begin{subequations}\label{gaugezetaret}
	\begin{align}
		& \zeta^0_{(1)} = \frac{4}{c^3}\sum_{\ell=0}^{+\infty} \frac{(-)^\ell}{\ell!}\,\partial_L \Wt_{\text{ret}\,L}\, ,\\
		& \zeta^i_{(1)} =  -\frac{4}{c^4}\sum_{\ell=0}^{+\infty} \frac{(-)^\ell}{\ell!}\,\partial_{iL} \Xt_{\text{ret}\,L}
		- \frac{4}{c^4}\sum_{\ell=1}^{+\infty} \frac{(-)^\ell}{\ell!}\,\biggl[ \partial_{L-1} \Yt_{\text{ret}\,iL-1} + \frac{\ell}{\ell+1}\partial_{L-1}
		\Zt_{\text{ret}\,i|L-1}\biggr]\,.
\end{align}\end{subequations}
The mass-type moments $\dI_L$, $\dW_L$, $\dX_L$ and $\dY_{L}$ are just symmetric-trace-free (STF) in all their indices so their symmetry is given by a symmetric Young tableau. However, the current-type moments $\dJ_{i\vert L}$ and $\dZ_{i|L}$ have the more complicated symmetry of a mixed Young tableau~\cite{HFB21}. Furthermore, we note the presence in $d$ dimensions of the additional ``Weyl'' moment $\dK_{ij\vert L}$ which vanishes in 3 dimensions (the number of its independent components is proportional to $d-3$).

We report the explicit expressions in $d$ dimensions of all these moments, completing the results of~\cite{HFB21} with the expressions of the gauge moments. We refer to~\cite{HFB21} for all details on the irreducible decomposition leading to such expressions.
We introduce the convenient notations
\begin{align}\label{eq:Sigma}
	\Sigma \equiv \frac{2}{d-1}\,
	\frac{(d-2)\overline{\tau}^{00}+\overline{\tau}^{ii}}{c^2}\,,\qquad
	\Sigma^i \equiv \frac{\overline{\tau}^{i0}}{c}\,,\qquad
	\Sigma^{ij} \equiv \overline{\tau}^{ij}\,,
\end{align}
where $\overline{\tau}^{\mu\nu}$ is the PN expansion of the stress-energy pseudo-tensor $\tau^{\mu\nu}$. Furthermore, we define for any of the previous quantities $\Sigma(\mathbf{x},t)$:
\begin{subequations}\label{exactform}
	\begin{align}
		\Sigma_{[\ell]}(\mathbf{x},t) &\equiv \int_{-1}^1 \dd z
		\,\delta_\ell(z)
		\,\Sigma(\mathbf{x},t+z r/c)\,,\\
		\text{with}\quad
		\delta_\ell(z) &\equiv
		\frac{\Gamma\left(\frac{d}{2}+\ell\right)}{
			\sqrt{\pi}\Gamma
			\left(\frac{d-1}{2}+\ell\right)}
		\,(1-z^2)^{\frac{d-3}{2}+\ell}\,, \qquad\int_{-1}^{1}
		\dd z\,\delta_\ell(z) = 1\,,
\end{align}
and in practice we shall use the (formally equivalent) infinite PN series
\begin{align}\label{eq:Sigma_ell_PNseries}
		\Sigma_{[\ell]}(\mathbf{x},t) = \sum_{k=0}^{+\infty}\frac{1}{2^{2k}k!}\frac{\Gamma\left(\frac{d}{2}+\ell\right)}{\Gamma\left(\frac{d}{2}+\ell+k\right)} \left(\frac{r}{c}\frac{\partial}{\partial t}\right)^{2k}\Sigma(\mathbf{x},t)\,.
\end{align}\end{subequations}
Then, the general expressions of the source moments in $d$ dimensions are 
\begin{subequations}\label{eq:sourcemts}
	\begin{align}\label{eq:sourcemts_IL}
	\dI_L &= \frac{d-1}{2(d-2)}\mathop{\mathrm{FP}}_{B=0} \int
	\dd^d\mathbf{x}\,\tilde{r}^B
	\Biggl\{\hat{x}^L\,\Sigma_{[\ell]}-\frac{4(d+2\ell-2)}
	{c^2(d+\ell-2)(d+2\ell)}\,\hat{x}^{iL}\,
	\dot{\Sigma}^{i}_{[\ell+1]}\nn\\
	&\qquad +\frac{2(d+2\ell-2)}
	{c^4(d+\ell-1)(d+\ell-2)(d+2\ell+2)}
	\,\hat{x}^{ijL}\,
	\ddot{\Sigma}^{ij}_{[\ell+2]} -
	\frac{4(d-3)(d+2\ell-2)}{c^2(d-1)(d+\ell-2)(d+2\ell)}
	B \,\hat{x}^{iL}\,\frac{x^j}{r^2}
	\,\Sigma^{ij}_{[\ell+1]}
	\Biggr\}\,,\\
	\dJ_{i|L} &= \mathop{\mathcal{A}}_{i i_{\ell}} \mathop{\text{FP}}_{B=0} \int
	\dd^d \mathbf{x} 
	\,\tilde{r}^{B} \Biggl\{-2
	\biggl[ \hat{x}^L \,\Sigma^{i}_{[\ell]} - 
	\frac{\ell (2\ell 
		+d-4)}{(\ell+d-3)(2\ell+d-2)} \delta^{i\langle i_\ell} \hat{x}^{L-1\rangle
		a} \Sigma^{a}_{[\ell]} \biggr] \nn\\& \qquad + \frac{2 (2\ell +d-2)}{c^2(\ell+d-1)(2\ell+d)}
	\biggl[ \hat{x}^{aL}\,\dot{\Sigma}^{ia}_{[\ell+1]} - \frac{\ell (2\ell 
		+d-4)}{(\ell+d-3)(2\ell+d-2)} \delta^{i\langle i_\ell} \hat{x}^{L-1\rangle
		ab} \dot{\Sigma}^{ab}_{[\ell+1]} \biggr]\Biggr\}\,\\
	\dK_{ij|L} &= 4 \mathop{\mathcal{A}}_{j i_{\ell}} \,\mathop{\mathcal{A}}_{i i_{\ell-1}}
	\mathop{\text{STF}}_{ij} \mathop{\text{STF}}_{L}  \mathop{\text{FP}}_{B=0} \int
	\dd^d \mathbf{x} \,\tilde{r}^{B} \Biggl\{ \hat{x}^L
	\,\Sigma^{ij}_{[\ell]} \nn \\ & \qquad - \frac{2\ell (2 \ell + d - 4)
		\Big[ (2(d - 2) \ell + d^2 -2 d + 4) \delta^{i i_\ell} \hat{x}^{L-1 a}
		\,\Sigma^{ja}_{[\ell]} + 4 (\ell - 1) \delta^{i i_\ell} \hat{x}^{L-2ja}
		\,\Sigma^{i_{\ell - 1} a}_{[\ell]} \Big]}{(d - 2) (\ell + d - 2) (2 \ell
		+ d - 2) (2 \ell + d)} \nn \\ & \qquad + \frac{(\ell - 1) \ell (2 \ell + d - 4) (2 (d
		- 2) \ell + d^2 - 4 d + 12)}{(d - 2) (\ell + d -3) (\ell + d - 2) (2 \ell + d - 2)
		(2 \ell + d)}\delta^{j i_\ell} \delta^{i i_{\ell-1}} \hat{x}^{L-2ab}
	\,\Sigma^{ab}_{[\ell]} \nn \\ & \qquad + \frac{2 \ell (2 \ell + d -4)}{(d
		-2) (\ell + d -2) (2 \ell + d)} \delta^{i i_\ell} \hat{x}^{L-1j}
	\,\Sigma^{kk}_{[\ell]} \Biggl\} \,.\label{eq:sourcemts_KL}
\end{align}
\end{subequations}
Here $\mathcal{A}_{i j}$ denotes the antisymmetrization with respect to the pair of indices $(i,j)$ (with the factor $1/2$ included) and the notation $\text{FP}_{B=0}$~refers to an IR regularization based on Hadamard's \textit{partie finie} (see~\cite{HFB21} for full details).
The expressions for the four gauge moments, which are new to this work, are:
\begin{subequations}\label{eq:gaugemts}
	\begin{align}
		\dW_L &= \FP \int \dd^d\mathbf{x}\,\widetilde r^B \Biggl\{ \frac{2\ell+d-2}{(\ell+d-2)(2\ell+d)} \hat{x}_{iL}
		\Sigma^i_{[\ell+1]} -
		\frac{2\ell+d-2}{2c^2(\ell+d-2)(\ell+d-1)(2\ell+d+2)} \hat{x}_{ijL} \dot{\Sigma}^{ij}_{[\ell+2]} \Biggr\}\,,\\
		\dX_L &= \FP \int \dd^d\mathbf{x}\,\widetilde r^B \Biggl\{ \frac{2\ell+d-2}{2(\ell+d-2)(\ell+d-1)(2\ell+d+2)} \hat{x}_{ijL} \Sigma^{ij}_{[\ell+2]} \Biggr\}\,,\\
		\dY_{L} &= \FP \int \dd^d\mathbf{x}\,\widetilde r^B \Biggl\{ -\frac{(\ell+3d-8)(2\ell+d)-4(d-3)}{(d-2)(\ell+d-2)(2\ell+d)} \hat{x}_L \Sigma^{ii}_{[\ell]} + \frac{3(2\ell+d-2)}{(\ell+d-2)(2\ell+d)} \hat{x}_{iL} \dot{\Sigma}^i_{[\ell+1]}
		\nn\\& \qquad  - \frac{(4\ell+3d-3)(2\ell+d-2)}{
			c^2(\ell+d-2)(\ell+d-1)(2\ell+d)(2\ell+d+2)} \hat{x}_{ijL}
		\ddot{\Sigma}^{ij}_{[\ell+2]} 
		- \frac{2(d-3)\ell (2\ell+d-4)}{(d-2)(\ell+d-2)(2\ell+d)} \hat{x}_{i\langle L-1} \Sigma^{i_\ell\rangle i}_{[\ell]} \Biggr\}\,,\\
		\dZ_{i|L} &=  \FP \int \dd^d\mathbf{x}\,\widetilde r^B  \, \frac{6\ell \delta^{[i}_a \delta^{i_{\ell}}_b \delta^{j]}_{\langle j}}{(\ell+d-3)} 
		\biggl\{- \frac{2\ell+d-2}{(\ell+d-1)(2\ell+d)} 
		\hat{x}_{L-1 \rangle bc} \Sigma^{ac}_{[\ell+1]} \biggr\}\,.
	\end{align}
Equivalently, the moment $\dY_L$ may be written in the more convenient form
\begin{align}\label{Yi_d_dim_∑}
\dY_L &= \frac{1}{d-2} \Bigg[ \FP \int \dd^d\mathbf{x}\,\widetilde r^B \Bigg\{ - \hat{x}_L \Sigma^{ii}_{[\ell]} + \frac{d(2\ell+d-2)}{(\ell+d-2)(2\ell+d)} \hat{x}_{iL} \dot{\Sigma}^i_{[\ell+1]}\nn \\& \qquad\qquad  - \frac{(d-1)(2\ell+d-2)}{
			c^2(\ell+d-2)(\ell+d-1)(2\ell+d+2)} \hat{x}_{ijL} \ddot{\Sigma}^{ij}_{[\ell+2]} + \frac{2(d-3)(2\ell+d-2)}{(\ell+d-2)(2\ell+d)} B\, \hat{x}_{iL} \frac{x^j}{r^2} \Sigma^{ij}_{[\ell+1]} \Bigg\} \Bigg]\,.
\end{align}
In this alternative expression the volume integral over the source piece proportional to $\Sigma^{i_\ell i}_{[\ell]}$ has been traded for a surface integral (carrying an explicit factor $B$), simpler to deal with, plus extra terms that combine with the other existing ones and affect their coefficients for $d\neq 3$. We also note the alternative form of the moment $\dZ_{i|L}$, where as before $\mathcal{A}_{i i_\ell}$ represents the antisymmetrization operator over indices $i$ and $i_\ell$:
\begin{align}
		\dZ_{i|L} &= \mathop{\mathcal{A}}_{i i_\ell} \FP \int \dd^d\mathbf{x}\,\widetilde r^B  
		\Biggl\{- \frac{2(2\ell+d-2)}{(\ell+d-1)(2\ell+d)} 
		\hat{x}_{aL} \Sigma^{ai}_{[\ell+1]} 
		+ \frac{2\ell (2\ell+d-4)}{(\ell+d-3)(\ell+d-1)(2\ell+d)} \delta_{i\langle i_\ell} \hat{x}_{L-1\rangle ab}  \Sigma^{ab}_{[\ell+1]} \Biggr\}\,.
	\end{align}
\end{subequations}

As can be seen from the leading-order expression \eqref{eq:aBT}, deriving the 4.5PN RR acceleration requires control of the gauge monopole moment $\dW$ and gauge dipole moment $\dY_i$ up to 2PN order. All the $d$-dimensional machinery we have developed here is only required to compute these two moments for compact binaries up to 2PN order, where DR will turn out to be crucial. At 2PN order, more explicit expressions for $\Sigma$, $\Sigma^i$ and $\Sigma^{ij}$ in terms of elementary PN potentials are provided in Appendix~\ref{app:ΣandPotentials}.

Once we know the multipole moments~\eqref{eq:sourcemts}--\eqref{eq:gaugemts}, the linearized approximation to the MPM metric~\eqref{eq:MPM} is complete and we can proceed to the non-linear iterations following the MPM algorithm~\cite{BD86}. Such non-linear iterations are only relevant for higher-order PN terms and can here be performed ordinarily in 3 dimensions. The MPM metric in harmonic coordinates is valid outside the source and matches (in the sense of matching of asymptotic expansions) the PN expansion of the metric inside the source in harmonic coordinates~\cite{B98mult, PB02, BFN05}. The resulting fully-fledged metric (globally valid inside and outside the source) corresponds, we argue, to the unique global harmonic coordinate system in which we shall compute the 4.5PN RR acceleration of compact binaries.\footnote{The harmonic coordinates are unique under two conditions: (1) the metric is generated by an isolated source which is regular, say a smooth [\textit{i.e.} $C^\infty(\mathbb{R}^3)$] matter distribution; (2) the no-incoming radiation condition holds at past null infinity $\mathcal{J}^{-}$: $r\to+\infty$ with $t+r/c =$ const. Suppose that two metrics differ by a linear gauge transformation, ${h'}{}^{\alpha\beta} = {h}^{\alpha\beta} + \partial\phi^{\alpha\beta}$ where $\partial\phi^{\alpha\beta}\equiv\partial^\alpha\phi^{\beta} + \partial^\beta\phi^{\alpha} - \eta^{\alpha\beta}\partial_\gamma\phi^{\gamma}$. If the two metrics are harmonic, $\partial_\beta {h'}{}^{\alpha\beta} = \partial_\beta {h}^{\alpha\beta} = 0$, the gauge vector relating them must be a harmonic function since $\partial_\beta\partial\phi^{\alpha\beta}=\Box\phi^{\alpha} = 0$. For a regular source we can write at any field point $(\mathbf{x}, t)$ (outside or inside the source) the Fresnel-Kirchhoff formula~\cite{BornWolf} 
\begin{align*}
	\phi^\alpha(\mathbf{x}, t) = \int\frac{\dd\Omega'}{4\pi}\left[\left(\frac{\partial}{\partial
			r}+\frac{\partial}{c\partial t}\right)\left(r
		\phi^\alpha\right)\right]\left(\mathbf{x}', t - \frac{\vert\mathbf{x}-\mathbf{x}'\vert}{c}\right)\,,
\end{align*}
where the surface integral extends on an arbitrary retarded sphere with generic source point $(\mathbf{x}', t')$ with $t' = t - \vert\mathbf{x}-\mathbf{x}'\vert/c$. Imposing the no-incoming radiation condition at $\mathcal{J}^{-}$:
\begin{align}
	\lim_{r\to+\infty\atop t+r/c = \text{const}} \left(\frac{\partial}{\partial
		r}+\frac{\partial}{c\partial t}\right)\left(r
	\phi^\alpha\right) = 0\,,
\end{align}
we conclude that $\phi^\alpha = 0$, hence the harmonic gauge is unique. The argument is easily extended to a true coordinate transformation by invoking a non-linear perturbation over $\phi^\alpha$. The uniqueness of the metric in harmonic coordinates has also to be understood up to the remaining Poincaré invariance at infinity.}

\section{Coordinate transformation from BT to harmonic coordinates} \label{sec:BTtoharmonic}

\begin{figure}
		\centering
		\begin{tikzpicture}[
		node/.style={
			rectangle,
			draw,
			rounded corners=2pt,
			minimum width=10mm,
			minimum height=8mm,
			align=center
		},
		arrow/.style={->, thick},
		lab/.style={draw=none},
		>=Stealth,
		shorten >=2pt,
		shorten <=2pt
		]

		\node[node] (A) at (-1.2,0) {$h_{\text{BT}}^{\alpha \beta}(\dM,\dS)$};
		\node[node] (B) at (3,0) {$h^{\alpha \beta}_{\text{BT}\,(1)}(\dI,\dJ)$};
		\node[node] (E) at (5.7,0) {$h^{\alpha\beta}_{\text{can}\,(1)}(\dI,\dJ)$};
		\node[node] (F) at (9,0) {$h^{\alpha\beta}_{(1)}(\dI,\dJ,\dW,\dX,\dY,\dZ)$};
		\node[node] (L) at (13.5,0) {$h^{\alpha \beta}(\dI,\dJ,\dW,\dX,\dY,\dZ)$};

		\node[node] (C) at (3,-2) {$h^{\alpha \beta}_{\text{BT}\,{(2)}}(\dI,\dJ)$};
		\node[node] (G) at (9,-2) {$h^{\alpha \beta}_{(2)}(\dI,\dJ,\dW,\dY)$};
		
		\node[lab] (H) at (6,3.6) {\textbf{Gauge transformation}};

		\node[lab] (Q) at (6.5,-3.8) {\textbf{Coordinate shift} of the body A};
		
		\node[node] (R) at (-1.2,-4.3) {$a_{\text{RR}~A}^i\Big|_\text{BT}$};
		\node[node] (S) at (13.5,-4.3) {$a_{\text{RR}~A}^i\Big|_\text{harm}$};

		\draw[arrow] (A.east) -- (B.west);
		\draw[arrow, align=center] (B.east) -- node[lab, below] {\\ \\ $\xi_{(1)}^\alpha(\dI,\dJ)$} (E.west);
		\draw[arrow, align=center] (E.east) -- node[lab, below] {\\ \\ $\zeta_{(1)}^\alpha(\dW,\dX,\dY,\dZ)$} (F.west);
		\draw[arrow] (F.east) -- (L.west);

		\coordinate (split) at ($(A.east)!0.5!(B.west)$);
		\draw[arrow, rounded corners=4pt] (split) |- (C.west);

		\coordinate (split2) at ($(F.east)!0.3!(L.west)$);
		\draw[arrow, rounded corners=4pt] (G.east) -| (split2);

		\draw[arrow, align=center] (C.east) -- node[lab, below] {\\ $\varphi_{(2)}^{\alpha}(\dI,\dW,\dY)$} (G.west);
		
		\node[node, align=center] (D) at (1.3,1.2) {$\dM \approx \dI$,~~$\dS \approx \dJ$ \\ PM expansion
		};
		\draw[arrow] (0.85,0.8) -- (0.85,0);
		
		\node[lab] (Inv1) at (10.9,1.2) {};
		\node[lab] (Inv2) at (6.5,-2.5) {};
		\node[lab] (Inv3) at (0.5,1.7) {};

		\draw[arrow, align=center, rounded corners=4pt]
		(B.north) -- (3,1.8)  --  node[lab,below] {\\ $\varphi_{(1)}^\alpha = \xi_{(1)}^\alpha + \zeta_{(1)}^\alpha$} (9,1.8)  -- (F.north);

		\draw[arrow, align=center, rounded corners=6pt]
		(A.north) -- (-1.2,3.1)  --  node[lab,below] {\\ $\varphi^\alpha=\varphi_{(1)}^\alpha+\varphi_{(2)}^\alpha +\mathcal{O}(G^3)$} (13.5,3.1)  -- (L.north);

		\draw[arrow, align=center] (R.east) -- node[lab, below] {\\ $\psi_A^{i} = \varphi_A^{i} - v_A^{i} \varphi_A^{0}/c$} (S.west);
		
		\begin{scope}[on background layer]
			\node[
			fill=yellow!10,
			draw,
			rounded corners,
			fit=(B) (C) (Inv3) (E) (F) (G) (Inv1) (Inv2),
			inner sep=12pt
			] {};
		\end{scope}
		
	\end{tikzpicture}
		\caption{Summary of the relations between the different quantities for the gauge transformation from BT to harmonic coordinates introduced  in Sec.~\ref{sec:BTtoharmonic}. The yellow box represents the total gauge transformation from the BT metric $h_{\text{BT}}^{\alpha \beta}$ to the harmonic metric $h^{\alpha \beta}$ and is composed of three parts $\xi^{\alpha}_{(1)}$, $\zeta^{\alpha}_{(1)}$ and $\varphi_{(2)}^{\alpha}$ given respectively by \eqref{xi1}, \eqref{gaugezetaasym2} and \eqref{2PMcontributionShift}. The coordinate shift $\psi^{i}$ is then obtained from the total gauge transformation $\varphi^{\alpha}$.}
        \label{fig:summary_gauge_transformation}
	\end{figure}

We have obtained in~\cite{BFT24} the RR acceleration at 4.5PN order in extended Burke-Thorne (BT) coordinates. We shall now determine a coordinate shift and the corresponding ``contact'' transformation of the particles' trajectories so that the 4.5PN RR acceleration is now in harmonic coordinates. Here by harmonic coordinates, we refer to the unique coordinate system covering the source (and valid everywhere outside the source) satisfying $\partial_\beta h^{\alpha\beta}=0$, as investigated in Sec.~\ref{sec:metricharmonic}. We recall that the transformation of the gothic metric deviation under a coordinate change $x^\alpha\longrightarrow {x'}^\alpha(x)$ reads
\begin{align}
\label{Transformation_metric_BT_to_harmonic}
	h^{\alpha \beta}(x') = \frac{1}{\vert J\vert} \frac{\partial {x'}^{\alpha}}{\partial {x}^{\gamma}} \frac{\partial {x'}^{\beta}}{\partial {x}^{\lambda}} \bigl(h^{\gamma \lambda}_{\text{BT}}(x) + \eta^{\gamma \lambda}\bigr) - \eta^{\alpha \beta}\,,
\end{align}
where $\{x^\alpha\}$ corresponds to the (initial) BT coordinates and $\{{x'}^\alpha\}$ to the final harmonic coordinates, and where \mbox{$J \equiv \text{det}(\partial x'/ \partial x)$} is the determinant of the Jacobian matrix of the transformation. In both sides of~\eqref{Transformation_metric_BT_to_harmonic}, the metrics are assumed to contain the same conservative parts and to differ only through the dissipative RR terms at orders 2.5PN, 3.5PN and 4.5PN. We look for a coordinate shift $\varphi^{\alpha}$ defined by ${x'}^{\alpha} = x^{\alpha} + \varphi^{\alpha}(x)$ and which contains only dissipative RR terms up to 4.5PN order. The modification of the BT metric thus contains only RR terms up to that order, namely
\begin{align}
	\label{Any_PM_metric_harmonic_to_BT}
	h^{\alpha \beta}(x) = h_{\text{BT}}^{\alpha \beta}(x) + \delta_{\varphi} h_{\text{BT}}^{\alpha \beta}(x)\,,
\end{align}
where $\delta_{\varphi} h_{\text{BT}}^{\alpha \beta}$ can be computed perturbatively\footnote{Since we work perturbatively for both $\varphi^{\alpha}$ and the metric, we can formally Taylor-expand the left-hand side of Eq.~\eqref{Transformation_metric_BT_to_harmonic} as 
\begin{align*}
		h^{\alpha \beta}(x') = \sum_{n=0}^{+\infty} \frac{1}{n!} \varphi^{\lambda_{1}} \!\cdots \varphi^{\lambda_{n}} \partial_{\lambda_{1}} \!\cdots \partial_{\lambda_{n}} h^{\alpha \beta}(x) \,.
\end{align*}} using Eq.~\eqref{Transformation_metric_BT_to_harmonic}. To linear order, the metric shift then reads \mbox{$\partial\varphi^{\alpha\beta}\equiv\partial^\alpha\varphi^\beta+\partial^\beta\varphi^\alpha-\eta^{\alpha\beta}\partial_\gamma\varphi^\gamma$,} thus we naturally pose~\cite{BFIS08,BFL22,TLB22}
\begin{align} \label{eq:gauge_transform}
\delta_{\varphi} h_{\text{BT}}^{\alpha \beta} = \partial \varphi^{\alpha \beta} + \Omega^{\alpha \beta}[\varphi, h]\,,
\end{align}
where $\Omega^{\alpha \beta}$ denotes a functional of the shift and the metric which represents the difference between a linear gauge transformation and a full non-linear coordinate transformation. Since the shift is composed only of RR terms up to 4.5PN order, the metric in the second slot of $\Omega^{\alpha \beta}[\varphi, h]$ can be approximated by the conservative part up to 2PN order (with the current approximation), which is identical in both the BT and harmonic coordinate systems.

Working initially in the exterior part of the source (before matching to the PN expansion inside the source),
we recall that $h^{\alpha\beta}$ satisfies the harmonic gauge condition not only in the ``unique'' harmonic coordinate system, but also in the BT gauge. 
However, unlike the ``true'' harmonic metric which satisfies $\partial_\beta h^{\alpha\beta}=0$ everywhere, the BT metric cannot be matched to the inner PN metric in harmonic coordinates without performing first a coordinate transformation. Hence, in the source's exterior, applying the harmonic gauge condition to both metrics and remembering the identity $\partial_{\beta} \partial \varphi^{\alpha \beta} = \Box \varphi^{\alpha}$, we get
\begin{align}
	\label{Relation_psi_omega_any_PM}
	\Box \varphi^{\alpha} + \partial_{\beta} \Omega^{\alpha \beta} = 0\, .
\end{align}

In practice, we look for the solution for $\varphi^\alpha$ in the form of a non-linearity (or PM) expansion. In this paper we shall need only the linear and quadratic contributions to $\varphi^\alpha$, namely
\begin{align}
	\label{expphi}
   \varphi^\alpha = G \varphi_{(1)}^\alpha + G^2 \varphi_{(2)}^\alpha + \mathcal{O}(G^3)\,.
\end{align}
Expanding the two metrics similarly, we obtain the hierarchy of relations
\begin{subequations}\label{expquadratic}
	\begin{align}
		h^{\alpha \beta}_{(1)} &= h^{\alpha \beta}_{\text{BT}\,(1)} + \partial \varphi_{(1)}^{\alpha \beta}\,,\\
		h^{\alpha \beta}_{(2)} &= h^{\alpha \beta}_{\text{BT}\,{(2)}} + \partial \varphi_{(2)}^{\alpha \beta} + \Omega_{(2)}^{\alpha \beta}[\varphi_{(1)}, h_{(1)}]\,,\\[-0.1cm]&\cdots\nn
	\end{align}
\end{subequations}
where the quadratic piece $\Omega_{(2)}^{\alpha \beta}$ is found by induction, as it depends both on the dissipative RR gauge vector $\varphi_{(1)}$, which we shall specify below, and on $h_{(1)}$, which is the conservative (2PN) part of the linear metric. The above procedure is easily extended to any non-linear order and applied to various cases~\cite{BFIS08,BFL22,TLB22}.

\subsection{The linear gauge transformation}

We specify here the linear gauge vector $\varphi_{(1)}^\alpha$. As we have seen in Sec.~\ref{sec:metricharmonic}, the harmonic metric (in the external zone of the source) depends on six families of multipole moments: two source moments $\dI_L$ and $\dJ_L$ (we can ignore the Weyl moment which has no pole), 
and four gauge moments $\dW_L$, $\dX_L$, $\dY_L$ and $\dZ_L$. All these moments are given in $d$ dimensions by Eqs.~\eqref{eq:sourcemts} and~\eqref{eq:gaugemts}. On the other hand the RR part of the BT metric computed in~\cite{BFT24} is a functional of only two ``canonical'' moments, $\dM_L$ and $\dS_L$. However, we know that the canonical moments differ from the source moments only by small 2.5PN corrections, namely $\dM_L=\dI_L+\mathcal{O}(c^{-5})$ and $\dS_L=\dJ_L+\mathcal{O}(c^{-5})$~\cite{BFIS08}, which contribute to the acceleration only at 5PN order. Hence, up to 4.5PN order, we simply replace the canonical moments $\dM_L$ and $\dS_L$ in the end result of~\cite{BFT24} by the source moments $\dI_L$ and $\dJ_L$, which defines our initial BT metric. Next, we have to \textit{undo} the linear gauge transformation which was applied in Eqs.~(3.2) of~\cite{BFT24}, by applying
\begin{subequations}\label{xi1}
	\begin{align}
		\xi_{(1)}^{0} =& -\frac{2}{c}\sum_{\ell=2}^{+\infty}
		\frac{(-)^\ell}{\ell !} \frac{2 \ell +1}{\ell (\ell -1)} \partial_L \left\{ \frac{\dI_{L}^{(-1)}(t-r/c)-\dI_{L}^{(-1)}(t+r/c)}{2r}\right\}\,, \\
		\xi_{(1)}^{i} ={}& 2 \sum_{\ell=2}^{+\infty} \frac{(-)^{ \ell}}{\ell !} \frac{(2 \ell + 1)(2 \ell +3)}{\ell (\ell-1)} \partial_{i L} \left\{\frac{\dI_{L}^{(-2)}(t-r/c)-\dI_{L}^{(-2)}(t+r/c)}{2r}\right\} \nn \\
		&- \frac{4}{c^2}\sum_{\ell=2}^{+\infty} \frac{(-)^{ \ell}}{\ell !} \frac{2 \ell + 1}{\ell-1} \partial_{L-1} \left\{\frac{\dI_{iL-1}(t-r/c)-\dI_{iL-1}(t+r/c)}{2r}\right\} \nn \\
		&- \frac{4}{c^2}\sum_{\ell=2}^{+\infty} \frac{(-)^{ \ell} \ell}{(\ell+1)!} \frac{2 \ell + 1}{\ell-1} \varepsilon_{iab} \partial_{aL-1} \left\{\frac{\dJ_{bL-1}^{(-1)}(t-r/c)-\dJ_{bL-1}^{(-1)}(t+r/c)}{2r}\right\} \,,
	\end{align}
\end{subequations}
which is defined from the antisymmetric wave component (retarded minus advanced) appropriate to RR effects. Note that in Eqs.~\eqref{xi1} we use the ordinary 3-dimensional notation, because at the relative 2PN level, the source moments $\dI_L$ and $\dJ_L$ have no poles. 

The expressions~\eqref{xi1} were originally constructed in the exterior zone of an isolated source using the MPM formalism~\cite{B93}. But since they are regular functions when $r\to 0$ they can be extended by matching to the full near zone, even including the region inside the matter source. In this paper we shall always implicitly consider that expressions such as~\eqref{xi1} are actually extended to the near zone. We can even think of them as formal regular (PN) expansions when $r\to 0$, and in principle we should add an overbar to mean the formal near zone expansion as in Eq.~\eqref{ULhom} below. In Appendix~\ref{appendix:innerzone} we present a different point of view (but equivalent concerning RR effects) in which the derivation is performed in the near zone, taking into account the stress-energy tensor of the matter source. Accordingly we systematically add the overbar indicating the near zone expansion in Appendix~\ref{appendix:innerzone}.

The second thing we have to do in order to construct the harmonic coordinates is to add the gauge transformation associated with the gauge moments $\dW_L$, $\dX_L$, $\dY_L$ and $\dZ_L$ and given by~\eqref{eq:gaugemts}. Here it is crucial to keep the expression in $d$ dimensions because the moment $\dY_L$ will turn out to have a pole $\propto 1/\varepsilon$ already at 2PN order. Consistently with RR effects, we must consider the antisymmetric component and hence we define [with a slight abuse of notation with respect to~\eqref{gaugezetaret} where we still denote $\zeta$ the transformation associated with antisymmetric moments],
\begin{subequations}\label{gaugezeta}
	\begin{align}
		& \zeta^0_{(1)} = \frac{4}{c^3}\sum_{\ell=0}^{+\infty} \frac{(-)^\ell}{\ell!}\,\partial_L \Wt_{\text{asym}\,L}\, ,\\
		& \zeta^i_{(1)} =  -\frac{4}{c^4}\sum_{\ell=0}^{+\infty} \frac{(-)^\ell}{\ell!}\,\partial_{iL} \Xt_{\text{asym}\,L}
		- \frac{4}{c^4}\sum_{\ell=1}^{+\infty} \frac{(-)^\ell}{\ell!}\,\biggl[ \partial_{L-1} \Yt_{\text{asym}\,iL-1} + \frac{\ell}{\ell+1}\partial_{L-1}
		\Zt_{\text{asym}\,i|L-1}\biggr]\,,
\end{align}\end{subequations}
where we have defined the antisymmetric part of the multipole moments in $d$ dimensions as $\It_{\text{asym}\,L}\equiv\frac{1}{2}\bigl(\It_{\text{ret}\,L}-\It_{\text{adv}\,L}\bigr)$ and the retarded and advanced pieces are given by~\eqref{notationtilde}.

In Sec.~\ref{sec:MultipoleMoments}, we shall show that the gauge moments $\dY_L$ develop a pole at 2PN order; actually, for the 4.5PN RR level we aim for, we  only need the dipole gauge moment $\dY_{i}$ at relative 2PN order. Thus, in the following, we shall split the moment $\dY_{i} = \dY_{i}^{\text{3D}} + \dY_{i}^{\text{pole}}$, where $\dY_{i}^{\text{pole}}$ is a pole part that enters at 2PN order and includes all the contributions proportional to $\varepsilon^{-1}$ (as well as some associated regular $\mathcal{O}(\varepsilon^0)$ contribution which are chosen to simplify the expression), and where $\dY_{i}^{\text{3D}}$ is a regular part which can be treated as an ordinary 3-dimensional (3D) quantity. 
Hence, for the purpose of computing the 4.5PN RR acceleration, we can single out the pole contribution due to $\dY_{i}^{\text{pole}}$, and write 
\begin{subequations}\label{gaugezetaasym2}
	\begin{align}
		\zeta^0_{(1)} &= \frac{4}{c^3}\sum_{\ell=0}^{+\infty} \frac{(-)^\ell}{\ell!}\,\partial_L \left\{\frac{\dW_{L}(t-r/c)-\dW_{L}(t+r/c)}{2r}\right\}\, ,\\
		\zeta^i_{(1)} &=  \frac{4}{c^4} \biggl[\Yt_{\text{asym}\,i}^{\text{pole}} + \frac{\dY_{i}^{\text{3D}}(t-r/c)-\dY_{i}^{\text{3D}}(t+r/c)}{2r}\biggr] \nn\\ & -\frac{4}{c^4}\sum_{\ell=0}^{+\infty} \frac{(-)^\ell}{\ell!}\,\partial_{iL} \left\{\frac{\dX_{L}(t-r/c)-\dX_{L}(t+r/c)}{2r}\right\}\nn\\
		&- \frac{4}{c^4}\sum_{\ell=2}^{+\infty} \frac{(-)^\ell}{\ell!}\,\partial_{L-1} \left\{\frac{\dY_{iL-1}(t-r/c)-\dY_{iL-1}(t+r/c)}{2r}\right\} \nn\\
		&- \frac{4}{c^4}\sum_{\ell=1}^{+\infty} \frac{(-)^\ell}{\ell!}\, \frac{\ell}{\ell+1}\varepsilon_{iab}\partial_{aL-1}
		\left\{\frac{\dZ_{bL-1}(t-r/c)-\dZ_{bL-1}(t+r/c)}{2r}\right\}\,,
\end{align}\end{subequations}
where the PN expansion of the pole part $\Yt_{\text{asym}\,i}^{\text{pole}}$ is explicitly given by Eq.~(A13) in~\cite{BBBFMc} and reads
\begin{align}
	\Yt_{\text{asym}\,i}^{\text{pole}} = -\frac{\Gamma(\frac{1+\varepsilon}{2})}{4 c \Gamma(1-\frac{\varepsilon}{2})} \sum_{j=0}^{\infty}\frac{{(r/c)}^{2j}}{2^{2j} j!\Gamma(j+\frac{3+\varepsilon}{2})} \left(\frac{\dd}{\dd t}\right)^{2j+2}\int_{0}^{+\infty} \dd\tau {\Big(\frac{c \tau \sqrt{\pi}}{\ell_{0}}\Big)}^{-\varepsilon}\Bigl[\dY_{i}^\text{pole}(t-\tau)-\dY_{i}^\text{pole}(t+\tau)\Bigr]\,.
\end{align}
The pole in $\dY_{i}$ enters the expression at 2PN order. Hence we can restrict ourselves to the leading PN order and neglect any term $\mathcal{O}(\varepsilon)$. This yields
\begin{align}\label{Yitildepole}
	\Yt_{\text{asym}\,i}^{\text{pole}} = \frac{1}{c} \Bigl(-1+\varepsilon\Bigl[ 1+\frac{\gamma_{\text{E}}}{2}\Bigr] \Bigr)\frac{\dd \dY_{i}^\text{pole}}{\dd t} + \frac{\varepsilon}{2 c } \frac{\dd^2}{\dd t^2} \int_{0}^{+\infty} \dd \tau \ln{\Big(\frac{c \tau \sqrt{\pi}}{\ell_{0}}\Big)}\Bigl[\dY_{i}^\text{pole}(t-\tau)-\dY_{i}^\text{pole}(t+\tau)\Bigr] + \mathcal{O}(\varepsilon)\, .
\end{align}
Finally, the looked-for linearized gauge transformation $\varphi_{(1)}^\alpha$ in Eqs.~\eqref{expphi}--\eqref{expquadratic} which will permit constructing the harmonic coordinates RR terms is obtained by summing the two previously constructed contributions, namely
\begin{align}
 	\label{linearvarphi1}
 	\varphi_{(1)}^\alpha = \xi_{(1)}^\alpha + \zeta_{(1)}^\alpha\,.
\end{align}

\subsection{Non-linear correction to the gauge transformation} \label{NL_gauge}

The non-linear part of the calculation can be done in 3 dimensions. The quadratic piece of the shift vector, $\varphi_{(2)}^\alpha$, is computed by solving the wave equation 
\begin{align}
	\label{box equation}
	\Box \varphi_{(2)}^{\alpha} = \Delta_{(2)}^\alpha \,,
\end{align}
which is the consequence of~\eqref{Relation_psi_omega_any_PM}, where the source term  $\Delta_{(2)}^\alpha = -\partial_{\beta} \Omega_{(2)}^{\alpha \beta}$ reads explicitly\footnote{Recall from Refs.~\cite{BFIS08,BFL22,TLB22} that
\begin{align*}
\Omega_{(2)}^{\alpha \beta} &= \partial_{\gamma}\bigl[\varphi_{(1)}^{\gamma}\bigl(h^{\alpha \beta}_{(1)} + \partial \varphi_{(1)}^{\alpha \beta}\bigr)\bigr] + 2 \partial_{\gamma} \varphi_{(1)}^{(\alpha} h^{\beta) \gamma}_{(1)} 
+ \partial^{\gamma} \varphi_{(1)}^{(\alpha} \partial_{\gamma} \varphi_{(1)}^{\beta)} + \frac{1}{2} \eta^{\alpha \beta}\bigl[\partial_{\gamma} \varphi _{(1)}^{\lambda} \partial_{\lambda} \varphi_{(1)}^{\gamma} - \partial_{\gamma} \varphi_{(1)}^{\gamma} \partial_{\lambda} \varphi_{(1)}^{\lambda}\bigr]\,.
\end{align*}}
\begin{align}\label{Delta2}
	\Delta_{(2)}^{\alpha}& = - h^{\beta \gamma}_{(1)} \partial_{\beta \gamma}	\varphi_{(1)}^{\alpha}\, .
\end{align}
We obtain $\varphi_{(2)}^{\alpha}$ directly in the form of a near-zone (or PN) expansion when $r \to 0$, starting from the near-zone expansion of the source term $\Delta_{(2)}^{\alpha}$. Denoting retarded time as $u\equiv t-r/c$ and defining $\hat{n}_{L}$ as the STF projection of the product of $\ell$ unit vectors $n^i=x^i/r$, let us recall~\cite{BD89,B93} that for the general d'Alembertian equation 
\begin{align}\label{Boxeq}
	\Box \varphi(\mathbf{x},t) = \hat{n}_{L} S(r,u)\,,
\end{align}
where the source term has multipolarity $\ell$ and tends to zero sufficiently rapidly when $r\to 0$, the near-zone expansion of the solution is given by 
\begin{align}
	\label{general_solution_inverse_Bow}
	\varphi(\mathbf{x},t) = \hat{\partial}_{L}\biggl\{\frac{\overline{G(t-r/c)-G(t+r/c)}}{r}\bigg\} + \Box^{-1}_{\text{inst}}\Bigl[\hat{n}_{L} \overline{S(r, u)}\Bigr]\,,
\end{align}
where we have added some overbars to insist that the result is valid in the sense of an expansion when $r\to 0$. The first term constitutes a homogeneous solution and can be expressed as the near-zone expansion of a ``retarded minus advanced'' multipolar wave, parametrized by the function
\begin{subequations}\label{fonctionG}
\begin{align}
	G(v) &= c \int_{-\infty}^{v} \dd s \,R\biggl(\frac{v-s}{2},s\biggr)\,,\\
	\text{where}\quad R(\rho,s) &= \rho^{\ell} \int_{0}^{\rho} \dd \lambda \,\frac{{(\rho-\lambda)}^{\ell}}{\ell !} \left(\frac{2}{\lambda}\right)^{\ell-1} S(\lambda,s)\,.
\end{align}
\end{subequations}
For the case at hand, we apply the previous formulas to a generic term in the PN expansion, say\footnote{\label{foot:FP} As usual (see \textit{e.g.}~\cite{B93}) we must introduce a regularization factor $\widetilde r^B = (r/r_0)^B$ to ensure that $S$ tends to zero more rapidly than any given power when $r\to 0$. In the end we have to apply the finite part procedure $\FP$.}
\begin{align}\label{Sourcedecompositionoverk}
	S(r,u) = \frac{F(t)}{r^k} = \frac{F(u+r/c)}{r^k}\, .
\end{align}
The function $F(t)$ is a quadratic product of two multipole moments, and $k\in\mathbb{Z}$. By working out Eqs.~\eqref{fonctionG} to this case, performing appropriate change of variables and integrations by part, we obtain a nice ``resummed'' expression for the function $G$ in the form of
\begin{subequations}\label{resultG}
\begin{align}
	G(v)= \FP \,C^{B}_{k,\ell} \int_{- \infty}^{v} \dd \tau \,\frac{F^{(k- \ell -2)}(\tau)}{c^{k-\ell-3}} \,\Big(\frac{v-\tau}{r_{0}}\Big)^{B}\,.
\end{align}
Note that the function $F$ can be time-differentiated or time-integrated multiple times inside the integral, depending of the sign of $k-\ell-2$ which can be positive or negative. The formula involves the closed-form ``factorized'' coefficient 
\begin{align}
	C^{B}_{k,\ell} = (-)^k \sqrt{\pi} \frac{2^{B-k+2} \,\Gamma(1-B)}{B \,\Gamma(\frac{\ell + k -B}{2})\Gamma(\frac{k-\ell -1 -B}{2})} \, .
\end{align}
\end{subequations}

The homogeneous solution, \textit{i.e.}, the first term in~\eqref{general_solution_inverse_Bow}, is given by
\begin{align}\label{ULhom}
	 \varphi^\text{hom} \equiv \hat{\partial}_{L}\biggl\{\frac{\overline{G(t-r/c)-G(t+r/c)}}{r}\bigg\} = - 2\hat{x}_L \sum_{p=0}^{+\infty}\frac{r^{2p}}{(2p)!!(2p+2\ell+1)!!} \frac{G^{(2\ell+2p+1)}(t)}{c^{2\ell+2p+1}}\,,
\end{align}
but we will now show that it actually does not contribute to RR effects, and is instead associated with purely conservative effects. To prove this, we simplify need to count the relevant powers of $1/c$. Hence we use the symbol $\sim$ to sketch only the generic structure of a term in the PN expansion, setting to one all numerical factors, and showing only the relevant power of $1/c$. From~\eqref{ULhom}, we have the structure
\begin{subequations}\label{ULhomsim}
	\begin{align}
	\varphi^\text{hom} \sim \frac{\hat{x}_L r^{2p}}{c^{2\ell+2p+1}} \,G^{(2\ell+2p+1)}(t)\,.
\end{align}
Combining it with Eqs.~\eqref{resultG} we obtain 
\begin{align}\label{ULhomsim2}
	\varphi^\text{hom} \sim \frac{\hat{x}_L r^{2p}}{c^{k+\ell+2p-2}} \int_{-\infty}^t \dd\tau \,(t-\tau)^B F^{(k+\ell+2p-1)}(\tau)\,.
\end{align}
\end{subequations}
In practice, we decompose the source term $\Delta_{(2)}^{\alpha}$ given by~\eqref{Delta2} in the near zone to match the form given by Eq.~\eqref{Boxeq}. We want to prove that the homogeneous solution~\eqref{ULhomsim2} has a half-integer PN order in the case of the zeroth component of the source term $\Delta_{(2)}^{0}$, and an integer PN order in the case of the spatial component $\Delta_{(2)}^{i}$. 

Let us take the example of a particular term in $\Delta_{(2)}^{0}$, say $h_{(1)}^{00}\partial_{00}\varphi_{(1)}^{0}$ (where $\partial_0=\frac{1}{c}\partial_t$). For $h_{(1)}^{00}$, we insert a typical multipolar retarded wave with source moment $\dI_L$ while for $\varphi_{(1)}^{0}$ we insert a typical antisymmetric (retarded minus advanced) wave, for instance with moment $\dI_N$ corresponding to the part~\eqref{xi1} of the gauge vector.
After the PN expansion, and taking into account the dimensionality of the term, the structure is
\begin{subequations}\label{struct}
	\begin{align}
	h_{(1)}^{00} &\sim \frac{1}{c^2}\partial_L\Bigl[r^{-1}\dI_L(t-r/c)\Bigr] \sim \frac{r^{j-\ell-1}}{c^{j+2}} \,n_L \dI_L^{(j)}(t)\,,\\
	\varphi_{(1)}^{0} &\sim \frac{x^N}{c^{2n+2}} \,\dI_N^{(2n)}(t)\,.
    \end{align}
\end{subequations}
The multipolarity of the source in Eq.~\eqref{Boxeq} will be $\ell+n-2s$ (where $s\in\mathbb{N}$) from the identity $\hat{n}^{L}\hat{n}^{N} \sim \sum_{s} \hat{n}^{L+N-2 S}$. Therefore, since we aim at matching a term with multipolarity $\ell$, it is necessary to perform the change of notation $\ell\longrightarrow \ell-n+2s$. Next, we can identify the power $k$ in~\eqref{Sourcedecompositionoverk} as being $k=\ell-j-2n+2s+1$. Re-expressing the result in terms of $k$ rather than $j$, we obtain
\begin{align}
	F(t) \sim \frac{1}{c^{\ell-k+2s+7}} \,\dI_{L-N+2S}^{(\ell-k-2n+2s+1)}(t) \,\dI_N^{(2n+2)}(t)\,.
\end{align}
Plugging this into~\eqref{ULhomsim2} leads to the looked-for structure of the homogeneous solution:
\begin{align}\label{structvarphi}
	\varphi^\text{hom} \sim \frac{\hat{x}_L r^{2p}}{c^{2\ell+2p+2s+5}} \int_{-\infty}^t \dd\tau \,(t-\tau)^B \,\bigl[\dI_{L-N+2S}\,\dI_N\bigr]^{(2\ell+2p+2s+2)}(\tau)\,.
\end{align}
The notation we use for the time derivatives means that they are to be distributed in all possible combinations, possibly including anti-derivatives, among the two moments $\dI_{L-N+2S}$ and $\dI_N$. Finally, we see from~\eqref{structvarphi} that we are left with an odd power of $1/c$ in the case of the zeroth component of the source term; therefore, as discussed above, the term in question is conservative and will be ignored in our RR calculation. We have checked that the same reasoning yields the same conclusion for all the contributions\footnote{In the $\Delta_{(2)}^{i}$ case, we obtain an even power of $1/c$, which leads to the same conclusion.} in $\Delta_{(2)}^{0}$ and $\Delta_{(2)}^{i}$. 

The second term in~\eqref{general_solution_inverse_Bow} is a particular solution of the wave equation and is defined with the so-called operator of the ``instantaneous'' potentials, namely
\begin{align}\label{inst}
	\Box^{-1}_{\text{inst}}\bigl[\hat{n}_{L} S\bigr] \equiv \sum_{i=0}^{+\infty} {\Big(\frac{\partial}{c \partial{t}}\Big)}^{2i} \Delta^{-1-i}\bigl[\hat{n}_{L} S\bigr]\,,
\end{align} 
where $\Delta^{-1-i}$ is the iterated Poisson integral (or inverse Laplacian). The computation of this part is straightforward and yields the quadratic contributions to the coordinate shift at 4.5PN order (with $\chi \equiv G m_1 r_1 + G m_2 r_2$),
\begin{subequations}\label{2PMcontributionShift}
\begin{align}
	\varphi_{(2)}^{0} &= -\frac{8\,\chi}{c^8} \,\dW^{(3)}  \, , \\
	\varphi_{(2)}^{i} &= -\frac{G r_{1} m_{1}}{c^9}\Big( 8  \dY^{(3)}_{i} - 4  v_{1}^{a} \dI^{(4)}_{ia} - 2 y_{1}^{a} \dI^{(5)}_{ia} -  r_{1} n_{1}^{a} \dI^{(5)}_{ia}  \Big) + (1 \leftrightarrow 2) \, .
\end{align}
\end{subequations}
In addition to the control of the coordinate transformation $\varphi_{(2)}^{\alpha}$ we have to worry about the change in the expressions of the multipole moments in the two coordinate systems. In Appendix~\ref{appendix:innerzone} (following the method of~\cite{BFIS08,BFL22,TLB22}) we shall check that the redefinitions of multipole moments (source type and gauge type) from extended BT to harmonic coordinates do not affect any of our results at 4.5PN order. A simple way to see that is to note that any relevant modification of the moments will be at quadratic order, and that any quadratic product of source and gauge moments produces terms having an odd PN order $c^{-3}$, $c^{-5}$, $\cdots$ (as is readily seen from dimensionality). When inserted into the RR acceleration these terms become ``even'' and do not have to be considered here.

Finally, the shift of the trajectory of the particle~1 is given by applying to the position 1 the complete gauge vector~\eqref{expphi}. Denoting the gauge vector at the position of particle~1 as $\varphi_1^\alpha=(\varphi_1^0,\varphi_1^i)$, the trajectory's shift (or contact transformation) is given by 
\begin{subequations}\label{contact}
\begin{align}
	\psi_1^{i} = \varphi_1^{i} - \frac{v_1^{i}}{c} \varphi_1^{0} + \mathcal{O}(\varphi_1^2)\,,
\end{align}
where the remainder is of the order of the square of RR effects which is negligible for the present purpose. The contact transformation will modify the acceleration of the particle~1 in the way
\begin{align}\label{contactacc}
	\delta_{\psi} a_{1}^{i} = \ddot{{\psi}_{1}^{i}} -\psi^{j}_{1}\frac{\partial a_{1}^{i}}{\partial y_1^{j}}-\psi^{j}_{2}\frac{\partial a_{1}^{i}}{\partial y_2^{j}}-\dot{\psi}^{j}_{1}\frac{\partial a_{1}^{i}}{\partial v_1^{j}}-\dot{\psi}^{j}_{2}\frac{\partial a_{1}^{i}}{\partial v_2^{j}} + \mathcal{O}(\psi_1^2)\,,
\end{align}
where the acceleration $a_1^i$ in the right-hand side represents the 2PN accurate conservative acceleration which is a functional of the positions $y_{1,2}^i$ and velocities $v_{1,2}^i$. Thus the change of acceleration~\eqref{contact} is what should be added to the BT 4.5PN RR acceleration (end result of~\cite{BFT24}) in order to obtain the 4.5PN RR harmonic acceleration, namely
\begin{align}
a_{\text{RR}1}^i\Big|_\text{harm} =
a_{\text{RR}1}^i\Big|_\text{BT} + \delta_{\psi} a_{1}^{i} \,.
\end{align}
\end{subequations}

\section{Computation of the multipole moments}
\label{sec:MultipoleMoments}

As we have seen in Sec.~\ref{sec:metricharmonic}, the RR acceleration in harmonic coordinates is parametrized by two source multipole moments $\dI_L$ and $\dJ_L$ given by Eqs.~\eqref{eq:sourcemts}, and also by four specific sets of gauge moments $\dW_L$, $\dX_L$, $\dY_L$ and $\dZ_L$ given by Eqs.~\eqref{eq:gaugemts}. The precision at which we need to compute these moments is dictated by the 4.5PN (or relative 2PN) accuracy of the final RR acceleration.
The source moments $\dI_L$ and $\dJ_L$ are well known from previous works\footnote{The ``Weyl'' moment in Eq.~\eqref{eq:sourcemts_KL} plays no role in our calculation.}: for instance the mass-type quadrupole $\dI_{ij}$ has been computed to 4PN order~\cite{MHLMFB20,LHBF22,LBHF22} and the current-type quadrupole $\dJ_{ij}$ is known to 3PN order~\cite{HFB21}. Thus we will focus on the computation of the gauge moments. For this computation, all the techniques are the same as for the source moments and have already been documented in previous papers, in particular in Ref.~\cite{MHLMFB20}. 

The main feature of the computation of the gauge moments is the appearance of a pole already at the 2PN order in the moments $\dY_L$, while such pole does not arise in the source moments before the 3PN order. This is the reason why we have reviewed in Sec.~\ref{sec:metricharmonic} the expressions of the moments in $d$ dimensions, which are required for dimensional regularization (DR). Furthermore, we need DR only to treat the 2PN-accurate moment $\dY_i$ (\textit{i.e.}, the dipolar case $\ell=1$). We briefly review the computation of the terms with non-compact support, as the ones with compact support (proportional to delta functions) are straightforward to obtain. Following~\cite{BDEI05dr}, we first compute the Hadamard regularization of the non-compact integral (more precisely the ``pure-Hadamard-Schwartz'' regularization~\cite{BDE04}) and then supplement it by the ``difference'' between DR and the Hadamard regularization, which consists of the pole $\propto 1/\varepsilon$ followed by the finite part $\propto\varepsilon^0$ (we always neglect terms vanishing in the limit $\varepsilon\to 0$). 

The integrand of a typical non-compact support term in $d$ dimensions is given by a function $F^{(d)}(\mathbf{x})$, which is smooth except at the source points $\bm{y}_1$ and $\bm{y}_2$ around which it admits a power-like singular expansion. In practice, this function results from the PN iteration parametrized by retarded potentials $V$, $V_i$, $\hat{W}_{ij}$, \textit{etc.}, as defined in the Appendix~\ref{app:ΣandPotentials}. When $r_1\equiv\vert\mathbf{x}-\bm{y}_1\vert\rightarrow 0$ and for any $N\in {\mathbb{N}}$, we have 
\begin{equation}\label{Fexpd}
	F^{(d)}(\mathbf{x}) = \sum_{\substack{p_0 \leqslant p
			\leqslant N}}\,\sum_{q_0 \leqslant q\leqslant q_1} r_1^{p+q\varepsilon}
	\ell_0^{-q\varepsilon}\mathop{f}_1{}_{p,q}^{(\varepsilon)}(\mathbf{n}_1) +
	o(r_1^{N})\,,
\end{equation}
where $\ell_0$ is the scale associated with DR, and where the coefficients $\mathop{f}_1\!{}_{p,q}^{(\varepsilon)}(\mathbf{n}_1)$, supposed to be finite in the limit $\varepsilon\to 0$, depend both on $\varepsilon = d-3$ and on \mbox{$\mathbf{n}_1=(\mathbf{x}-\bm{y}_1)/r_1$,} the unit direction of approach to the singularity. The powers of $r_1$ involve the integers $p,q \in \mathbb{Z}$ whose values are limited by some $p_0,\,q_0,\,q_1$ $\in\mathbb{Z}$ and $N$ as indicated. When $\varepsilon=0$, the corresponding function in 3 dimensions reads
\begin{equation}\label{Fexp}
F(\mathbf{x}) =
	\!\!\!\!\!  \sum_{p_0\leqslant p\leqslant N} \!\!\!  r_1^p
	\, \mathop{f}_{1}{}_{\!\!p}(\mathbf{n}_1)+o(r_1^{N})\,,\qquad\text{where}\qquad
		\mathop{f}_1{}_{\!p}(\mathbf{n}_1) = \sum_{q=q_0}^{q_1}\mathop{f}_1{}_{p,q}^{(\varepsilon=0)}
		(\mathbf{n}_1)\,.
\end{equation}
The generic non-compact contribution in the multipole (gauge) moment is the spatial integral of $F^{(d)}$ in the limit $\varepsilon\to 0$. To obtain it, one first evaluates the Hadamard partie finie (Pf) of the integral in 3 dimensions~\cite{BFreg}, which depends on two regularization scales $s_1$ and $s_2$, and can be given an explicit form based on a double analytic continuation with two complex parameters $\alpha$, $\beta\in {\mathbb C}$:
\begin{align}\label{intH}
	H \equiv \Pf \int \dd^3\mathbf{x} \, F(\mathbf{x}) = \mathop{\mathrm{FP}}_{\alpha \to 0 \atop
		\beta \to 0} \int \dd^3 \mathbf{x}
	\left(\frac{r_1}{s_1} \right)^\alpha \!\left(\frac{r_2}{s_2} \right)^\beta F \,.
\end{align}
The operation $\mathrm{FP}$ means taking the finite parts in the Laurent expansions when $\alpha\to 0$ and $\beta\to 0$ successively (in any order). Now the analogue of the term~\eqref{intH} in $d$ dimensions just reads
\begin{equation}\label{intHd}
	H^{(d)} = \int \dd^d\mathbf{x}\,F^{(d)}(\mathbf{x})\,,
\end{equation}
where the singularities at points $\bm{y}_1$ and $\bm{y}_2$ are treated with DR, evaluated close to the physical limit $\varepsilon=0$. As expected, this limit is related to the Hadamard partie finie~\eqref{intH}, but differs from it by the difference mentioned above \mbox{$\mathcal{D}H\equiv H^{(d)} - H$.} The result, keeping the pole part $\propto\varepsilon^{-1}$ and the finite part, is
\begin{subequations}\label{DH}
	\begin{align}
		\mathcal{D}H = \frac{1}{\varepsilon}\!\sum_{q_0\leqslant
			q\leqslant q_1}
		\biggl[\frac{1}{q+1}+\varepsilon \ln \left(\frac{s_1}{\ell_0}\right) \biggr]
		\bigl<\mathop{f}_1{}_{-3,q}^{(\varepsilon)}\bigr> + \calO(\varepsilon) + \big( 1 \leftrightarrow 2 \big)\,.
	\end{align}
The angular average is defined in $d$ dimensions as
\begin{equation}\label{IntAngul}
	\bigl<\mathop{f}_1{}_{p,q}^{(\varepsilon)}\bigr>
	\equiv\int\dd\Omega_1^{(d-1)} \mathop{f}_1{}_{p,q}^{(\varepsilon)}(\mathbf{n}_1)\,,
\end{equation}
\end{subequations}
where $\dd\Omega_1^{(d-1)}$ denotes the solid angle element around the direction $\mathbf{n}_1$.
As a check of the result~\eqref{DH}, the two Hadamard scales $s_1$ and $s_2$ in $H$ cancel out when adding the difference $\mathcal{D}H$, and $\ell_0$ is the only scale left over in the DR result. 

Another important check is made possible by the presence of many terms that are in the form of a total divergence; they can thus be computed either as a volume integral, as described previously, or as a surface integral at infinity. 
Both calculations agree with DR, but provided that we take into account the purely distributional part of the derivatives of potentials ($V$, $V_i$, \textit{etc.}) appearing in the non-compact terms in $d$ dimensions. For that purpose, we use the Schwartz distributional derivative~\cite{Schwartz}, or, equivalently, the Gel'fand-Shilov formula~\cite{gelfand} in $d$ dimensions. The distributional derivative of the function $F^{(d)}$, which admits the singular expansion~\eqref{Fexpd}, is 
\begin{equation}\label{guelfandshilov}
\partial_{i} F^{(d)} = \partial_{i} F^{(d)}\big|_\text{ord} + D_{i}\bigl[F^{(d)}\bigr]\,,
\end{equation}
where $\partial_{i} F^{(d)}\big|_\text{ord}$ is the ``ordinary'' part of the derivative (say, computed algorithmically by \texttt{Mathematica}), while the purely distributional part reads (with the limit $\varepsilon\to 0$ taken in the end)~\cite{gelfand}
\begin{equation}\label{distrpart}
	D_{i}\bigl[F^{(d)}\bigr] = \sum_{\ell=0}^{+\infty} \frac{(-)^\ell}{\ell!} \,\partial_L\delta^{(d)}(\mathbf{x}-\bm{y}_1)\,\langle n_1^{iL} \mathop{f}_1{}_{\!-\ell-2,-1}^{(\varepsilon)} \rangle + \big( 1 \leftrightarrow 2 \big)\,.
\end{equation}
Note that the only contributions to~\eqref{distrpart} come from the singular terms with powers $p = -\ell -2$ and $q=-1$. Moreover, as $p_0\leqslant p$ in~\eqref{Fexpd} the sum in~\eqref{distrpart} is actually finite since we have $\ell \leqslant -2 - p_0$. To compute the distributional part of the time derivatives we also need to define the partial derivatives with respect to $\bm{y}_{1}$ and $\bm{y}_{2}$,
\begin{equation}\label{distrpart1}
	\mathop{D_{i}}_{1}\bigl[F^{(d)}\bigr] = - \sum_{\ell=0}^{+\infty} \frac{(-)^\ell}{\ell!} \,\partial_L\delta^{(d)}(\mathbf{x}-\bm{y}_1)\,\langle n_1^{iL} \mathop{f}_1{}_{\!-\ell-2,-1}^{(\varepsilon)} \rangle \quad \text{and} \quad  \big( 1 \leftrightarrow 2 \big)\, ,
\end{equation}
and the distributional time derivative is given by
\begin{align}
    D_{t}\bigl[F^{(d)}\bigr]=v_{1}^{i} {\mathop{D_i}_{1}}\bigl[F^{(d)}\bigr]+v_{2}^{i} {\mathop{D_i}_{2}}\bigl[F^{(d)}\bigr]\, .
\end{align}
Typically, the distributional derivatives contributes when computing the second derivative of a potential. In that case, the distributional term is given by~\cite{BFreg}
\begin{equation}\label{distrpartij}
	D_{\alpha \beta}\bigl[F^{(d)}\bigr] = D_{\alpha}\bigl[\partial_\beta F^{(d)}\bigr] + \partial_\alpha D_{\beta}\bigl[F^{(d)}\bigr]\,.
\end{equation}
where $D_\alpha=(\frac{1}{c}D_t, D_i)$. In practice, we compute the distributional contributions of simple elementary terms $n_1^{L}/r_1^m$, which is given in 3 dimensions when $\ell+m$ is an odd integer by
\begin{align}
\label{Gelfand3d}
    D_{i}\Big(\frac{n_{1}^{L}}{r_1^m}\Big) = \frac{{4 \pi (-2)}^m (\ell+1)!(\frac{\ell + m -1}{2})!}{(\ell+m)!} \sum_{p=p_{0}}^{[m/2]} \frac{\Delta^{p-1} \partial_{(M-2P}{\delta}_{iL+2P-M)} \delta_1(\mathbf{x})}{4^p (p-1)!(m-2p)!(\frac{\ell+1-m}{2}+p)!} \quad \text{and} \quad  \big( 1 \leftrightarrow 2 \big)\, ,
\end{align}
and is zero when $\ell + m$ is even; note that we have denoted $p_{0}=\text{max}(1,\frac{m-\ell-1}{2})$ and introduce $\delta_{2K}\equiv \delta_{j_{1}j_{2}}\cdots\delta_{j_{2k-1}j_{2k}}$, the product of $K$ Kronecker symbols. This formula is then applied to the expansion of singular functions around the singularities $\bm{y}_{1}$ and $\bm{y}_{2}$.\footnote{Note that, if we apply the formula~\eqref{distrpart} with terms diverging like $1/r_1^3$ in 3 dimensions when $r_{1}\rightarrow 0$, we obtain that there is a distributional contribution. However, these terms can only appear from the derivative $\partial_i\hat{X}$ of the potential $\hat{X}$ defined in the Appendix~\ref{app:ΣandPotentials}, and come by construction from the product of three Newtonian-like potentials $\sim 1/r_1^{1+\varepsilon}$. Thus, in $d$ dimensions, this diverges like $1/r_1^{3+3\varepsilon}$ and does not generate a distributional contribution because the singularity corresponds to a value $q=-3$ in Eq.~\eqref{Fexpd}, whereas $q=-1$ is the only possibility leading to a non zero contribution in Eq.~\eqref{distrpart}. Nevertheless, in our case, we have checked that $\partial_i\hat{X}$ is the only term where the $d \rightarrow 3$ limit does not correspond to the distributional contribution computed with~\eqref{Gelfand3d} in 3 dimensions. We can therefore consider the distributional contribution $n_1^{L}/r_1^3$ as ordinary [\textit{i.e.}, $D_{i}(n_1^{L}/r_1^3)=0$] and continue using Eq.~\eqref{Gelfand3d}. This also highlights the importance of performing the full computation in $d$ dimensions.}

The gauge moments required for this calculations are $\dW$ and $\dY_i$ at 2PN order, $\dX$, $\dX_i$, $\dW_i$, $\dY_{ij}$ and $\dZ_i$ at 1PN order, and $\dW_{ij}$, $\dX_{ij}$, $\dY_{ijk}$ and $\dZ_{ij}$ at Newtonian order. All the results are provided in Appendix~\ref{app:jauge}, while we refer to previous papers for the source moments. Among those moments only one contains a pole: namely $\dY_i$ at 2PN order, which is therefore computed with the $d$-dimensional expressions~\eqref{eq:gaugemts}. Following~\eqref{gaugezetaasym2}, we split $\dY_i$ rather arbitrarily into (i) a 3D part, which is free of $1/\varepsilon$ poles;  and (ii) a pole part, which contains the contributions proportional to $1/\varepsilon$, but also, for convenience, some~$\mathcal{O}(\varepsilon^0)$ terms which yield logarithmic contributions and a hereditary integral in the limit where $\varepsilon \rightarrow 0$. The appearance of a pole at 2PN order in the gauge moments is expected since the terms in the $d$-dimensional expressions of $\dY^{i}$ given by (\ref{Yi_d_dim_∑}) are similar at 2PN order to those in the source terms at 3PN order which are known to generate poles, namely cubic terms in the PN potentials. Regarding the expression of the source quantities $\Sigma$ as a function of the PN potentials given in Appendix~\ref{app:ΣandPotentials}  several terms in $\Sigma^{ii}_{[1]}$ and $\dot{\Sigma}^i_{[1]}$ [defined in \eqref{exactform}] could, in principle,  produce a pole contribution at 2PN order. However, we find that the only pole contribution arises from a specific combination of potentials in $\Sigma^{ii}_{[1]}$,  given by \red{}

\begin{align}
	\frac{(d-1)}{8 c^4 (d-2) G \pi} \bigg[(d-1)V \Big(\frac{d}{d-2} V \partial^2_t V -4 \partial_t V_{i} \partial^i V \Big)+ \hat W_{ij}\Big((d-4)\partial^i V \partial^j V + 2 d V \partial^i \partial^j V\Big)\bigg] \in  \Sigma^{ii}_{[1]} \, .
\end{align}  
Thus, we have $\dY_i = \dY_i^\text{3D} + \dY_i^\text{pole}$ and our explicit calculation of the pole part yields 
\begin{align}\label{Yipole}
	\dY_{i}^{\text{pole}} = \frac{G^3 m_{1}^{3} m_{2}}{c^4}\frac{n_{12}^{i}}{r_{12}^{2}}\left[ - \frac{1}{\varepsilon} + 3 \ln\bigg(\frac{\sqrt{\bar{q}} \, r_{12}}{\ell_{0}}\bigg) \right] + \big( 1 \leftrightarrow 2 \big)\,.
\end{align}
Here we have introduced $\bar{q} \equiv 4\pi\de^{\gamma_\text{E}}$, where $\gamma_\text{E}$ is the Euler constant, and we recall that $\ell_0$ is the characteristic scale of DR. The 3D part is given at 1PN order by Eq.~\eqref{Yi3D}, and the complete expression is relegated to the Supplemental Material~\cite{SuppMaterial}. From~\eqref{gaugezetaasym2} and~\eqref{Yitildepole} we find that the associated contribution to the linear gauge transformation is
\begin{align}\label{zeta1pole}
	\zeta^\text{pole}_{(1)\,i} &=  \frac{4}{c^4} \Yt_{\text{asym}\,i}^{\text{pole}} = \frac{4}{c^5}\left\{\Bigl(-1+\varepsilon\Bigl[ 1+\frac{\gamma_{\text{E}}}{2}\Bigr] \Bigr)\frac{\dd \dY_{i}^\text{pole}}{\dd t} + \frac{\varepsilon}{2} \frac{\dd^2}{\dd t^2} \int_{0}^{+\infty} \dd \tau \ln{\Big(\frac{c \tau \sqrt{\pi}}{\ell_{0}}\Big)}\Bigl[\dY_{i}^\text{pole}(t-\tau)-\dY_{i}^\text{pole}(t+\tau)\Bigr]\right\}\, .
\end{align}
We note that this is just a function of time, hence the contribution to the acceleration is given by the second time derivative of the expression~\eqref{zeta1pole}, namely
\begin{align}\label{a1pole}
	a^\text{pole}_{\text{RR1}\,i} &= G \frac{\dd^2 \zeta^\text{pole}_{(1)\,i}}{\dd t^2} = \frac{4G}{c^5}\left\{\Bigl(-1+\varepsilon\Bigl[ 1+\frac{\gamma_{\text{E}}}{2}\Bigr] \Bigr)\frac{\dd^3 \dY_{i}^\text{pole}}{\dd t^3} + \frac{\varepsilon}{2} \frac{\dd^4}{\dd t^4} \int_{0}^{+\infty} \dd \tau \ln{\Big(\frac{c \tau \sqrt{\pi}}{\ell_{0}}\Big)}\Bigl[\dY_{i}^\text{pole}(t-\tau)-\dY_{i}^\text{pole}(t+\tau)\Bigr]\right\}\, .
\end{align}
Remembering that $\dY_{i}^\text{pole}$ is a 2PN term~\eqref{Yipole}, we find that the pole part of the acceleration~\eqref{a1pole} is a RR 4.5PN effect.

\section{Acceleration in a general frame}
\label{sec:acceleration_general}

\subsection{Radiation-reaction acceleration in a general frame}

Starting from the 4.5PN RR acceleration of body 1 in BT coordinates~\cite{BFT24},\footnote{Where, as we said, the canonical moments $\dM_L$, $\dS_L$ must be replaced by the source moments $\dI_L$, $\dJ_L$.} we implement the contact transformation~\eqref{contact}, in which the explicit expressions of all the source and gauge multipole moments are given in Appendix~\ref{app:jauge}, and obtain the final 4.5PN RR in harmonic coordinates in a general frame:  
\begin{align}\label{aRR1harm}
	a_{\text{RR}1}^i\Big|_\text{harm} =
	a^i_{\text{RR1 2.5PN}} + a^i_{\text{RR1 3.5PN}} + a^i_{\text{RR1 4PN}} + a^i_{\text{RR1 4.5PN}} + \mathcal{O}\left(\frac{1}{c^{11}}\right)\,.
\end{align}
We do not discuss how to obtain dissipative RR contribution~\eqref{aRR1_4PN_tail} due to the tail term at the 4PN order, since it is known from previous works, \textit{e.g.},~\cite{BBFM17}. We have two expressions for the RR acceleration: one is given in terms of the multipole (source and gauge) moments left in abstract form, and one where the moments are replaced by their expressions for two body systems. The former form, whose leading 2.5PN order matches with Eq.~\eqref{eq:aharm}, is relegated to Appendix~\ref{app:accMoments} for the 3.5PN expression; the full 4.5PN expression is provided in the Supplemental Material~\cite{SuppMaterial}. The latter form, which is much shorter than in BT coordinates, is given here in full; see also the Supplemental Material~\cite{SuppMaterial}. The 2.5PN and 3.5PN terms are already known from previous works~\cite{PW02, NB05} and read\footnote{As the RR acceleration in harmonic coordinates is manifestly Lorentz invariant (see Sec.~\ref{sec:Lorentz}) it is advantageous to express the results in terms of the Galileo invariant relative velocity $v_{12}^{i}\equiv v_{1}^{i}-v_{2}^{i}$.}
\begin{subequations}\label{RR25PN35PN}
\begin{align}
a^i_{\text{RR1 2.5PN}} &=\frac{G^2 m_{1} m_{2} v_{12}^{i}}{c^5 r_{12}^4} \biggl [\Bigl(\frac{8}{5} m_{1}
 -  \frac{32}{5} m_{2}\Bigr) G
 -  \frac{4}{5} r_{12} v_{12}^{2}\biggl]
 + \frac{G^2 m_{1} m_{2} n_{12}^{i}}{c^5 r_{12}^4} \biggl [G \Bigl(- \frac{24}{5} m_{1} (n_{12}{} v_{12}{})\nn\\
& + \frac{208}{15} m_{2} (n_{12}{} v_{12}{})\Bigr)
 + \frac{12}{5} r_{12} (n_{12}{} v_{12}{}) v_{12}^{2}\biggl]\, ,\\
a^i_{\text{RR1 3.5PN}} &=\frac{G^2 m_{1} m_{2} v_{12}^{i}}{c^7 r_{12}^5} \biggl(\Bigl(- \frac{184}{21} m_{1}^2
 + \frac{6224}{105} m_{1} m_{2}
 + \frac{6388}{105} m_{2}^2\Bigr) G^2
 + r_{12}^2 \Bigl(60 (n_{12}{} v_{12}{})^4
 -  \frac{4}{5} (v_{12}{} v_{2}{})^2\nn\\
& -  \frac{348}{5} (n_{12}{} v_{12}{})^2 v_{12}^{2}
 -  \frac{12}{5} (n_{12}{} v_{12}{}) (n_{12}{} v_{2}{}) v_{12}^{2}
 + \frac{6}{5} (n_{12}{} v_{2}{})^2 v_{12}^{2}
 + \frac{334}{35} v_{12}^{4}
 -  \frac{2}{5} v_{12}^{2} v_{2}^{2}\Bigr)\nn\\
& + G r_{12} \biggl [m_{1} \Bigl(\frac{52}{15} (n_{12}{} v_{12}{})^2
 + \frac{16}{5} (n_{12}{} v_{12}{}) (n_{12}{} v_{2}{})
 -  \frac{16}{5} (n_{12}{} v_{2}{})^2
 -  \frac{16}{5} (v_{12}{} v_{2}{})
 -  \frac{132}{35} v_{12}^{2}
 -  \frac{4}{5} v_{2}^{2}\Bigr)\nn\\
& + m_{2} \Bigl(\frac{454}{15} (n_{12}{} v_{12}{})^2
 -  \frac{208}{15} (n_{12}{} v_{12}{}) (n_{12}{} v_{2}{})
 + \frac{64}{5} (n_{12}{} v_{2}{})^2
 + \frac{64}{5} (v_{12}{} v_{2}{})
 -  \frac{152}{21} v_{12}^{2}
 + \frac{16}{5} v_{2}^{2}\Bigr)\biggl]\biggl)\nn\\
& + \frac{G^2 m_{1} m_{2} n_{12}^{i}}{c^7 r_{12}^5} \biggl(G^2 \biggl [- \frac{3172}{21} m_{2}^2 (n_{12}{} v_{12}{})
 + m_{1}^2 \Bigl(\frac{3992}{105} (n_{12}{} v_{12}{})
 -  \frac{16}{5} (n_{12}{} v_{2}{})\Bigr)\nn\\
& + m_{1} m_{2} \Bigl(- \frac{13576}{105} (n_{12}{} v_{12}{})
 + \frac{112}{15} (n_{12}{} v_{2}{})\Bigr)\biggl]
 + r_{12}^2 \Bigl(-56 (n_{12}{} v_{12}{})^5
 + \frac{12}{5} (n_{12}{} v_{12}{}) (v_{12}{} v_{2}{})^2\nn\\
& + 60 (n_{12}{} v_{12}{})^3 v_{12}^{2}
 - 6 (n_{12}{} v_{12}{}) (n_{12}{} v_{2}{})^2 v_{12}^{2}
 + \frac{12}{5} (n_{12}{} v_{12}{}) (v_{12}{} v_{2}{}) v_{12}^{2}
 + \frac{12}{5} (n_{12}{} v_{2}{}) (v_{12}{} v_{2}{}) v_{12}^{2}\nn\\
& -  \frac{246}{35} (n_{12}{} v_{12}{}) v_{12}^{4}
 + \frac{6}{5} (n_{12}{} v_{12}{}) v_{12}^{2} v_{2}^{2}\Bigr)
 + G r_{12} \biggl [m_{2} \Bigl(- \frac{582}{5} (n_{12}{} v_{12}{})^3
 -  \frac{208}{5} (n_{12}{} v_{12}{}) (n_{12}{} v_{2}{})^2\nn\\
& -  \frac{208}{15} (n_{12}{} v_{12}{}) (v_{12}{} v_{2}{})
 + \frac{208}{15} (n_{12}{} v_{2}{}) (v_{12}{} v_{2}{})
 + \frac{3568}{105} (n_{12}{} v_{12}{}) v_{12}^{2}
 -  \frac{104}{15} (n_{12}{} v_{12}{}) v_{2}^{2}\Bigr)\nn\\
& + m_{1} \Bigl(48 (n_{12}{} v_{12}{})^3
 + \frac{24}{5} (n_{12}{} v_{12}{})^2 (n_{12}{} v_{2}{})
 + \frac{72}{5} (n_{12}{} v_{12}{}) (n_{12}{} v_{2}{})^2
 + \frac{24}{5} (n_{12}{} v_{12}{}) (v_{12}{} v_{2}{})\nn\\
& -  \frac{24}{5} (n_{12}{} v_{2}{}) (v_{12}{} v_{2}{})
 -  \frac{4888}{105} (n_{12}{} v_{12}{}) v_{12}^{2}
 + \frac{8}{5} (n_{12}{} v_{2}{}) v_{12}^{2}
 + \frac{12}{5} (n_{12}{} v_{12}{}) v_{2}^{2}\Bigr)\biggl]\biggl).
\end{align}
\end{subequations}
At the 4PN order, the tail generates a dissipative hereditary contribution to the acceleration that reads~\cite{BBFM17,GLPR16,T25}
\begin{align}\label{aRR1_4PN_tail}
    a^i_{\text{RR1 4PN}} &= - \frac{4 G (m_{1}+m_{2})}{5 c^8} y_1^j \int_0^{+\infty}\dd \tau \, \ln\left(\frac{c \tau}{2 b_0}\right) \Big[\dI_{ij}^{(7)}(t-\tau)+\dI_{ij}^{(7)}(t+\tau)\Big]\,,
\end{align}
where we recall that $\dI_{ij} = m_1 y_1^{\langle i} y_1^{j \rangle} + (1\leftrightarrow2) + \mathcal{O}(2)$, and the time derivatives are taken using the Newtonian acceleration.
At the 4.5PN order and in harmonic coordinates, the RR acceleration contains a pole $\propto 1/\varepsilon$ (computed with DR), which can be traced back to the pole arising at 2PN relative order in the gauge dipole moment $\dY_i$, see Sec.~\ref{sec:metricharmonic}. Accordingly we split the 4.5PN acceleration into some 3D part (finite when $\varepsilon\to 0$) and the pole part:
\begin{align}\label{aRR1polenonpole}
	a^i_{\text{RR1 4.5PN}} = a^i_{\text{RR1 4.5PN 3D}} + a^i_{\text{RR1 4.5PN pole}}\,.
\end{align}
Again, in the pole part, we shall include (somewhat arbitrarily) an associated finite part in the form of instantaneous logarithmic terms as well as a hereditary integral. The 3D part reads
\begin{align}\label{eq:a1i_RR_4p5PN_3D}
a^i_{\text{RR1 4.5PN 3D}} &= \frac{G^2 m_{1} m_{2} v_{12}^{i}}{c^9 r_{12}^6} \Biggl\{\Bigl(\frac{106972}{567} m_{1}^3
 -  \frac{5242}{135} m_{1}^2 m_{2}
 -  \frac{706996}{945} m_{1} m_{2}^2
 -  \frac{1075618}{2835} m_{2}^3\Bigr) G^3\nn\\
&\qquad + G^2 r_{12} \biggl [m_{2}^2 \Bigl(- \frac{22384}{35} (n_{12}{} v_{12}{})^2
 + \frac{16196}{105} (n_{12}{} v_{12}{}) (n_{12}{} v_{2}{})
 -  \frac{16054}{105} (n_{12}{} v_{2}{})^2
 -  \frac{12776}{105} (v_{12}{} v_{2}{})\nn\\
&\qquad + \frac{7412}{21} v_{12}^{2}
 -  \frac{3194}{105} v_{2}^{2}\Bigr)
 + m_{1} m_{2} \Bigl(\frac{48749}{105} (n_{12}{} v_{12}{})^2
 + \frac{21716}{105} (n_{12}{} v_{12}{}) (n_{12}{} v_{2}{})
 -  \frac{5896}{35} (n_{12}{} v_{2}{})^2\nn\\
&\qquad -  \frac{12448}{105} (v_{12}{} v_{2}{})
 + \frac{13273}{105} v_{12}^{2}
 -  \frac{3112}{105} v_{2}^{2}\Bigr)
 + m_{1}^2 \Bigl(\frac{261323}{315} (n_{12}{} v_{12}{})^2
 -  \frac{1296}{35} (n_{12}{} v_{12}{}) (n_{12}{} v_{2}{})\nn\\
&\qquad + \frac{2972}{105} (n_{12}{} v_{2}{})^2
 + \frac{368}{21} (v_{12}{} v_{2}{})
 -  \frac{55477}{315} v_{12}^{2}
 + \frac{92}{21} v_{2}^{2}\Bigr)\biggl]
 + G r_{12}^2 \biggl [m_{1} \Bigl(\frac{4073}{35} (n_{12}{} v_{12}{})^4\nn\\
&\qquad + \frac{276}{5} (n_{12}{} v_{12}{})^3 (n_{12}{} v_{2}{})
 -  \frac{64}{5} (n_{12}{} v_{12}{})^2 (n_{12}{} v_{2}{})^2
 -  \frac{48}{5} (n_{12}{} v_{12}{}) (n_{12}{} v_{2}{})^3
 + \frac{24}{5} (n_{12}{} v_{2}{})^4\nn\\
&\qquad + \frac{56}{15} (n_{12}{} v_{12}{}) (n_{12}{} v_{2}{}) (v_{12}{} v_{2}{})
 + \frac{48}{5} (n_{12}{} v_{2}{})^2 (v_{12}{} v_{2}{})
 -  \frac{76}{35} (v_{12}{} v_{2}{})^2
 -  \frac{18898}{105} (n_{12}{} v_{12}{})^2 v_{12}^{2}\nn\\
&\qquad -  \frac{5048}{105} (n_{12}{} v_{12}{}) (n_{12}{} v_{2}{}) v_{12}^{2}
 + \frac{124}{35} (n_{12}{} v_{2}{})^2 v_{12}^{2}
 + \frac{479}{15} v_{12}^{4}
 + \frac{26}{15} (n_{12}{} v_{12}{})^2 v_{2}^{2}
 + \frac{8}{5} (n_{12}{} v_{12}{}) (n_{12}{} v_{2}{}) v_{2}^{2}\nn\\
&\qquad -  \frac{8}{5} (n_{12}{} v_{2}{})^2 v_{2}^{2}
 -  \frac{8}{5} (v_{12}{} v_{2}{}) v_{2}^{2}
 -  \frac{66}{35} v_{12}^{2} v_{2}^{2}
 -  \frac{1}{5} v_{2}^{4}\Bigr)
 + m_{2} \Bigl(- \frac{93166}{105} (n_{12}{} v_{12}{})^4
 + \frac{582}{5} (n_{12}{} v_{12}{})^3 (n_{12}{} v_{2}{})\nn\\
&\qquad -  \frac{454}{5} (n_{12}{} v_{12}{})^2 (n_{12}{} v_{2}{})^2
 + \frac{208}{5} (n_{12}{} v_{12}{}) (n_{12}{} v_{2}{})^3
 -  \frac{96}{5} (n_{12}{} v_{2}{})^4
 + \frac{372}{5} (n_{12}{} v_{12}{}) (n_{12}{} v_{2}{}) (v_{12}{} v_{2}{})\nn\\
&\qquad -  \frac{592}{15} (n_{12}{} v_{2}{})^2 (v_{12}{} v_{2}{})
 -  \frac{1432}{105} (v_{12}{} v_{2}{})^2
 + \frac{36389}{30} (n_{12}{} v_{12}{})^2 v_{12}^{2}
 -  \frac{3736}{105} (n_{12}{} v_{12}{}) (n_{12}{} v_{2}{}) v_{12}^{2}\nn\\
&\qquad + \frac{1562}{105} (n_{12}{} v_{2}{})^2 v_{12}^{2}
 -  \frac{38933}{210} v_{12}^{4}
 + \frac{227}{15} (n_{12}{} v_{12}{})^2 v_{2}^{2}
 -  \frac{104}{15} (n_{12}{} v_{12}{}) (n_{12}{} v_{2}{}) v_{2}^{2}
 + \frac{32}{5} (n_{12}{} v_{2}{})^2 v_{2}^{2}\nn\\
&\qquad + \frac{32}{5} (v_{12}{} v_{2}{}) v_{2}^{2}
 -  \frac{76}{21} v_{12}^{2} v_{2}^{2}
 + \frac{4}{5} v_{2}^{4}\Bigr)\biggl]
 + r_{12}^3 \Bigl(308 (n_{12}{} v_{12}{})^6
 + 56 (n_{12}{} v_{12}{})^5 (n_{12}{} v_{2}{})\nn\\
&\qquad - 210 (n_{12}{} v_{12}{})^4 (n_{12}{} v_{2}{})^2
 + 120 (n_{12}{} v_{12}{})^4 (v_{12}{} v_{2}{})
 + 240 (n_{12}{} v_{12}{})^3 (n_{12}{} v_{2}{}) (v_{12}{} v_{2}{})\nn\\
&\qquad -  \frac{348}{5} (n_{12}{} v_{12}{})^2 (v_{12}{} v_{2}{})^2
 -  \frac{12}{5} (n_{12}{} v_{12}{}) (n_{12}{} v_{2}{}) (v_{12}{} v_{2}{})^2
 + \frac{6}{5} (n_{12}{} v_{2}{})^2 (v_{12}{} v_{2}{})^2
 - 470 (n_{12}{} v_{12}{})^4 v_{12}^{2}\nn\\
&\qquad - 60 (n_{12}{} v_{12}{})^3 (n_{12}{} v_{2}{}) v_{12}^{2}
 + 174 (n_{12}{} v_{12}{})^2 (n_{12}{} v_{2}{})^2 v_{12}^{2}
 + 6 (n_{12}{} v_{12}{}) (n_{12}{} v_{2}{})^3 v_{12}^{2}
 -  \frac{3}{2} (n_{12}{} v_{2}{})^4 v_{12}^{2}\nn\\
&\qquad -  \frac{696}{5} (n_{12}{} v_{12}{})^2 (v_{12}{} v_{2}{}) v_{12}^{2}
 -  \frac{708}{5} (n_{12}{} v_{12}{}) (n_{12}{} v_{2}{}) (v_{12}{} v_{2}{}) v_{12}^{2}
 -  \frac{12}{5} (n_{12}{} v_{2}{})^2 (v_{12}{} v_{2}{}) v_{12}^{2}\nn\\
&\qquad + \frac{668}{35} (v_{12}{} v_{2}{})^2 v_{12}^{2}
 + \frac{1230}{7} (n_{12}{} v_{12}{})^2 v_{12}^{4}
 + \frac{246}{35} (n_{12}{} v_{12}{}) (n_{12}{} v_{2}{}) v_{12}^{4}
 -  \frac{501}{35} (n_{12}{} v_{2}{})^2 v_{12}^{4}\nn\\
&\qquad + \frac{668}{35} (v_{12}{} v_{2}{}) v_{12}^{4}
 -  \frac{6473}{630} v_{12}^{6}
 + 90 (n_{12}{} v_{12}{})^4 v_{2}^{2}
 -  \frac{6}{5} (v_{12}{} v_{2}{})^2 v_{2}^{2}
 -  \frac{522}{5} (n_{12}{} v_{12}{})^2 v_{12}^{2} v_{2}^{2}\nn\\
&\qquad -  \frac{18}{5} (n_{12}{} v_{12}{}) (n_{12}{} v_{2}{}) v_{12}^{2} v_{2}^{2}
 + \frac{9}{5} (n_{12}{} v_{2}{})^2 v_{12}^{2} v_{2}^{2}
 + \frac{501}{35} v_{12}^{4} v_{2}^{2}
 -  \frac{3}{10} v_{12}^{2} v_{2}^{4}\Bigr)\Biggl\}\nn\\
& + \frac{G^2 m_{1} m_{2} n_{12}^{i}}{c^9 r_{12}^6} \Biggl\{G^3 \biggl [\frac{1096496}{945} m_{2}^3 (n_{12}{} v_{12}{})
 + m_{1} m_{2}^2 \Bigl(\frac{634292}{315} (n_{12}{} v_{12}{})
 -  \frac{4464}{35} (n_{12}{} v_{2}{})\Bigr)\nn\\
&\qquad + m_{1}^2 m_{2} \Bigl(- \frac{146686}{315} (n_{12}{} v_{12}{})
 -  \frac{9592}{105} (n_{12}{} v_{2}{})\Bigr)
 + m_{1}^3 \Bigl(- \frac{744454}{945} (n_{12}{} v_{12}{})
 + \frac{1584}{35} (n_{12}{} v_{2}{})\Bigr)\biggl]\nn\\
&\qquad + G^2 r_{12} \biggl [m_{2}^2 \Bigl(\frac{36428}{15} (n_{12}{} v_{12}{})^3
 + \frac{7894}{15} (n_{12}{} v_{12}{}) (n_{12}{} v_{2}{})^2
 + \frac{3172}{21} (n_{12}{} v_{12}{}) (v_{12}{} v_{2}{})\nn\\
&\qquad -  \frac{3172}{21} (n_{12}{} v_{2}{}) (v_{12}{} v_{2}{})
 -  \frac{24034}{15} (n_{12}{} v_{12}{}) v_{12}^{2}
 + \frac{32}{5} (n_{12}{} v_{2}{}) v_{12}^{2}
 + \frac{1586}{21} (n_{12}{} v_{12}{}) v_{2}^{2}\Bigr)\nn\\
&\qquad + m_{1} m_{2} \Bigl(- \frac{154853}{315} (n_{12}{} v_{12}{})^3
 -  \frac{6592}{21} (n_{12}{} v_{12}{})^2 (n_{12}{} v_{2}{})
 + \frac{7204}{15} (n_{12}{} v_{12}{}) (n_{12}{} v_{2}{})^2
 -  \frac{392}{15} (n_{12}{} v_{2}{})^3\nn\\
&\qquad + \frac{13576}{105} (n_{12}{} v_{12}{}) (v_{12}{} v_{2}{})
 -  \frac{5048}{35} (n_{12}{} v_{2}{}) (v_{12}{} v_{2}{})
 -  \frac{1879}{7} (n_{12}{} v_{12}{}) v_{12}^{2}
 + \frac{848}{105} (n_{12}{} v_{2}{}) v_{12}^{2}\nn\\
&\qquad + \frac{6788}{105} (n_{12}{} v_{12}{}) v_{2}^{2}
 -  \frac{56}{15} (n_{12}{} v_{2}{}) v_{2}^{2}\Bigr)
 + m_{1}^2 \Bigl(- \frac{2220173}{945} (n_{12}{} v_{12}{})^3
 + \frac{940}{21} (n_{12}{} v_{12}{})^2 (n_{12}{} v_{2}{})\nn\\
&\qquad -  \frac{404}{3} (n_{12}{} v_{12}{}) (n_{12}{} v_{2}{})^2
 + \frac{56}{5} (n_{12}{} v_{2}{})^3
 -  \frac{3992}{105} (n_{12}{} v_{12}{}) (v_{12}{} v_{2}{})
 + \frac{4664}{105} (n_{12}{} v_{2}{}) (v_{12}{} v_{2}{})\nn\\
&\qquad + \frac{342617}{315} (n_{12}{} v_{12}{}) v_{12}^{2}
 -  \frac{6796}{105} (n_{12}{} v_{2}{}) v_{12}^{2}
 -  \frac{1996}{105} (n_{12}{} v_{12}{}) v_{2}^{2}
 + \frac{8}{5} (n_{12}{} v_{2}{}) v_{2}^{2}\Bigr)\biggl]
 + r_{12}^3 \Bigl(-216 (n_{12}{} v_{12}{})^7\nn\\
&\qquad + 252 (n_{12}{} v_{12}{})^5 (n_{12}{} v_{2}{})^2
 - 168 (n_{12}{} v_{12}{})^5 (v_{12}{} v_{2}{})
 - 280 (n_{12}{} v_{12}{})^4 (n_{12}{} v_{2}{}) (v_{12}{} v_{2}{})\nn\\
&\qquad + 60 (n_{12}{} v_{12}{})^3 (v_{12}{} v_{2}{})^2
 - 6 (n_{12}{} v_{12}{}) (n_{12}{} v_{2}{})^2 (v_{12}{} v_{2}{})^2
 + \frac{12}{5} (n_{12}{} v_{12}{}) (v_{12}{} v_{2}{})^3
 + \frac{12}{5} (n_{12}{} v_{2}{}) (v_{12}{} v_{2}{})^3\nn\\
&\qquad + 280 (n_{12}{} v_{12}{})^5 v_{12}^{2}
 - 210 (n_{12}{} v_{12}{})^3 (n_{12}{} v_{2}{})^2 v_{12}^{2}
 + \frac{21}{2} (n_{12}{} v_{12}{}) (n_{12}{} v_{2}{})^4 v_{12}^{2}
 + 180 (n_{12}{} v_{12}{})^3 (v_{12}{} v_{2}{}) v_{12}^{2}\nn\\
&\qquad + 180 (n_{12}{} v_{12}{})^2 (n_{12}{} v_{2}{}) (v_{12}{} v_{2}{}) v_{12}^{2}
 - 6 (n_{12}{} v_{12}{}) (n_{12}{} v_{2}{})^2 (v_{12}{} v_{2}{}) v_{12}^{2}
 - 6 (n_{12}{} v_{2}{})^3 (v_{12}{} v_{2}{}) v_{12}^{2}\nn\\
&\qquad -  \frac{408}{35} (n_{12}{} v_{12}{}) (v_{12}{} v_{2}{})^2 v_{12}^{2}
 + \frac{12}{5} (n_{12}{} v_{2}{}) (v_{12}{} v_{2}{})^2 v_{12}^{2}
 - 70 (n_{12}{} v_{12}{})^3 v_{12}^{4}
 + \frac{123}{7} (n_{12}{} v_{12}{}) (n_{12}{} v_{2}{})^2 v_{12}^{4}\nn\\
&\qquad -  \frac{738}{35} (n_{12}{} v_{12}{}) (v_{12}{} v_{2}{}) v_{12}^{4}
 -  \frac{246}{35} (n_{12}{} v_{2}{}) (v_{12}{} v_{2}{}) v_{12}^{4}
 -  \frac{187}{210} (n_{12}{} v_{12}{}) v_{12}^{6}
 - 84 (n_{12}{} v_{12}{})^5 v_{2}^{2}\nn\\
&\qquad + \frac{18}{5} (n_{12}{} v_{12}{}) (v_{12}{} v_{2}{})^2 v_{2}^{2}
 + 90 (n_{12}{} v_{12}{})^3 v_{12}^{2} v_{2}^{2}
 - 9 (n_{12}{} v_{12}{}) (n_{12}{} v_{2}{})^2 v_{12}^{2} v_{2}^{2}
 + \frac{18}{5} (n_{12}{} v_{12}{}) (v_{12}{} v_{2}{}) v_{12}^{2} v_{2}^{2}\nn\\
&\qquad + \frac{18}{5} (n_{12}{} v_{2}{}) (v_{12}{} v_{2}{}) v_{12}^{2} v_{2}^{2}
 -  \frac{369}{35} (n_{12}{} v_{12}{}) v_{12}^{4} v_{2}^{2}
 + \frac{9}{10} (n_{12}{} v_{12}{}) v_{12}^{2} v_{2}^{4}\Bigr)
 + G r_{12}^2 \biggl [m_{2} \Bigl(\frac{61658}{21} (n_{12}{} v_{12}{})^5\nn\\
&\qquad + \frac{2328}{5} (n_{12}{} v_{12}{})^3 (n_{12}{} v_{2}{})^2
 + \frac{416}{5} (n_{12}{} v_{12}{}) (n_{12}{} v_{2}{})^4
 -  \frac{582}{5} (n_{12}{} v_{12}{})^3 (v_{12}{} v_{2}{})\nn\\
&\qquad -  \frac{1746}{5} (n_{12}{} v_{12}{})^2 (n_{12}{} v_{2}{}) (v_{12}{} v_{2}{})
 + \frac{208}{5} (n_{12}{} v_{12}{}) (n_{12}{} v_{2}{})^2 (v_{12}{} v_{2}{})
 -  \frac{208}{5} (n_{12}{} v_{2}{})^3 (v_{12}{} v_{2}{})\nn\\
&\qquad + \frac{3568}{105} (n_{12}{} v_{12}{}) (v_{12}{} v_{2}{})^2
 -  \frac{208}{15} (n_{12}{} v_{2}{}) (v_{12}{} v_{2}{})^2
 -  \frac{183863}{42} (n_{12}{} v_{12}{})^3 v_{12}^{2}
 -  \frac{3526}{35} (n_{12}{} v_{12}{}) (n_{12}{} v_{2}{})^2 v_{12}^{2}\nn\\
&\qquad + \frac{3568}{105} (n_{12}{} v_{12}{}) (v_{12}{} v_{2}{}) v_{12}^{2}
 + \frac{3568}{105} (n_{12}{} v_{2}{}) (v_{12}{} v_{2}{}) v_{12}^{2}
 + \frac{261973}{210} (n_{12}{} v_{12}{}) v_{12}^{4}
 -  \frac{16}{5} (n_{12}{} v_{2}{}) v_{12}^{4}\nn\\
&\qquad -  \frac{291}{5} (n_{12}{} v_{12}{})^3 v_{2}^{2}
 -  \frac{104}{5} (n_{12}{} v_{12}{}) (n_{12}{} v_{2}{})^2 v_{2}^{2}
 -  \frac{104}{15} (n_{12}{} v_{12}{}) (v_{12}{} v_{2}{}) v_{2}^{2}
 + \frac{104}{15} (n_{12}{} v_{2}{}) (v_{12}{} v_{2}{}) v_{2}^{2}\nn\\
&\qquad + \frac{1784}{105} (n_{12}{} v_{12}{}) v_{12}^{2} v_{2}^{2}
 -  \frac{26}{15} (n_{12}{} v_{12}{}) v_{2}^{4}\Bigr)
 + m_{1} \Bigl(- \frac{11353}{35} (n_{12}{} v_{12}{})^5
 -  \frac{536}{5} (n_{12}{} v_{12}{})^4 (n_{12}{} v_{2}{})\nn\\
&\qquad -  \frac{996}{5} (n_{12}{} v_{12}{})^3 (n_{12}{} v_{2}{})^2
 -  \frac{96}{5} (n_{12}{} v_{12}{})^2 (n_{12}{} v_{2}{})^3
 -  \frac{144}{5} (n_{12}{} v_{12}{}) (n_{12}{} v_{2}{})^4
 + 48 (n_{12}{} v_{12}{})^3 (v_{12}{} v_{2}{})\nn\\
&\qquad + 144 (n_{12}{} v_{12}{})^2 (n_{12}{} v_{2}{}) (v_{12}{} v_{2}{})
 -  \frac{24}{5} (n_{12}{} v_{12}{}) (n_{12}{} v_{2}{})^2 (v_{12}{} v_{2}{})
 + \frac{72}{5} (n_{12}{} v_{2}{})^3 (v_{12}{} v_{2}{})\nn\\
&\qquad -  \frac{4888}{105} (n_{12}{} v_{12}{}) (v_{12}{} v_{2}{})^2
 + \frac{32}{5} (n_{12}{} v_{2}{}) (v_{12}{} v_{2}{})^2
 + \frac{15102}{35} (n_{12}{} v_{12}{})^3 v_{12}^{2}
 + \frac{2292}{35} (n_{12}{} v_{12}{})^2 (n_{12}{} v_{2}{}) v_{12}^{2}\nn\\
&\qquad + \frac{5056}{35} (n_{12}{} v_{12}{}) (n_{12}{} v_{2}{})^2 v_{12}^{2}
 -  \frac{24}{5} (n_{12}{} v_{2}{})^3 v_{12}^{2}
 -  \frac{4888}{105} (n_{12}{} v_{12}{}) (v_{12}{} v_{2}{}) v_{12}^{2}
 -  \frac{4888}{105} (n_{12}{} v_{2}{}) (v_{12}{} v_{2}{}) v_{12}^{2}\nn\\
&\qquad -  \frac{1829}{15} (n_{12}{} v_{12}{}) v_{12}^{4}
 + \frac{40}{7} (n_{12}{} v_{2}{}) v_{12}^{4}
 + 24 (n_{12}{} v_{12}{})^3 v_{2}^{2}
 + \frac{12}{5} (n_{12}{} v_{12}{})^2 (n_{12}{} v_{2}{}) v_{2}^{2}\nn\\
&\qquad + \frac{36}{5} (n_{12}{} v_{12}{}) (n_{12}{} v_{2}{})^2 v_{2}^{2}
 + \frac{12}{5} (n_{12}{} v_{12}{}) (v_{12}{} v_{2}{}) v_{2}^{2}
 -  \frac{12}{5} (n_{12}{} v_{2}{}) (v_{12}{} v_{2}{}) v_{2}^{2}
 -  \frac{2444}{105} (n_{12}{} v_{12}{}) v_{12}^{2} v_{2}^{2}\nn\\
&\qquad + \frac{4}{5} (n_{12}{} v_{2}{}) v_{12}^{2} v_{2}^{2}
 + \frac{3}{5} (n_{12}{} v_{12}{}) v_{2}^{4}\Bigr)\biggl]\Biggl\}\, .
\end{align}
The pole part can be split into a ``residue'' part, proportional to $1/\varepsilon$ and which blows up when $\varepsilon \rightarrow 0$, a ``finite'' part, and a ``hereditary'' part. That is,
\begin{subequations}
\begin{align}\label{eq:ai_RR1_4p5PN_pole}
a^i_{\text{RR1 4.5PN pole}} &=
a^i_{\text{RR1 4.5PN res}} + a^i_{\text{RR1 4.5PN finite}} + a^i_{\text{RR1 4.5PN hered}} \,,
\end{align}
where the three parts read

\begin{align}\label{apolepart}
    a^i_{\text{RR1 4.5PN res}} &= \frac{G^4 m_1 m_2 m^2\Delta}{\varepsilon \,c^9 r_{12}^5}\Bigg\{v_{12}^i\Bigg[180 (n_{12}v_{12})^2-36 v_{12}^2 + 32 \frac{Gm}{r_{12}}\Bigg] \nonumber\\*
    &\qquad\ \ \quad+ n_{12}^i(n_{12}v_{12}) \Bigg[-420( n_{12} v_{12})^2 + 180 v_{12}^2  - 120\frac{G m }{r_{12}}\Bigg] \Bigg\} \,,\\
    a^i_{\text{RR1 4.5PN finite}} &= \frac{G^4 m_1 m_2 m^2\Delta}{c^9 r_{12}^5}\Bigg\{ v_{12}^i \bigg((108-90 \gamma_\text{E})(n_{12}v_{12})^2 + 18 \gamma_\text{E} v_{12}^2  + (-20-16 \gamma_\text{E}+2 \pi^2) \frac{Gm}{r_{12}} \bigg) \nonumber\\
    &\qquad\qquad+ n_{12}^i (n_{12} v_{12})\bigg((-432+210 \gamma_\text{E}) (n_{12} v_{12})^2 + (108-90\gamma_\text{E})v_{12}^2  + (-42+60 \gamma_\text{E}- \frac{15}{2} \pi^2) \frac{G m }{r_{12}}\bigg)\nonumber\\
    &\qquad+ \ln\Bigl(\frac{\sqrt{\bar{q}} r_{12}}{{\ell_{0}}}\Bigr) \Bigg[v_{12}^i \bigg(-540(n_{12}v_{12})^2+ 108 v_{12}^2 - 144 \frac{G m}{r_{12}}\bigg) \nn \\ &\qquad\qquad\qquad\qquad+ n_{12}^i (n_{12}v_{12}) \bigg(1260 (n_{12}v_{12})^2 - 540 v_{12}^2 + 548 \frac{Gm}{r_{12}}\bigg)\Bigg] \nonumber\\
    &\qquad + \ln^2\Bigl(\frac{\sqrt{\bar{q}} r_{12}}{{\ell_{0}}}\Bigr) \frac{G m}{r_{12}}\Bigg[ -60 n_{12}^i (n_{12}v_{12})+16v_{12}^i \Bigg] \Bigg\} \,,\\
a^i_{\text{RR1 4.5PN hered}} &= \frac{2G}{c^9} \int_{0}^{+\infty} \mathrm{d}\tau \ln{\left(\frac{c \tau \sqrt{\pi}}{\ell_{0}}\right)}\Bigl[\mathrm{R}_i^{(4)}(t-\tau)-\mathrm{R}_i^{(4)}(t+\tau)\Bigr]\,,
\end{align}
where we recall that $\bar{q}=4\pi \de^{\gamma_\mathrm{E}}$ and we have introduced for convenience the total mass $m=m_1+m_2$ and the relative mass difference \mbox{$\Delta = (m_1-m_2)/m$}. The last term of this expression is the non-local-in-time logarithmic integral associated with the DR pole, whose integrand is given by the residue when $\varepsilon\to 0$ of $\dY^{i}_{\text{pole}}$ given by~\eqref{Yipole}, namely
\begin{align}
	\mathrm{R}_i(t) &= c^4 \lim_{\varepsilon\to 0} \Bigl[\varepsilon \,\dY^{i}_{\text{pole}}\Bigr] = - G^3 m_1^3 m_2 \frac{ n_{12}^i}{r_{12}^2} + \big( 1 \leftrightarrow 2 \big)\,.
\end{align}
\end{subequations}

It is important to notice that the whole expression \eqref{eq:ai_RR1_4p5PN_pole} for $a^i_{\text{RR1 4.5PN pole}}$ is in fact symmetric by particle exchange $(1 \leftrightarrow 2)$ due to the symmetry of $\dY_i^\text{pole}$ in \eqref{Yipole}. Consequently, both the $1/\varepsilon$ pole and the non-local integral cancel out in the relative acceleration $a_{12}^i=a_1^i-a_2^i$ in any frame and for any orbit.
The pole can be gauged away by applying a shift associated with the offending gauge moment $\dY_i$ or, more generally, by ignoring the contribution of the gauge moments~\eqref{gaugezeta} in the contact transformation. But as we have argued in Sec.~\ref{sec:metricharmonic}, setting to zero the gauge moments we would end up outside the harmonic coordinates, and in particular we would loose the property of Lorentz invariance.

If we were to relax our requirement and allow for another Lorentz-invariant coordinate system (which would \textit{not} be the unique harmonic coordinates suitable for asymptotic matching), then we can repeat the whole procedure but \textit{choose} to  perform the replacement $\dY^i \rightarrow \dY_i^\text{3D}$. This simply consists in a different gauge transformation, which is specified by the choice of gauge moments, and the resulting acceleration is simply $a^i_{\text{RR1 4.5PN 3D}} $. Since we have found that the pole part is separately Lorentz-invariant, it follows that this new acceleration would be Lorentz-invariant and would not feature a $1/\varepsilon$ pole nor a 4.5PN hereditary term. It could therefore be more useful in practice for applications in the lines of Ref.~\cite{Bonetti:2016eif}.

\subsection{Flux-balance laws in a general frame}
\label{eq:flux_balance_general}
With the previous expression for the 4.5PN RR acceleration in BT gauge~\cite{BFT24}, we had verified the complete set of flux-balance equations associated with the energy, the angular and linear momenta, and the CM position. They read
\begin{subequations}\label{eq:fluxbalance}\begin{align}
	\frac{\dd E}{\dd t} &= - \calF_{E} \,,\qquad\qquad\qquad
	\frac{\dd J^i}{\dd t} = - \calF^i_{\bm{J}} \,,\\
	\frac{\dd P^i}{\dd t} &= - \calF^i_{\bm{P}} \,,\qquad\qquad\qquad
	\frac{\dd G^i}{\dd t} = P^i - \calF^i_{\bm{G}} \,,
\end{align}
\end{subequations}
where all the fluxes are given with the required 2PN relative precision by Eqs.~(5.2) in~\cite{BFT24} and are given explicitly in the Supplemental Material~\cite{SuppMaterial}; recall that at this order, the canonical moments $(\dM_L, \dS_L)$ can be identified with the source moments $(\dI_L, \dJ_L)$.  Here we prove the same flux-balance laws, but in harmonic coordinates, using the acceleration we have obtained above.

The calculation is similar to the one of our previous paper in BT coordinates~\cite{BFT24}. For any conserved quantity, collectively denoted $\bm{Q}_\text{cons} \in \{E,\bm{J}, \bm{P}, \bm{G}\}$, we compute $\dd \bm{Q}_{\text{cons}}/\dd t + \calF_{\bm{Q}} = - \delta \calF_{\bm{Q}}$, then attempt to express the right-hand side as a total derivative $\delta \calF_{\bm{Q}} = \dd \bm{Q}_\text{RR}/\dd t$, where $\bm{Q}_\text{RR}$ is the dissipative correction to the conserved charge, commonly referred to as a Schott term~\cite{Schott}. In our previous calculation, we took advantage of the fact that the dissipative acceleration and the fluxes could be expressed linearly in the multipole moments. We then integrated by parts to put $\delta F_\mathbf{Q}$ into a canonical form, and showed that all terms could be explicitly expressed as total derivatives, which were also expressed linearly in the moments. In the present computation, we need to separate contributions to the acceleration that have a purely 3-dimensional form, from those that arise from the DR and contain the pole $\propto\varepsilon^{-1}$ and the associated non-local term. From~\eqref{aRR1harm} and~\eqref{aRR1polenonpole} we can write
\begin{align}
	\bm{a}_{\text{RR}1} =
	\bm{a}_{\text{RR1 3D}} + \bm{a}_{\text{RR1 pole}}\,,
\end{align}
where the 4.5PN pole part (also containing the non-local term) is given by Eqs.~\eqref{apolepart}, and the 3D part contains both the instantaneous contributions of Eqs.~\eqref{RR25PN35PN} and \eqref{eq:a1i_RR_4p5PN_3D} and the tail part of Eq.~\eqref{aRR1_4PN_tail}.

In the 3-dimensional sector, the acceleration is still linear in the multipole moments, but there are now six (source and gauge) moments rather than two (canonical) moments as in~\cite{BFT24}. The fluxes are expressed at this order only in terms of the two source moments. Adapting the algorithm of Ref.~\cite{BFT24} to the case of six moments, we are able to show that
\begin{align}\label{eq:Schott_equation_3D}
	\frac{\dd \bm{Q}_\text{cons}}{\dd t}\bigg|_{\bm{a}_{\text{cons} + \text{RR\,3D}}} + \calF_{\bm{Q}} = - \frac{\dd \bm{Q}^{\text{3D}}_{\text{RR}}}{\dd t} \,,
\end{align}
where the time derivative on the right-hand side of~\eqref{eq:Schott_equation_3D} is performed using the 3D acceleration and where $\bm{Q}^{\text{3D}}_{\text{RR}}$ are the 3-dimensional Schott terms that can be expressed linearly in the multipole moments. We subdivide these Schott terms into local and tail terms, $\bm{Q}^{\text{3D}}_{\text{RR}} = \bm{Q}^{\text{3D loc}}_{\text{RR}} +\bm{Q}^{\text{3D tail}}_{\text{RR}} $. The local 3D Schott terms for $\{E,\bm{J}, \bm{P}, \bm{G}\}$ are long and their full expressions are provided in the Supplemental Material~\cite{SuppMaterial}. 
The tail contributions enter exclusively at 4PN and were computed in Eq.~(3.13) of~Ref.~\cite{T25}; after substituting the canonical moments by source moments, they read
\begin{subequations}\label{eq:E_J_P_G_diss_4PN}\begin{align}\label{eq:E_diss_4PN}
    E^{\text{3D tail}}_{\text{RR}}  &=  \frac{2 G^2 \dM}{5 c^8} \Bigg\{
     \int_0^\infty \dd \rho \, \dI_{ij}^{(4)}(t-\rho) \int_{0}^\infty \dd \tau \ln\left(\frac{c\tau}{2 b_0}\right) \left[\dI_{ij}^{(4)}(t-\rho-\tau) - \dI_{ij}^{(4)}(t-\rho+\tau)\right]\nn\\*
     &\qquad\qquad -\dI_{ij}^{(3)} \int_{0}^\infty  \dd\tau \ln\left(\frac{c\tau}{2 b_0}\right) \left[\dI_{ij}^{(4)}(t-\tau) -  \dI_{ij}^{(4)}(t+\tau) \right] -\dI_{ij}^{(2)} \int_{0}^\infty  \dd\tau \ln\left(\frac{c\tau}{2 b_0}\right) \left[\dI_{ij}^{(5)}(t-\tau) +
    \dI_{ij}^{(5)}(t+\tau) \right] \nn\\*
    &\qquad\qquad\ +\dI_{ij}^{(1)} \int_{0}^\infty  \dd\tau \ln\left(\frac{c\tau}{2 b_0}\right) \left[\dI_{ij}^{(6)}(t-\tau) +
    \dI_{ij}^{(6)}(t+\tau) \right] - \frac{11}{12} \dI_{ij}^{(3)} \dI_{ij}^{(3)}    
      \Bigg\} \,, \\
\label{eq:J_diss_4PN}
J^{i\,\text{3D tail}}_{\text{RR}} &=  \frac{4 G^2 \dM}{5 c^8} \varepsilon_{ijk} \Bigg\{
     \int_0^\infty \dd \rho \, \dI_{jl}^{(3)}(t-\rho) \int_{0}^\infty \dd \tau \ln\left(\frac{c\tau}{2 b_0}\right) \left[\dI_{kl}^{(4)}(t-\rho-\tau) - \dI_{kl}^{(4)}(t-\rho+\tau)\right]\nn\\*
     &\qquad\qquad +\dI_{jl}^{(2)} \int_{0}^\infty  \dd\tau \ln\left(\frac{c\tau}{2 b_0}\right) \dI_{kl}^{(4)}(t+\tau)   -\dI_{jl}^{(1)} \int_{0}^\infty  \dd\tau \ln\left(\frac{c\tau}{2 b_0}\right) \left[\dI_{kl}^{(5)}(t-\tau) +
    \dI_{kl}^{(5)}(t+\tau) \right] \nn\\*
    &\qquad\qquad\ +\dI_{jl} \int_{0}^\infty  \dd\tau \ln\left(\frac{c\tau}{2 b_0}\right) \left[\dI_{kl}^{(6)}(t-\tau) +
    \dI_{kl}^{(6)}(t+\tau) \right]   - \frac{11}{12} \dI_{jl}^{(2)} \dI_{kl}^{(3)} 
      \Bigg\} \,,\\
\label{eq:P_diss_4PN}
P^{i\,\text{3D tail}}_{\text{RR}}   &= \frac{4 G^2 \dM}{5 c^8} \dI_{j} \int_0^\infty \dd \tau \ln\left(\frac{c\tau}{2 b_0}\right) \left[\dI_{ij}^{(6)}(t-\tau) + \dI_{ij}^{(6)}(t+\tau) \right]\,,
     \\
\label{eq:G_diss_4PN}
G^{i\,\text{3D tail}}_{\text{RR}}  &= 0 \,.
\end{align}\end{subequations}
Notice that $G^{i\,\text{3D hered}}_{\text{RR}}$ is vanishing, and that $P^{i\,\text{3D hered}}_{\text{RR}}$ vanishes in the center-of-mass frame, where $I_j = 0$ at this order.  Thus the hereditary contribution to the Schott term does not contribute when going to the CM frame.
We then add the contributions arising from DR and the pole, which are needed to account for the full acceleration. We have already written in Eq.~\eqref{a1pole} the 4.5PN dimensional regularization contribution to the acceleration, where $\dY_i^\text{pole}$ is given by~\eqref{Yipole}. Due to the presence of the $1/\varepsilon$ pole, it is necessary to keep terms of order $\mathcal{O}(\varepsilon)$ in the acceleration~\eqref{a1pole}, and the time derivatives of $\dY_i^\text{pole}$ must crucially be taken using the $d$-dimensional Newtonian acceleration.\footnote{\label{footnote:acc_N_d_dim}The Newtonian acceleration reads~\cite{BDE04}, with $\tilde{k}=\frac{\Gamma(\frac{d}{2}-1)}{\pi^{\frac{d}{2}-1}}$ and $\bar{q}=4\pi \de^{\gamma_\text{E}}$,
\begin{align*}
 a_{1} ^i = 2\tilde{k} \,\frac{d-2}{d-1} \ell_{0}^{d-3} \,\partial_i \!\left(\frac{G m_2}{r_{12}^{d-2}}\right) =  -\frac{G m_2}{r_{12}^2} n_{12}^i\biggl(1 - \varepsilon \left[\ln\left(\frac{\sqrt{\bar{q}}\,r_{12}}{\ell_0}\right)- \frac{3}{2}\right]\biggr) + \mathcal{O}(\varepsilon^2)\,.
\end{align*}
}
Since these contributions enter at the highest 4.5PN order, we can: (i) take the time derivatives of the moments using the ($d$-dimensional) Newtonian equations of motion, without introducing extra poles; and (ii) add the corrections to the flux-balance law linearly, without introducing any cross terms. Note that the flux contributions have been entirely absorbed in the 3D calculation. We find that 
\begin{align}
    \frac{\dd \left(\bm{Q}_\text{cons}+\bm{Q}_\text{RR}^\text{3D}\right)}{\dd t}\bigg|_{\bm{a}_{\text{cons} + \text{RR\,3D} + \text{RR pole}}} \,+ \calF_{\bm{Q}} =  -   \delta F_{\bm{Q}}^{ \text{RR pole}}   \,.
\end{align}
In the explicit expressions for $\delta F_{\bm{Q}}^{ \text{RR pole}}$, we recognize\footnote{Note that, at Newtonian order, the expressions for the monopole and dipole do \textit{not} acquire any $\mathcal{O}(\varepsilon)$ corrections in $d$ dimensions, because the $\frac{2 (d-2)}{d-1}$ factor in the expression~\eqref{eq:Sigma} for $\Sigma$ exactly cancels out with the prefactor in Eq.~\eqref{eq:sourcemts_IL}; notice also that the $\Gamma$ functions cancel out in Eq.~\eqref{eq:Sigma_ell_PNseries} at Newtonian $(k=0)$ order.} the monopole $\dM= m_1 + m_2$, the dipole $\dI_i=m_1 y_1^i+m_2 y_2^i$, and its derivative ${\dI}^{(1)}_i=m_1 v_1^i+m_2 v_2^i$. Using the conservation laws ${\dM}^{(1)} = 0$ and ${\dI}^{(2)}_i=0$, which are valid for any dimension~$d$, we are able to integrate by parts and show that
\begin{align}
	\delta F_{\bm{Q}}^{ \text{RR pole}} = \frac{\dd \bm{Q}_\text{RR}^\text{pole}}{\dd t}\,
\end{align}
is a total derivative with respect to the $d$-dimensional Newtonian acceleration. Thus we find that final Schott quantity entering the flux-balance law $\dd \bm{Q}/\dd t = - \calF_{\bm{Q}}$ is given by $\bm{Q} = \bm{Q}_\text{cons}+\bm{Q}_\text{RR}^\text{3D}+\bm{Q}_\text{RR}^\text{pole}$. The dimensional regularization contributions to the Schott terms are given explicitly by
\begin{subequations}\label{eq:Schott_dimreg}
    \begin{align}
    {E}_\text{RR}^\text{pole} &= \frac{4 G}{c^5}{\dI}^{(1)}_i\Bigg\{ \Big[1- \varepsilon \left(1+ \frac{\gamma_E}{2}\right)\Big] \frac{\dd^2 \dY_i^{\text{pole}}}{\dd t^2} - \frac{\varepsilon}{2} \frac{\dd^3}{\dd t^3}\int_0^{+\infty} \dd \tau \ln \left(\frac{c \tau \sqrt{\pi}}{\ell_0}\right)\Big[\dY_i^{\text{pole}}(t - \tau) - \dY_i^{\text{pole}}(t + \tau) \Big] \Bigg\} \,,\\
    {J}_\text{RR}^{i\,\text{pole}} &= \frac{4 G}{c^5} \varepsilon_{ijk}\Bigg\{   \Big[1- \varepsilon\left(1+ \frac{\gamma_E}{2}\right) \Big] \Big( \dI_j\,\frac{\dd^2 \dY_i^{\text{pole}}}{\dd t^2} - {\dI}^{(1)}_j\,\frac{\dd \dY_i^{\text{pole}}}{\dd t} \Big) \nn \\*
    &\qquad\qquad\quad- \frac{\varepsilon}{2}\,\dI_j\, \frac{\dd^3}{\dd t^3}\int_0^{+\infty} \dd \tau \ln \left(\frac{c \tau \sqrt{\pi}}{\ell_0}\right)\Big[\dY_k^{\text{pole}}(t - \tau) - \dY_k^{\text{pole}}(t + \tau) \Big] \Bigg) \nn \\*
    &\qquad\qquad\quad + \frac{\varepsilon}{2}\,{\dI}^{(1)}_j\, \frac{\dd^2}{\dd t^2}\int_0^{+\infty} \dd \tau \ln \left(\frac{c \tau \sqrt{\pi}}{\ell_0}\right)\Big[\dY_k^{\text{pole}}(t - \tau) - \dY_k^{\text{pole}}(t + \tau) \Big]  \Bigg\} \,, \\
    {P}_\text{RR}^{i\,\text{pole}} &= \frac{4 G}{c^5} \dM\, \Bigg\{ \Big[1- \varepsilon \left(1+ \frac{\gamma_E}{2}\right) \Big] \frac{\dd^2 \dY_i^{\text{pole}}}{\dd t^2} - \frac{\varepsilon}{2} \frac{\dd^3}{\dd t^3}\int_0^{+\infty} \dd \tau \ln \left(\frac{c \tau \sqrt{\pi}}{\ell_0}\right)\Big[\dY_i^{\text{pole}}(t - \tau) - \dY_i^{\text{pole}}(t + \tau) \Big] \Bigg\} \,,\\
    {G}_\text{RR}^{i\,\text{pole}} &= 0\,,
\end{align}\end{subequations}
where $\dY_i^{\text{pole}}$ is given by~\eqref{Yipole} and carries a factor $1/c^4$; thus, the latter Schott terms are indeed of order 4.5PN, despite the $1/c^5$ prefactor.
Note that the presence of the dipole moment means that ${E}_\text{RR}^\text{pole}$ and ${J}_\text{RR}^{i\,\text{pole}}$ will identically vanish in the center-of-mass frame, where $\dI_i = 0$. Moreover, there is no pole contribution to the CM position. In these expressions, it is crucial that the time derivatives of $\dY_i^{\text{pole}}$ be taken using the $d$-dimensional Newtonian acceleration (see${}^{\text{\ref{footnote:acc_N_d_dim}}}$ for the explicit expression).
Thus we have proved the existence of some energy $E$, angular momentum $J^i$, linear momentum $P^i$ and CM position $G^i$, which can be attributed to the matter system, such that the balance Eqs.~\eqref{eq:fluxbalance} are satisfied, and where the fluxes in the right-hand side are attributed to the radiation. Note that the Schott terms that we have computed correspond to harmonic coordinates and differ from the Schott terms in BT coordinates obtained in~\cite{BFT24}. In harmonic coordinates, the Schott terms include contributions arising from the DR scheme, \textit{i.e.} a pole $1/\varepsilon$, a logarithm involving the DR scale $\ell_0$ and a hereditary time-integral; see Eqs.~\eqref{eq:Schott_dimreg}.

\section{Acceleration in the center-of-mass frame}
\label{sec:CM}

\subsection{Passage to the center-of-mass frame}

Using the expressions for the 4.5PN RR contributions to the integral of the center of mass (CM) $G^i$, we can compute the 4.5PN-accurate formulas for the passage to the CM variables. Following Ref.~\cite{BFT24}, we introduce the integrated fluxes of linear momentum and CM position as
\begin{subequations}\label{eq:def_Pi_Gamma}
	\begin{align}
    \Pi^i &\equiv \int_{-\infty}^{t} \dd t' \,\mathcal{F}_{\bm{P}}^i(t')\,,\label{eq:def_Pi}\\
    \Gamma^i &\equiv \int_{-\infty}^{t} \dd t' \int_{-\infty}^{t'} \dd t''\,\mathcal{F}_{\bm{P}}^i(t'') + \int_{-\infty}^{t} \dd t'\, \mathcal{F}_{\bm{G}}^i(t')\,,
\end{align}
\end{subequations}
which are 3.5PN quantities. As explained in Sec.~VI A of~\cite{BFT24}, the CM position is then defined by the condition $G^i + \Gamma^i=0$, where $G^i$ corresponds to the contribution of the ``matter'', while $\Gamma^i$ is the contribution of the ``radiation''. If one overlooks the contribution from the radiation the CM integral is no longer conserved.\footnote{Thus the ``matter part'' in Eq~\eqref{eq:P_Q_R_CoM_RR} is obtained by solving $G^i=0$; the ``radiation'' part then is defined as the corrections that needs to be added to solve for the physically meaningful equations $G^i+\Gamma^i=0$. Moreover, note there are several possible definitions for the flux associated to the center-of-mass frame. For instance, in the present work we define the flux $\mathcal{F}_{\bm{G}}^i$ associated to the CM position as in Ref.~\cite{BF19}, which stems from a MPM calculation. However, there exists another definition, introduced in~\cite{KNQ18, COS20}, that arises from a Bondi-Sachs calculation at future null infinity; see Eqs.~(5.2d) and~(5.4) in~\cite{BFT24} for the explicit expressions. Because the two expressions differ by a total time derivative, they correspond to a redefinition of the CM position through the associated Schott terms. Using one or the other definition is physically equivalent, as it simply corresponds to a different split between ``matter'' and ``radiation'' contributions. Thus neither the full transformation~\eqref{eq:P_Q_R_CoM_RR} nor the equations of motion in the CM~\eqref{eq:aRR_CM_total} depend on this choice.} Here the CM position is decomposed into a conservative and a RR contribution, $G^i = G^i_\text{cons} + G^{i}_\text{RR}$. In fact, the RR contribution is given purely by the \textit{local} 3D part, i.e., $G^{i}_\text{RR} = G^{i\,\text{3D loc}}_\text{RR}$, because we have found that $G^{i\,\text{pole}}_\text{RR} = 0$ and $G^{i\,\text{3D hered}}_\text{RR} = 0$. This is convenient, since $G^{i\,\text{3D}}_\text{RR loc}$ is expressed in terms of $\dY_i^\text{3D}$ and does not contain any poles or hereditary contributions.
Solving iteratively, we obtain an expression for $\bm{y}_1$ and $\bm{y}_2$ in terms of the relative positions and velocities $(\bm{x}, \bm{v})$ and the source and gauge moments. Replacing the source and gauge moments by their explicit expressions in the CM frame, we obtain the formula for the passage to the center-of-mass frame which takes the form\footnote{Following the notation of~\cite{BFT24}: $x^i=y_1^i-y_2^i$ and $v^i=v_1^i-v_2^i=\dd x^i/\dd t$ for the relative position and velocity, and mass parameters defined by $m=m_1+m_2$, $X_1=m_1/m$, $X_2=m_2/m$, along with the symmetric mass ratio $\nu=X_1 X_2$ and mass difference $\Delta=X_1-X_2$.}
\begin{subequations}\label{eq:P_Q_R_CoM_RR}
	\begin{align}
    y_1^i &= x^i\Bigl(X_2 + \nu \,\Delta \,\mathcal{P} \Bigr) + \nu \,\Delta \,\mathcal{Q} \,v^i + \mathcal{R}^i \,,\\
    y_2^i &= x^i\Bigl(-X_1 + \nu \,\Delta \,\mathcal{P} \Bigr) + \nu \,\Delta \,\mathcal{Q} \,v^i + \mathcal{R}^i \,.
    \end{align}
\end{subequations}
The contribution of the radiation is identical to that in the BT gauge and reads
\begin{align}\label{eq:radcontr}
    \mathcal{R}^i &= - \frac{\Gamma^i}{m} + \frac{\nu}{m c^2}\bigg[\left(\frac{v^2}{2} - \frac{G m}{r}\right)\,\Gamma^i + v^j \left(\Pi^j + \calF^j_{\bm{G}}\right)x^i \bigg] + \mathcal{O}\left(\frac{1}{c^{11}}\right) \,,
\end{align}
where the explicit expressions for $\calF_{\bm{G}}^i$ and $\calF_{\bm{P}}^i$ (and thus $\Pi^i$ and $\Gamma^i$) are provided in Eq.~(6.8) of~\cite{BFT24}. On the other hand the matter contributions are subdivided into conservative and RR contributions, $\mathcal{P} = \mathcal{P}_\text{cons} + \mathcal{P}_\text{RR}$ and $\mathcal{Q} = \mathcal{Q}_\text{cons} + \mathcal{Q}_\text{RR}$. The conservative parts are given up to 4PN order by Eqs.~(B5-B6) of Ref.~\cite{BBFM17}. By solving for $G^i + \Gamma^i=0$, and using the notation $\dot{r}=(nv)$, we find that the dissipative sector reads
\begin{subequations}\label{eq:mattercontr}\begin{align}
    \mathcal{P}_\text{RR} &= \frac{G^2 m^2 \dot{r}}{r^2} \Bigg\{\frac{1}{c^7}\Bigg[\frac{G m}{r}\left(\frac{52}{15}-\frac{116}{35} \nu\right)-\frac{8}{5} \dot{r}^2+\left(-\frac{2}{15}+\frac{68}{35} \nu \right) v^2\Bigg] \nn\\
    &\qquad\qquad + \frac{1}{c^9}\Bigg[\frac{G^2 m^2}{r^2} \left(-\frac{62191}{945} -\frac{538}{45} \nu  -\frac{116}{9}\nu^2 \right)+ \frac{G m }{r}\dot{r}^2\left(-\frac{223}{15} +\frac{116141}{6615}\nu -\frac{165716}{6615}  \nu ^2 \right) \nn\\
    &\qquad\qquad\qquad  +\dot{r}^2 v^2 \left(-\frac{143}{70}+\frac{9602}{245}\nu +\frac{1098}{245} \nu ^2\right)
   +\frac{G m}{r} v^2\left(\frac{1202}{105}-\frac{39059}{2205} \nu +\frac{83458}{2205} \nu ^2\right)+\dot{r}^4\left(\frac{4}{7}-20 \nu \right) \nn\\
    &\qquad\qquad\qquad  +v^4 \left(-\frac{1}{42}-\frac{5878}{441} \nu -\frac{11798}{735}\nu ^2\right) \Bigg]\Bigg\}  + \mathcal{O}\left(\frac{1}{c^{11}}\right) \,,\\
    \mathcal{Q}_\text{RR} &= G  m \Bigg\{\frac{1}{c^5}\Bigg[\frac{4}{5} v^2-\frac{8 G m}{5 r}\Bigg] 
    \nn\\
    & \qquad 
    +\frac{1}{c^7} \Bigg[\frac{G^2 m^2}{r^2} \left(-\frac{172}{105} -\frac{64}{35}  \nu  \right) +\frac{G m }{r}\dot{r}^2 \left(\frac{44}{15}+\frac{68}{35}\nu\right)  +\frac{G m}{r}  v^2 \left(\frac{6}{35}+\frac{132}{35} \nu \right) +  v^4  \left(\frac{6}{7}-\frac{22}{7} \nu\right)\Bigg]
     \nn\\
    & \qquad
   + \frac{1}{c^9}\Bigg[\frac{G^3 m^3}{r^3} \left(\frac{55477}{2835}+\frac{2650}{189}\nu-\frac{304}{135} \nu ^2 \right) +  \frac{G^2  m^2}{r^2} \dot{r}^2\left(\frac{49}{5}+\frac{34514}{2205}\nu+\frac{5116}{441} \nu^2 \right) \nn\\
    & \qquad\qquad  +\frac{G m v^2}{r} \dot{r}^2 \left(\frac{13}{42}-\frac{30754}{2205}\nu -\frac{11974}{735}\nu^2\right)  +\frac{G^2 m^2}{r^2} v^2\left(-\frac{323}{35}-\frac{686}{45}\nu +\frac{3184}{315} \nu^2\right)  \nn\\
    & \qquad\qquad   
    +\frac{G m}{r} \dot{r}^4 \left(\frac{10}{7}+\frac{2113}{735}\nu +\frac{1121}{245} \nu ^2\right) +\frac{G m}{r}v^4\left(\frac{149}{30}+\frac{2309}{2205} \nu -\frac{14629}{735} \nu ^2\right) \nn\\
    & \qquad\qquad   + v^6 \left(\frac{533}{630}-\frac{3343}{490} \nu +\frac{20609}{1470}\nu^2\right) \Bigg]\Bigg\} + \mathcal{O}\left(\frac{1}{c^{11}}\right)\,.
\end{align}\end{subequations}
This result in harmonic coordinates differs from the result in BT coordinates given by Eqs.~(6.7) in~\cite{BFT24}. Up to 3.5PN order the matter sector~\eqref{eq:mattercontr} agrees with previous literature in harmonic coordinates~\cite{BF19}.
Note that the 4PN tails do not affect the passage to the CM frame. This was shown for the conservative part of the tail at the end of Sec.~IV of~\cite{BBFM17}, and for the dissipative Schott term in Sec.~III of~\cite{T25}.

\subsection{RR acceleration in the CM frame}

We now reduce the relative acceleration $a_{\text{RR}}^i\equiv a_{\text{RR}1}^i-a_{\text{RR}2}^i$ to its expression in the CM frame using the relations derived in the previous section. One should account for two types of contributions: (i) direct contributions, which arise by reducing the RR acceleration $a_{\text{RR}}^i$ to the CM frame using the conservative formulas of Ref.~\cite{BBFM17} at 2PN order; (ii) indirect contributions, which arise by reducing the 2PN-accurate conservative acceleration $a^{i}_\text{cons}$ using the dissipative formulas of Eq.~\eqref{eq:P_Q_R_CoM_RR}. The relative 4.5PN RR acceleration in the CM frame takes the form 
\begin{align}\label{eq:aRR_CM_total}
	a_{\text{RR}}^i\Big|_\text{harm} =
	a^i_{\text{RR 2.5PN}} + a^i_{\text{RR 3.5PN}} + a^i_{\text{RR 4PN}}  + a^i_{\text{RR 4.5PN}} + \mathcal{O}\left(\frac{1}{c^{11}}\right)\,.
\end{align}
The 2.5PN and 3.5PN pieces read in harmonic coordinates
\begin{align}\label{eq:accCM2p5PN}
    a_\text{RR\,2.5PN}^i &= \frac{G^2 m^2 \nu }{c^5 r^3}\Bigg\{n^i \left(\frac{136}{15} \frac{G m}{r}\dot{r} +\frac{24}{5}\dot{r} v^2\right)-v^i \left(\frac{24}{5}\frac{G m}{r}+\frac{8}{5} v^2\right)\Bigg\} \,, \\
   \label{eq:accCM3p5PN}
    a_\text{RR\,3.5PN}^i &= \frac{G^2 m^2 \nu }{c^7 r^3} \Bigg\{n^i \Bigg[\frac{G^2 m^2}{r^2} \dot{r} \left(-\frac{3956}{35}-\frac{184}{5} \nu \right)+\frac{G m}{r}  \dot{r}^3 \left(-\frac{294}{5}-\frac{376}{5} \nu \right)+\frac{G m}{r} \dot{r} v^2\left(-\frac{692}{35}+\frac{724}{15} \nu \right) \nn\\
    &\qquad\qquad\qquad\quad\quad -112\, \dot{r}^5 +
   \dot{r}^3 v^2 (114+12 \nu ) + \dot{r} v^4 \left( -\frac{366}{35}-12 \nu\right) \Bigg]\nn\\
    &\qquad\qquad +v^i \Bigg[\frac{G^2 m^2}{r^2}\left(\frac{1060}{21}+\frac{104}{5} \nu \right) + \frac{G m}{r} \dot{r}^2 \left(\frac{82}{3}+\frac{848}{15} \nu\right) + \frac{G m}{r} v^2 \left(-\frac{164}{21}-\frac{148}{5} \nu \right) \nn\\
    &\qquad\qquad\qquad\quad +120\,
   \dot{r}^4+ \dot{r}^2 v^2 \left(-\frac{678}{5}-\frac{12}{5} \nu \right)+  v^4\left(\frac{626}{35}+\frac{12}{5} \nu\right)\Bigg]\Bigg\} \,,
\end{align}
and agree with Eqs.~(5.8) and (5.9) in Ref.~\cite{NB05}. At the 4PN order, the dissipative tail  contribution becomes~\cite{BBFM17,GLPR16,T25}
\begin{align}\label{aitail}
    a^i_{\text{RR 4PN}} &= - \frac{4 G m}{5 c^8} x^j \int_0^{+\infty}\dd \tau \, \ln\left(\frac{c \tau}{2 P}\right) \Big[\dI_{ij}^{(7)}(t-\tau)+\dI_{ij}^{(7)}(t+\tau)\Big]\,,
\end{align}
involving the seventh time-derivative of the quadrupole in the CM frame $\dI_{ij}=m\nu \hat{x}^{ij}+\mathcal{O}(c^{-2})$. The arbitrary constant lengthscale $P$ cancels
out from the two terms of~\eqref{aitail}.
The 4.5PN piece is more subtle because of the contribution from the radiation in the CM definition, as given by~\eqref{eq:radcontr}. Thus we split the 4.5PN CM acceleration into a matter part and a radiation part, namely 
\begin{equation}
	a_{\text{RR 4.5PN}}^i = a_{\text{RR 4.5PN mat}}^i + a_{\text{RR 4.5PN rad}}^i  \,.
\end{equation}
However we recall that the 4.5PN pole (which arose from the application of DR) cancels out in the relative acceleration, even before going to the CM frame. The radiation part has been computed in~\cite{BFT24} and reads
\begin{align}\label{eq:accCMrad}
	a_{\text{RR 4.5PN rad}}^i = \frac{G \Delta}{r^2c^2}\left(2 n^iv^j+n^jv^i\right)\Bigl[\Pi^j + \calF_{\bm{G}}^j\Bigr]\,,
\end{align}
which is a functional of the fluxes of linear momentum and CM position; in fact, it is non-local-in-time\footnote{More precisely, it is \textit{semi-hereditary}, because it becomes local-in-time after applying a finite number of time derivatives.} as it involves the integrated flux of linear momentum or source's recoil, see Eq.~\eqref{eq:def_Pi_Gamma}.
After replacement of the multipole moments, these fluxes read at the required 3.5PN accuracy 
\begin{subequations}\label{eq:FP_FG_COM}
	\begin{align} 
	  \calF_{\bm{P}}^i&= \frac{G^3 m^4 \Delta \nu^2}{c^7 r^4} \Bigg\{ \dot{r} n^i \Bigg(\frac{32}{35} \frac{G m}{r}  -  \frac{24}{7} \dot{r}^2 + \frac{88}{21} v^2\Bigg) + v^i\Bigg(- \frac{64}{105} \frac{G m}{r}  + \frac{304}{105} \dot{r}^2 -  \frac{80}{21} v^2\Bigg)   \Bigg\} + \mathcal{O}\left(\frac{1}{c^9}\right) \,,\\
\label{eq:FG_COM} 
	\calF_{\bm{G}}^i&= \frac{G^2 m^3 \nu^2 \Delta }{c^7 r^2} \Bigg\{n^i \Bigg(\frac{272}{105}\frac{G m}{r} \dot{r}^2 -  \frac{64}{15} \frac{G m}{r}  v^2 + \frac{48}{35} \dot{r}^2 v^2 -  \frac{32}{35} v^{4}\Bigg) +  \dot{r} v^i \Bigg(\frac{16}{21} \frac{G m}{r} -  \frac{24}{7} \dot{r}^2 + \frac{24}{7} v^2\Bigg) \Bigg\} + \mathcal{O}\left(\frac{1}{c^9}\right)  \,.
\end{align}
\end{subequations}
Finally, the 4.5PN RR contribution from the matter (whose expression is specific to harmonic coordinates) reads
\begin{align}\label{eq:accCM4pPNmat}
    a_\text{RR\,4.5PN\,mat}^i &= \frac{G^2 m^2 \nu}{c^9 r^3}  \Bigg\{n^i \Bigg[\frac{G^3 m^3}{r^3}  \dot{r} \left(\frac{336922}{945}+\frac{20644}{35} \nu-\frac{3632}{105} \nu ^2\right)  +\frac{G^2 m^2}{r^2}  \dot{r}^3 \left(-\frac{83177}{945}+\frac{41524}{135} \nu -\frac{30076}{105} \nu ^2\right)  \nn\\
    &\qquad\qquad\qquad\quad +\frac{G^2 m^2}{r^2}\dot{r} v^2 \left(-\frac{129769}{315}-\frac{58468}{315} \nu +\frac{4636}{105} \nu^2\right) +\frac{G m}{r}  \dot{r}^5  \left(\frac{261883}{105}+\frac{5316}{5} \nu -\frac{1776}{5}\nu ^2\right)  \nn\\
    &\qquad\qquad\qquad\quad  +\frac{G m}{r}   \dot{r}^3 v^2 \left(-\frac{22133}{6}-\frac{47643}{35} \nu +\frac{16028}{35} \nu^2\right)  +\frac{G m}{r}  \dot{r} v^4 \left(\frac{14307}{14}+\frac{46337}{105} \nu -\frac{2659}{21} \nu^2\right)  \nn\\
    &\qquad\qquad\qquad\quad   + \dot{r}^7(-180-504 \nu)
   + \dot{r}^5 v^2 \left(\frac{329}{2}+1106 \nu+21 \nu ^2\right) + \dot{r}^3 v^4 \left(\frac{88}{7}-\frac{4362}{7} \nu-66 \nu ^2\right) \nn\\
    &\qquad\qquad\qquad\quad  + \dot{r} v^6 \left(-\frac{1643}{210}+\frac{1248}{35} \nu +45 \nu^2\right)
   \Bigg] \nn\\
   &\qquad\qquad +v^i \Bigg[\frac{G^3 m^3}{r^3} \left(-\frac{499286}{2835}-\frac{1376}{5} \nu+\frac{272}{35}\nu^2\right)+\frac{G^2 m^2}{r^2}\dot{r}^2\left(\frac{85991}{315}-\frac{97228}{315} \nu+\frac{4604}{21} \nu^2\right) \nn\\
    &\qquad\qquad\qquad\quad +\frac{G^2 m^2}{r^2}  v^2 \left(\frac{47459}{315}+\frac{66632}{315} \nu -\frac{1796}{35} \nu ^2\right) +
     \frac{G m}{r} \dot{r}^4  \left(-\frac{86323}{105}-\frac{10646}{35} \nu+\frac{1192}{7} \nu^2\right)\nn\\
    &\qquad\qquad\qquad\quad 
   + \frac{G m}{r} \dot{r}^2 v^2 \left(\frac{224351}{210}+\frac{7177}{35} \nu-\frac{2816}{15} \nu ^2\right) 
   + \frac{G m}{r} v^4 \left(-\frac{10747}{70}-\frac{539}{15} \nu+\frac{1769}{35} \nu ^2\right) 
   \nn\\
    &\qquad\qquad\qquad\quad  +  \dot{r}^6(350+420 \nu )+ \dot{r}^4 v^2 \left(-\frac{1007}{2} -984 \nu -3 \nu ^2\right)+ \dot{r}^2 v^4\left(\frac{5778}{35}+\frac{4350}{7} \nu+\frac{54}{5} \nu^2\right)  \nn\\
    &\qquad\qquad\qquad\quad + v^6\left(-\frac{4873}{630} -\frac{298}{5} \nu-\frac{39}{5} \nu ^2\right) \Bigg]\Bigg\} \,.
\end{align}
\subsection{Flux-balance laws in the CM frame}

We now verify that the flux-balance laws still hold in the CM frame. For this, we reduce the total energy and angular momentum to the CM frame. Just like the acceleration, the Schott terms in the CM will feature (i) direct contributions, which arise from the reduction of the Schott terms in a general frame using the (conservative) 2PN-accurate passage to the CM; and (ii) indirect contributions, which arise from the dissipative contributions to the passage to the CM applied on the conservative energy and angular momentum. 
As mentioned in Sec.~\ref{eq:flux_balance_general}, the DR contributions to the RR energy and angular momentum (which involve a pole, a logarithm, and a hereditary term) vanish in the CM frame because they are multiplied by the dipole. We decompose the Schott terms for the energy and angular momentum into matter, radiation, and tail contributions:
\begin{subequations}\begin{align}
    E_\text{RR} &=  E_\text{RR\,mat} + E_\text{RR\,rad} +E_\text{RR\,tail} \,,\\
    J^i_\text{RR} &= J^i_\text{RR\,mat} + J^i_\text{RR\,rad}   +J^i_\text{RR\,tail} \,.
\end{align}\end{subequations}
The 4.5PN matter contributions read
\begin{subequations}\begin{align}
    E_\text{RR\,mat} &= \frac{G^2 m^3 \nu ^2 \dot{r} }{r^2 c^5} \Bigg\{ \frac{8}{5} v^2  \nn \\*
   &\qquad\qquad\quad +\frac{1}{c^2}\Bigg[\frac{G^2 m^2}{r^2} \left(\frac{32}{105}-\frac{128}{105} \nu \right) + \frac{G m}{r}\dot{r}^2 \left(-\frac{4}{21}-\frac{1936}{105} \nu \right)  +\dot{r}^2  v^2\left(\frac{154}{5}+\frac{12}{5} \nu\right) \nn \\*
   & \qquad\qquad\qquad\qquad   +\frac{G m}{r}v^2
   \left(-\frac{316}{35}+\frac{736}{35} \nu\right) -16 \dot{r}^4+v^4\left(-\frac{542}{35}-\frac{48}{5} \nu \right) \Bigg] \nn \\*
   &\qquad\qquad\quad +\frac{1}{c^4}\Bigg[
   \frac{G^3  m^3}{r^3}\left(-\frac{2896}{945} +\frac{1376}{105} \nu  -\frac{640}{189} \nu ^2 \right)
   + \frac{G^2 m^2}{r^2} \dot{r}^2   \left(-\frac{48476}{315} -\frac{2482}{315} \nu   - \frac{69568}{945}  \nu ^2 \right) 
    \nn\\
   &\qquad\qquad\qquad\qquad  +\frac{G m}{r} \dot{r}^2  v^2  \left(-\frac{340174}{945}-\frac{196598}{945}\nu  +\frac{23536}{135}  \nu ^2 \right)
   + \dot{r}^2  v^4  \left(\frac{53}{35}-\frac{10233}{35}\nu - \frac{1044}{35} \nu ^2\right)   \nn\\
   &\qquad\qquad\qquad\qquad 
   +\frac{G^2 m^2 }{r^2} v^2  \left(\frac{516}{5}  +\frac{1912}{315} \nu +\frac{7264}{105} \nu ^2 \right)
   + \frac{G m}{r} \dot{r}^4   \left(\frac{14863}{63} +\frac{9974}{63} \nu - \frac{17432}{315} \nu ^2 \right) \nn\\
   &\qquad\qquad\qquad\qquad 
   +\dot{r}^4  v^2  \left(\frac{59}{2}+234 \nu  + 3\nu^2 \right) 
   +\frac{G m}{r} v^4 \left(\frac{26561}{315}  +\frac{25504}{315}\nu  -\frac{7088}{45}\nu^2 \right) \nn\\
   &\qquad\qquad\qquad\qquad 
   +\dot{r}^6 \left(-20 - 56 \nu \right) 
   + v^6  \left(-\frac{10139}{630}+\frac{3954}{35} \nu +\frac{2368}{35} \nu ^2\right)\Bigg]\Bigg\}\,, \\
    J^i_\text{RR\,mat} &= \frac{G^2 m^3 \nu ^2 \dot{r}}{r^2 c^5}  \,\varepsilon_{ijk} \,x^j v^k  \Bigg\{
    -\frac{8}{5} \nn\\
    &\qquad\qquad\quad+\frac{1}{c^2}\Bigg[\frac{G m}{r} \left(\frac{136}{105}+\frac{1352}{105} \nu \right) + \dot{r}^2 \left(\frac{878}{35} -\frac{144}{35}\nu \right) +v^2\left(-\frac{64}{3}+\frac{88}{105}\nu \right) \Bigg] \nn\\
   &\qquad\qquad\quad+
    \frac{1}{c^4}\Bigg[\frac{G^2 m^2}{r^2} \left(\frac{30958}{315}+\frac{8026}{105} \nu +\frac{15908}{315} \nu ^2\right)
    + \frac{G m}{r}\dot{r}^2 \left(-\frac{53687}{945} -\frac{31028}{945} \nu +\frac{1178}{15} \nu ^2\right) \nn\\
   &\qquad\qquad\qquad\qquad
    +\dot{r}^2 v^2 \left(-\frac{107}{30} -\frac{88957}{315} \nu+\frac{2419}{315} \nu ^2\right) 
   +\frac{G m}{r} v^2 \left(\frac{28657}{315}+\frac{3112}{105} \nu -\frac{4988}{45}\nu ^2 \right)  \nn\\
   &\qquad\qquad\qquad\qquad  + \dot{r}^4 \left(\frac{662}{21} +\frac{7456}{63} \nu -\frac{388}{63} \nu ^2\right)
   + v^4 \left(-\frac{15469}{630} +\frac{47941}{315} \nu + \frac{2806}{105} \nu ^2 \right) \Bigg] \Bigg\} \,,
\end{align}\end{subequations}
whereas the radiation contributions read~\cite{BFT24}
\begin{subequations}\begin{align}
    E_\text{RR\,rad} &= \frac{\nu \Delta}{c^2} v^2 v^i\Big[\Pi^i + \mathcal{F}^i_{\bm{G}}\Big] \,,\\
    J^i_\text{RR\,rad} &= \frac{\nu \Delta}{c^2} \varepsilon_{ijk} \,x^j v^k v^l \Big[\Pi^l + \mathcal{F}^l_{\bm{G}}\Big]\,.
\end{align}\end{subequations}
Finally, the tail-induced 4PN pseudo-Schott terms $E_\text{RR\,tail}$ and $J^i_\text{RR\,tail}$ are given, respectively, by Eqs.~\eqref{eq:E_diss_4PN} and \eqref{eq:J_diss_4PN}; here, it suffices to replace the quadrupole source moment by its expression in the center-of-mass.

We have checked that the flux-balance laws for the energy and angular momentum are still satisfied after reduction to the CM frame. As explained in detail in Sec.~VI of~\cite{BFT24}, the contributions due to the recoil (both in the equations of motion and in the Schott terms) are crucial in preserving these flux-balance laws. This verification is also important in light of the parametrized approach for obtaining the equations of motion, described in the following section, which relies on the validity of these flux balance laws in the CM frame. It is also a robust test of the validity of the 4.5PN-accurate formula for the passage to the CM frame derived in Eq.~\eqref{eq:P_Q_R_CoM_RR}.

\subsection{The Iyer-Will-Gopakumar parameters in harmonic gauge}

At 2.5PN and 3.5PN orders, Refs.~\cite{IW93,IW95} obtained the RR acceleration in the CM frame directly from the flux-balance laws. They wrote down the most general parametrized expression for the energy, angular momentum, and equations of motion which was local in time. Then, by imposing that the flux-balance laws for the energy and angular momentum hold, they found relations between the parameters. The residual free parameters corresponded to the gauge freedom; through a direct computation in generalized BT gauge~\cite{B97}, they found specific values for the coefficients corresponding to that choice of gauge. In Refs.~\cite{PW02, NB05}, a different set of parameters was found corresponding to the harmonic gauge.

At 4.5PN order, Ref.~\cite{GII97} extended the flux-balance method, and proposed a purely local expression for the CM acceleration at this order given by
\begin{align}\label{eq:accGII}
	a\RR^{i\,\text{GII}} = - \frac{8}{5}\frac{G^2m^2\nu}{c^3 r^3}\Bigl[-\bigl(A_\text{2.5PN}+A_\text{3.5PN}+A_\text{4.5PN}\bigr) \dot{r} n^i + \left(B_\text{2.5PN}+B_\text{3.5PN}+B_\text{4.5PN}\right) v^i\Bigr] + \calO\left(\frac{1}{c^{11}}\right)\,,
\end{align}
where the coefficients $A_\text{$n$PN}$ and $B_\text{$n$PN}$ are given by Eqs.~(2.8), (2.9), (2.11) and (2.16) in~\cite{GII97}. These coefficients depend on arbitrary gauge parameters which reflect the freedom left in the choice of coordinate system up to 4.5PN order, and are denoted
\begin{align}\label{IWGparameters}
	\mathcal{P} = \{\alpha_3, \beta_2, \xi_1, \xi_2, \xi_3, \xi_4, \xi_5, \rho_5, \psi_1, \psi_2, \psi_3, \psi_4, \psi_5, \psi_6, \psi_7, \psi_8, \psi_9, \chi_6, \chi_8, \chi_9\}\,.
\end{align}
At the 2.5PN order, only the two parameters $\{\alpha_3, \beta_2\}$ are sufficient; at 3.5PN, one must add $\{\xi_1, \xi_2, \xi_3, \xi_4, \xi_5, \rho_5\}$; and at 4.5PN order the full set of 20 parameters~\eqref{IWGparameters} are needed to parametrize the most general coordinate transformation. However, Ref.~\cite{BFT24} showed that at the 4.5PN order the equations of motion cannot be local-in-time in the CM frame. Indeed, due to the gravitational-wave recoil, semi-hereditary contributions arise at 4.5PN; see~Eq.~\eqref{eq:accCMrad}. Thus Eq.~\eqref{eq:accGII} cannot be the physical expression for the CM acceleration at 4.5PN order. Instead one must correct it by adding a piece $\delta a\RR^i$, such that the CM acceleration at 4.5PN order in an arbitrary gauge is actually given by  
\begin{align}\label{eq:accCM_GII_corrige}
	a\RR^{i} =	a\RR^{i\,\text{GII}} + \delta a\RR^i + \calO\left(\frac{1}{c^{11}}\right)\,.
\end{align}
This correction term is entirely given in terms of the fluxes of linear momentum and CM position as~\cite{BFT24}
\begin{align}\label{eq:deltaaccCM}
	\delta a\RR^i &= \frac{\Delta}{c^2}\left\{ \frac{G}{r^2}\left(2 n^iv^j+n^jv^i\right)\Bigl[\Pi^j + \calF_{\bm{G}}^j\Bigr] - \frac{v^i v^j}{m}  \Bigl[\calF_{\bm{P}}^j + \dot{\calF}_{\bm{G}}^j\Bigr]\right\}\,,
\end{align}
which involves besides the local fluxes~\eqref{eq:FP_FG_COM} the non-local integral of the linear momentum or recoil~\eqref{eq:def_Pi}.

Once this correction has been made, we can compare our end result for the CM 4.5PN RR acceleration computed in Sec.~\ref{sec:CM} 
with the parametrized expression~\eqref{eq:accCM_GII_corrige}, including the explicit correction~\eqref{eq:deltaaccCM}. We find a match if, and only if, the gauge parameters are given by
\begin{equation}\label{eq:paramGII-BT}
	\begin{array}{ll} \displaystyle \alpha_3 = 0\,,& \displaystyle \qquad\qquad \beta_2 = -1\,,\\[0.3cm]
		\displaystyle \xi_1 = \frac{271}{28}+6 \nu \,,& \displaystyle \qquad\qquad \xi_2 = - \frac{77}{4} - \frac{3}{2} \nu \,,\\[0.4cm]
		\displaystyle \xi_3 = \frac{79}{14}-\frac{92}{7} \nu \,,& \displaystyle \qquad\qquad \xi_4 = 10 \,,\\[0.4cm]
		\displaystyle \xi_5 = \frac{5}{42}+ \frac{242}{21} \nu \,,& \displaystyle \qquad\qquad \rho_5 = -\frac{439}{28} + \frac{18}{7} \nu \,,\\[0.4cm]
		\displaystyle \psi_1 = \frac{10139}{1008}- \frac{1977}{28}\nu-\frac{296}{7} \nu^2 \,,& \displaystyle \qquad\qquad \psi_2 = -\frac{53}{56}+\frac{10233}{56} \nu + \frac{261}{14} \nu^2 \,,\\[0.4cm]
		\displaystyle \psi_3 = -\frac{26561}{504} -\frac{3188}{63}\nu + \frac{886}{9} \nu^2 \,,& \displaystyle \qquad\qquad \psi_4 = -\frac{295}{16}-\frac{585}{4} \nu - \frac{15}{8} \nu^2 \,,\\[0.4cm]
		\displaystyle \psi_5 = -\frac{129}{2}-\frac{239}{63} \nu - \frac{908}{21} \nu^2 \,,& \displaystyle \qquad\qquad \psi_6 = \frac{170087}{756}+\frac{98299}{756}\nu - \frac{2942}{27} \nu^2 \,,\\[0.4cm]
		\displaystyle \psi_7 = \frac{25}{2}+35 \nu \,,& \displaystyle \qquad\qquad \psi_8 = -\frac{74315}{504}-\frac{24935}{252} \nu +\frac{2179}{63} \nu^2 \,,\\[0.4cm]
		\displaystyle \psi_9 = \frac{12119}{126}+\frac{1241}{252}\nu + \frac{8696}{189} \nu^2 \,,& \displaystyle \qquad\qquad \chi_6 = \frac{107}{48}+ \frac{88957}{504} \nu - \frac{2419}{504} \nu^2 \,,\\[0.4cm]
		\displaystyle \chi_8 = -\frac{1655}{84}-\frac{4660}{63} \nu + \frac{485}{126} \nu^2 \,,& \displaystyle \qquad\qquad \chi_9 = \frac{53687}{1512} + \frac{7757}{378}\nu - \frac{589}{12} \nu^2 \,.
	\end{array}
\end{equation}
This unique set of parameters corresponds to the unique harmonic coordinate system as we have argued from the construction \textit{via} matching of the metric in Sec.~\ref{sec:metricharmonic}. To 3.5PN order we recover the parameters computed in~\cite{PW02, NB05}.

\section{Verification of the Lorentz invariance}
\label{sec:Lorentz}

We have described in Sec.~\ref{sec:metricharmonic} the MPM-PN metric whose multipole expansion in the external zone is given by~\eqref{eq:MPM} and matches the PN expansion in the source's near zone. The metric satisfies globally the harmonic coordinate condition $\partial_\beta h^{\alpha\beta}=0$, where the harmonic coordinate system covers the general regular isolated source in the near zone and is asymptotically flat at infinity from the source. Consequently, the two-body acceleration (including the RR terms therein) derived in the previous sections should be manifestly Lorentz invariant. This is what we check now.  

The PN expanded RR acceleration of the particle~1 in a general frame is given in Sec.~\ref{sec:acceleration_general} as a functional of the individual positions of the particles $\bm{y}_1$ and $\bm{y}_2$ and their velocities $\bm{v}_1$ and $\bm{v}_2$. Actually, in harmonic coordinates many of the terms (but not all) can be combined into the relative position $\bm{r}_{12} = \bm{y}_1-\bm{y}_2$ (and relative separation $r_{12}=\vert\bm{y}_1-\bm{y}_2\vert$, with $\bm{n}_{12}=\bm{r}_{12}/r_{12}$) and velocity $\bm{v}_{12}=\bm{v}_1-\bm{v}_2$. The problem of checking the Lorentz invariance of the PN expanded acceleration (in a perturbative PN sense) has been solved in Ref.~\cite{BFregM}, and we summarize here the method. As the RR acceleration given by Eqs.~\eqref{aRR1harm}--\eqref{apolepart} is manifestly invariant by translation, we restrict attention to homogeneous proper Lorentz transformations or boosts, say $x'^\alpha = \Lambda^\alpha_{~\beta} x^\beta$, where the boost $\Lambda^\alpha_{~\beta}$ depends on the constant boost velocity $\bm{V}=(V_i)_{i=1,2,3}$ and is given by
\begin{subequations}\label{boost}
\begin{align}
	\Lambda^0_{~0} &= \gamma\,,\\
	\Lambda^i_{~0} &= \Lambda^0_{~i} = -\gamma {V_i\over c}\,,\\
	\Lambda^i_{~j} &= \delta_{ij}+\frac{\gamma^2}{\gamma+1}\frac{V_i V_j}{c^2}\,,
\end{align}
\end{subequations}
where the Lorentz factor reads $\gamma \equiv (1-\bm{V}^2/c^2)^{-1/2}$. In our convention, the boost is such that a particle which has velocity $\bm{V}$ at time $t$ in the frame $\{x^\alpha\}$ is at rest in the frame $\{x'^\alpha\}$ at time $t'$. With $x^\alpha=(c t, \mathbf{x})$ we have
\begin{subequations}\label{tx}
\begin{align}
	\label{eq:fieldtransformation}
	t' &= \gamma \Big(t- \frac{1}{c^2}({\bm V}\cdot{\mathbf{x}})\Big)\, , \\
	\mathbf{x}' &= \mathbf{x}- \gamma \bm{V}\Big(t-\frac{\gamma}{c^2(\gamma+1)}({\bm V}\cdot{\mathbf x})\Big)\, ,
\end{align}
\end{subequations}
where the dot denotes the ordinary scalar product.

We now give the end results for the changes of all the variables associated with the two-body system when subject to the Lorentz boost. All the formulas are consistently in the form of formal PN expansions, which need to be truncated at the requested PN order when $c\to\infty$, while keeping the boost velocity $\bm V$ independent of $c$. The transformations of the individual trajectories of each particles under the boost are 
\begin{subequations}\label{eq:y1y2}
	\begin{align}
		{\bm y}'_1 &= {\bm y}_1 - \gamma {\bm V} \left(t-\frac{1}{c^2}\frac{\gamma}{\gamma+1}({\bm V}\cdot{\mathbf{x}})\right) + \sum_{n=1}^{+\infty}\frac{(-)^n}{c^{2n} n!}\left(\frac{\partial}{\partial t}\right)^{\!n-1}\!\!\left[({\bm V}\cdot{\bm r}_1)^n \left({\bm
			v}_1-\frac{\gamma}{\gamma+1}{\bm V}\right)\right]\,,\\
		{\bm y}'_2 &= {\bm y}_2 - \gamma {\bm V} \left(t-\frac{1}{c^2}\frac{\gamma}{\gamma+1}({\bm V}\cdot{\mathbf{x}})\right) + \sum_{n=1}^{+\infty}\frac{(-)^n}{c^{2n} n!}\left(\frac{\partial}{\partial t}\right)^{\!n-1}\!\!\left[({\bm V}\cdot{\bm r}_2)^n \left({\bm
			v}_2-\frac{\gamma}{\gamma+1}{\bm V}\right)\right]\,,
	\end{align}
\end{subequations}
where we pose ${\bm r}_1 \equiv {\mathbf{x}}-{\bm y}_1$, ${\bm r}_2 \equiv {\mathbf{x}}-{\bm y}_2$, and the positions in the frame at time $t'$, \textit{i.e.} ${\bm y}'_1(t')$ and ${\bm y}'_2(t')$, are expressed as functions of $\mathbf{x}$ and the quantities at time $t$, \textit{i.e.} ${\bm y}_1(t)$ and ${\bm y}_2(t)$. The derivatives $\partial/\partial t$ on
the right-hand side are partial time derivatives with respect to the coordinate time $t$, with the spatial coordinate $\mathbf{x}$ being held constant. They act on $\bm{r}_1$ through the trajectory $\bm{y}_1$: \textit{e.g.}, we have $\partial\bm{r}_1/\partial t=-\bm{v}_1$ and $\partial(\bm{V}\cdot\bm{r}_1)/\partial t=-\bm{V}\cdot\bm{v}_1$. They also act on velocities and (derivatives of) accelerations: thus $\partial {\bm v}_1/\partial
t={\bm a}_1$, $\partial {\bm a}_1/\partial t={\bm b}_1$, and so on, where ${\bm a}_1$, ${\bm b}_1$ represent the acceleration and the derivative of the acceleration (in such cases the partial derivative is a total derivative, \textit{e.g.} $\dd{\bm v}_1/\dd t={\bm a}_1$). In Eq.~\eqref{eq:y1y2}, it is understood that all accelerations and derivatives of accelerations are systematically order reduced up to the required PN order by means of the equations of motion. For ${\bm r}_1$ and ${\bm r}_2$, we have
\begin{subequations}\label{eq:r1r2}
	\begin{align}
		{\bm r}'_1 &= {\bm r}_1-\sum_{n=1}^{+\infty}\frac{(-)^n}{c^{2n}
			n!}\left(\frac{\partial}{\partial t}\right)^{\!n-1}\!\!\left[({\bm
			V}\cdot{\bm r}_1)^n \left({\bm v}_1-\frac{\gamma}{\gamma+1}{\bm
			V}\right)\right]\,,\\ 
		{\bm r}'_2 &= {\bm
			r}_2-\sum_{n=1}^{+\infty}\frac{(-)^n}{c^{2n} n!}\left(\frac{\partial}{\partial t}\right)^{\!n-1}\!\!\left[({\bm V}\cdot{\bm r}_2)^n \left({\bm
			v}_2-\frac{\gamma}{\gamma+1}{\bm V}\right)\right]\,.
\end{align}
\end{subequations}$\!\!$
We also obtain the changes in the coordinate velocities ${\bm v}'_1(t')=\dd{\bm y}'_1/\dd t'$ and ${\bm v}'_2(t')=\dd{\bm y}'_2/\dd t'$, which follow from the law of transformation of the partial time derivative: $\partial_t'=\gamma(\partial_t+V_i\partial_i)$, where $\partial_i=\partial/\partial x^i$. We obtain
\begin{subequations}\label{eq:v1v2}
	\begin{align}
	{\bm v}'_1 &= \frac{1}{\gamma}{\bm v}_1-{\bm V}+\frac{1}{\gamma}\sum_{n=1}^{+\infty}\frac{(-)^n}{c^{2n} n!}\left(\frac{\partial}{\partial t}\right)^{\!n}\!\!\left[({\bm V}\cdot{\bm r}_1)^n \left({\bm
		v}_1-\frac{\gamma}{\gamma+1}{\bm V}\right)\right]\,,\\
	{\bm v}'_2 &= \frac{1}{\gamma}{\bm v}_2-{\bm V}+\frac{1}{\gamma}\sum_{n=1}^{+\infty}\frac{(-)^n}{c^{2n} n!}\left(\frac{\partial}{\partial t}\right)^{\!n}\!\!\left[({\bm V}\cdot{\bm r}_2)^n \left({\bm
	v}_1-\frac{\gamma}{\gamma+1}{\bm V}\right)\right]\,.
\end{align}
\end{subequations}
Similarly the transformation of the accelerations ${\bm a}'_1(t')=\dd{\bm v}'_1/\dd t'$ and ${\bm a}'_2(t')=\dd{\bm v}'_2/\dd t'$ reads
\begin{subequations}\label{eq:a1a2}
	\begin{align}
		{\bm a}'_1 &= \frac{1}{\gamma^2}\left\{{\bm a}_1 +\sum_{n=1}^{+\infty}\frac{(-)^n}{c^{2n} n!}\left(\frac{\partial}{\partial t}\right)^{\!n+1}\!\!\left[({\bm V}\cdot{\bm r}_1)^n \left({\bm
			v}_1-\frac{\gamma}{\gamma+1}{\bm V}\right)\right]\right\}\,,\label{eq:a1a21}\\
		{\bm a}'_2 &= \frac{1}{\gamma^2}\left\{{\bm a}_2 +\sum_{n=1}^{+\infty}\frac{(-)^n}{c^{2n} n!}\left(\frac{\partial}{\partial t}\right)^{\!n+1}\!\!\left[({\bm V}\cdot{\bm r}_2)^n \left({\bm
		v}_2-\frac{\gamma}{\gamma+1}{\bm V}\right)\right]\right\}\,.
	\end{align}
\end{subequations}

The acceleration is a functional of positions and velocities and we denote for time $t$ and $t'$ with a slight abuse of notation $\bm{a}_1(t) \equiv \bm{a}_1[\bm{y}_{1}(t),\bm{y}_{2}(t),\bm{v}_{1}(t),\bm{v}_{2}(t)]$ and $\bm{a}'_1(t') \equiv \bm{a}'_1[\bm{y}'_{1}(t'),\bm{y}'_{2}(t'),\bm{v}'_{1}(t'),\bm{v}'_{2}(t')]$. Applying the previous transformation laws we compute the difference of acceleration due to the boost, namely
\begin{align}
	\delta_{\Lambda}\bm{a}_1 \equiv \bm{a}'_1(t')-\bm{a}_1(t')=\bm{a}'_1\big[\bm{y}'_{1},\bm{y}'_{2},\bm{v}'_{1},\bm{v}'_{2}\big]-\bm{a}_1\big[\bm{y}'_{1},\bm{y}'_{2},\bm{v}'_{1},\bm{v}'_{2}\big]\, ,  
\end{align}
where the first term is given by the right-hand side of Eq.~\eqref{eq:a1a21} while the second term is obtained by the replacements in the initial harmonic-coordinates acceleration of the positions and velocities using Eqs.~\eqref{eq:y1y2} and~\eqref{eq:v1v2}, followed by the PN expansion and order reduction. After an explicit computation, we obtain
\begin{align}
	\delta_{\Lambda}\bm{a}_1 = 0\, ,  
\end{align}
for the conservative terms at order 2PN and the RR terms at relative 2PN order, \textit{i.e.}, up to 4.5PN order. In conclusion the harmonic-coordinates acceleration at 4.5PN in a general frame is indeed manifestly Lorentz invariant.\footnote{The conservative sector was previously shown to be Lorentz invariant up to 4PN order~\cite{BBFM17}.} This confirms that the general MPM-PN solution described in Sec.~\ref{sec:metricharmonic} corresponds to the unique global harmonic coordinate system covering source and radiation field. Note that in order to obtain this result we crucially need the inclusion of the gauge moments $\dW_L$, $\dX_L$, $\dY_L$ and $\dZ_L$ in the formalism. Omitting the gauge moments would result in a physically equivalent RR acceleration, but it would not be in harmonic coordinates and thus not be manifestly Lorentz invariant. Finally, we note that the DR pole contribution to the harmonic acceleration in Eq.~\eqref{aRR1polenonpole} is already manifestly invariant under the Lorentz transformation and plays no role for the present check.

\section{Conclusions}
\label{sec:conclusions}

In this paper, we have computed the radiation-reaction (RR) force for non-spinning compact binary systems up to the 2PN relative order, \textit{i.e.}, including the leading 2.5PN effect as well as the next-to-leading 3.5PN and next-to-next-to-leading 4.5PN corrections. We use the uniquely defined harmonic coordinate system covering the source and asymptotically flat at infinity. Our result applies to general binary orbits, in an arbitrary reference frame, and it completes the equations of motion of compact binaries up to 4.5PN order, since the conservative part has already been derived in harmonic coordinates to 4PN order~\cite{DJS14,BBFM17,FS19}, and previous computations of the dissipative part have obtained the RR force in harmonic coordinates at the 1PN relative order~\cite{PW02, NB05}.

The different steps leading to this result are illustrated in the figure~\ref{fig:summary_gauge_transformation} and summarized as follows. First, we performed a coordinate transformation of the RR acceleration previously obtained in the Burke-Thorne (BT) coordinate system at 4.5PN order~\cite{BFT24}. This step required computing, in the form of a PN expansion in the near zone, the linear and quadratic contributions of the gauge vector leading to the harmonic coordinates. We then obtained an expression for the RR force at 4.5PN order in harmonic coordinates, expressed solely in terms of the multipole moments of the compact binary source in an unexpanded form, \textit{i.e.}, without substituting explicit expressions for the moments or their time derivatives. The expression for the RR force in terms of the particles' positions and velocities involved the computation of source moments and, in addition, of different gauge moments, up to 2PN relative order. In particular we found that one of the gauge moment at 2PN order required the use of the full dimensional regularization procedure. Consequently, the RR acceleration at 4.5PN order in harmonic coordinates includes a pole in the physical limit $d \rightarrow 3$.

We draw our main results in Secs.~\ref{sec:acceleration_general} and~\ref{sec:CM} where we present the explicit acceleration for the dissipative  parts in harmonic coordinates up to 4.5PN order. A feature of this formulation is its relative simplicity compared to the BT RR force~\cite{BFT24}, together with the presence of a non-local-in-time contribution in the explicit acceleration. We have verified (and proven) the complete set of flux-balance equations corresponding to energy, angular momentum, linear momentum and center-of-mass position; in particular, we have explicitly obtained the expressions for the associated Schott terms. We have also derived the relative acceleration in the center-of-mass (CM) frame and obtained (after a correction to account for the binary's CM displacement and recoil by radiation~\cite{BFT24}) the Iyer-Will-Gopakumar~\cite{IW93,IW95,GII97} gauge parameters corresponding to the harmonic coordinate system. We explicitly confirm that the harmonic-coordinate acceleration is manifestly Lorentz invariant; this is one of the reasons for the simplicity of the final expression, since the number of possible terms is heavily constrained. Although the RR force has been derived in the main text using an exterior-zone approach, we show in Appendix~\ref{appendix:innerzone} that this derivation is equivalent (\textit{via} matching) to the near-zone approach. Our final result should be useful for comparisons with other approaches such as the gravitational self force (GSF) and the post-Minkowskian (PM) effective field theory. In particular, it agrees with the recent 2PM derivation of the RR force using effective field theory~\cite{BDG24}. The 4.5PN RR force could be implemented in current EOB waveforms~\cite{BuonD99,DNorleans}, and the methods developed here could be in principle extended to include spin-orbit and spin-spin effects at comparable PN order. Pushing the RR at such high 4.5PN level is also necessary to prepare for the full 4.5PN template waveforms (beyond circular orbits) to be used for supermassive binary black holes (SMBBH) observed by LISA, and to help the assessment of the estimation of the systematic errors due to the PN modeling in the data analysis of LISA (see \textit{e.g.}~\cite{Owen23}). 

\acknowledgments

L.B. thanks Thibault Damour and Donato Bini for a discussion on Burke-Thorne coordinates \textit{vs.} harmonic coordinates. We thank Donato Bini for having checked that our 4.5PN results agree with the 2PM RR results (up to order $G^2$) given in Ref.~\cite{BDG24}. D.T. acknowledges the support of the ERC Consolidator/UKRI Frontier Research Grant GWModels (selected by the ERC and funded by UKRI [grant number EP/Y008251/1]). Most computations as well as the formatting of lengthy equations were performed using the \textit{xAct} library~\cite{xtensor} for \emph{Mathematica}.

\appendix

\section{Construction of the coordinate transformation in the near zone}
\label{appendix:innerzone}

\subsubsection{General considerations}

The construction of the coordinate transformation between BT and harmonic coordinates in Sec.~\ref{sec:BTtoharmonic} was implemented in the vacuum region outside the source, and then formally developed in the near zone, when $r\to 0$. However, as all the computations and results of Ref.~\cite{BFT24} are presented in the near zone, including inside the matter source, an alternative approach consists in computing the coordinate transformation directly in the near zone following~\cite{BFT24}. The PN expansion of relevant quantities will be henceforth indicated by an overbar, \textit{e.g.}, the near-zone gravitational field will be denoted as \smash{$\overline{h}\ab$}. In harmonic gauge and in BT gauge such that \smash{$\partial_\beta \overline{h}\ab=0$}, the ``relaxed'' Einstein field equations to be solved for \smash{$\overline{h}{}\ab$} read
\begin{align} \label{eq:relaxed_EE}
  \Box \overline{h}\ab = \frac{16\pi G}{c^4} \overline{\tau}\ab \equiv  \frac{16\pi G}{c^4} \vert\overline{g}\vert \overline{T}\ab + \Lambda\ab[\overline{h}]\,,
\end{align}
where $\overline{g}$ is the determinant of \smash{$\overline{g}{}_{\alpha\beta}$} and $\Lambda\ab[\overline{h}]$ is a non-linear functional of \smash{$\overline{h}{}\ab$}. The important change compared with the exterior-zone calculation in Sec.~\ref{sec:BTtoharmonic}, is that we consider the matter stress energy tensor \smash{$\overline{T}\ab$}. The equation~\eqref{eq:relaxed_EE} can be solved iteratively using various schemes. One is based on the truncated version of Eq.~(3.7) in~\cite{BFN05}. Instead, here we shall resort to the near-zone iteration described in Sec.~2.2 of~\cite{BFT24}, valid in both the extended BT and the standard harmonic coordinates. We will remind it below, but rephrased in a way that makes more apparent the similarities and differences with the exterior-zone calculation. Next, it will be straightforward to infer the near-zone algorithm by transposing the exterior-zone algorithm developed in~\cite{BFL22} for computing the coordinate transformation. We start with the linearized solution, see Eq.~(2.16) in~\cite{BFT24},
\begin{align}\label{eq:h1def}
	\mathop{\overline{h}}_{}\!{}\ab_{(1)} \equiv \frac{16\pi}{c^4}
	\,\Box^{-1}_\text{inst}\big[\mathop{\overline{T}}_{}\!{}\ab\big] + \mathop{\overline{h}}_{}\!{}\ab_{\mathrm{RR}\, (1)}\,,
\end{align}
where \smash{$\Box^{-1}_\text{inst}$} is the ``instantaneous'' symmetric integral operator defined by Eq.~\eqref{inst}, while the radiation-reaction piece is generically defined as a certain functional \smash{$\overline{\mathcal{H}}{}\ab_{\mathrm{RR\, gen}\,{(1)}}$} of the source and, possibly, gauge moments:
\begin{align}
	\overline{h}{}\ab_{\mathrm{RR}\,{(1)}} \equiv \overline{\mathcal{H}}{}\ab_{\mathrm{RR\, gen}\,{(1)}}[\dI_L,\dJ_L, \dW_L, \dX_L, \dY_L, \dZ_L]\,,
\end{align}
Here for convenience we adopt a different notation for the object and for the functional defining this object in terms of the multipole moments. In extended BT coordinates~\cite{BFT24}, the functional only depends on the source moments $\dI_L$, $\dJ_L$ and reads 
\begin{subequations}
\begin{align}
\overline{\mathcal{H}}{}^{00}_{\mathrm{RR}\,\text{BT}\,{(1)}}[\dI_J,\dJ_L] &=  -\frac{4}{c^2}\sum_{\ell=2}^{+\infty}
		\frac{(-)^\ell}{\ell !} \frac{(\ell+1)(\ell+2)}{\ell (\ell-1)} \partial_L \Biggl\{\frac{\overline{\dI_L(t-r/c) - \dI_L(t+r/c)}}{2r} \Biggr\}\,, \\
\overline{\mathcal{H}}{}^{0i}_{\mathrm{RR}\,\text{BT}\,{(1)}}[\dI_J,\dJ_L]  &=
		\frac{4}{c}\sum_{\ell=2}^{+\infty} \frac{(-)^\ell}{\ell !} \frac{\ell+2}{\ell-1}
		\Biggl[ \frac{2\ell+1}{\ell} \hat{\partial}_{iL} \Biggl\{
		\frac{\overline{\dI_{L}^{(-1)}(t-r/c) - \dI_{L}^{(-1)}(t+r/c)}}{2r} \Biggr\}  \nn \\ & \qquad \qquad \qquad \qquad - \frac{\ell}{(\ell+1)\, c^2}
		\varepsilon_{iab} \partial_{aL-1} \Biggl\{ \frac{\overline{\dJ_{bL-1}(t-r/c) - \dJ_{bL-1}(t+r/c)} }{2r}\Biggr\}\Biggr]\,, \\
\overline{\mathcal{H}}{}^{ij}_{\mathrm{RR}\,\text{BT}\,{(1)}}[\dI_J,\dJ_L] &=
		- 4 \sum_{\ell=2}^{+\infty} \frac{(-)^\ell}{\ell !} \frac{2\ell+1}{\ell-1}
	 \Biggl[ \frac{2\ell+3}{\ell} \hat{\partial}_{ijL} \Biggl\{ \frac{\overline{\dI_{L}^{(-2)}(t-r/c) - \dI_{L}^{(-2)}(t+r/c)}}{2r} \Biggr\} \nn \\ &  \qquad \qquad \qquad \qquad \quad - \frac{2\ell}{(\ell+1)\, c^2}
		\varepsilon_{ab(i} \hat{\partial}_{j)aL-1} \Biggl\{ 
		\frac{\overline{\dJ_{bL-1}^{(-1)}(t-r/c) - \dJ_{bL-1}^{(-1)}(t+r/c)}}{2r} \Biggr\}\Biggr]\,.
\end{align}  
\end{subequations}
In harmonic coordinates, the functional is given by Eq.~\eqref{eq:h1} together with~\eqref{h1can} and~\eqref{gaugezeta}, except that all retarded moments in~\eqref{h1can} must be replaced by their PN expanded time antisymmetric (retarded minus advanced) combinations.
Denoting as \smash{$\overline{\mathcal{H}}{}\ab_{\text{RR}\,{(1)}}[\dI_L, \dJ_L]$} the latter contribution coming from Eq.~\eqref{h1can}, which depends on the source moments, and by \smash{$\overline{\mathfrak{Z}}{}^\alpha_{(1)}[\dW_L, \dX_L, \dY_L, \dZ_L]$} the one which coincides with the near-zone expansion of the expression~\eqref{gaugezeta} for the gauge vector $\zeta{}^\alpha_{(1)}$, depending on the gauge moments, we can write in harmonic coordinates
\begin{align}\label{expressionharm}
  \overline{\mathcal{H}}\ab_{\text{RR\,harm}\,{(1)}}[\dI_L, \dJ_L, \dW_L, \dX_L, \dY_L, \dZ_L] = \overline{\mathcal{H}}\ab_{\text{RR}\,{(1)}}[\dI_L,\dJ_L] +  \partial\overline{\mathfrak{Z}}\ab_{{(1)}}[\dW_L, \dX_L, \dY_L, \dZ_L] \,.
\end{align}
The $\overline{\mathcal{H}}{}\ab_{\mathrm{RR}\,\text{gen}\,{(1)}}$'s are regular functions in the near zone and satisfy the divergence-free condition \smash{$\partial_{\beta} \overline{\mathcal{H}}{}\ab_{\mathrm{RR}\,\text{gen}\,{(1)}} =0$}. A property of the linear solution~\eqref{eq:h1def} is that, due to the conservation of the stress-energy matter tensor, \smash{$\partial_\beta \overline{T}{}\ab = \mathcal{O}(G)$}, we have \smash{$\partial_{\beta} \mathop{\overline{h}}_{}\!{}\ab_{(1)} =\mathcal{O}(G)$}: the gauge condition is fulfilled at linear order, but only up to $\mathcal{O}(G)$ corrections.

Higher order contributions are constructed iteratively, in powers of $G$, namely
\begin{align}
	\overline{h}\ab = \sum_{m=1}^{+\infty} G^m\overline{h}\ab_{(m)} \,.
\end{align}
Beware that here, contrary to the usual MPM expansion, each of the coefficients \smash{$\overline{h}\ab_{(m)}$} is in the form of a PN expansion.
Knowing all the \smash{$\mathop{\overline{h}}_{}\!{}\ab_{(m)}$}'s for $1 \leqslant m\leqslant n-1$, the solution of order $n$
is obtained by solving
\begin{subequations}
  \begin{align}
    \label{eq:relaxed_EEn}
     & \Box \mathop{\overline{h}}_{}\!{}\ab_{(n)} \equiv 
	\overline{\Sigma}{}\ab_{(n)} \,, \\
    \label{eq:harmonicity}
   &\partial_\beta \mathop{\overline{h}}_{}\!{}\ab_{(n)} =- \partial_\beta \mathop{\overline{h}}_{}\!{}\ab_{(\leqslant n-1)} + \mathcal{O}(G^{n+1})\,,
  \end{align}
\end{subequations}
where \smash{$\overline{\Sigma}{}\ab_{(n)}$} is a function of the previous \smash{$\overline{h}\ab_{(\leqslant n-1)}$}'s and stands for the $n$-th order contribution to the non-linear part of the effective source in Eq.~\eqref{eq:relaxed_EE}.\footnote{More explicitly we have, see Eq.~(2.19) in~\cite{BFT24}:
\begin{align*}
	\overline{\Sigma}\ab = \frac{16\pi G}{c^4}\bigl(|\overline{g}|-1\bigr) \overline{T}\ab + \Lambda\ab[\overline{h}]\,.
\end{align*}
}
The right-hand sides of both equations are entirely determined by previous orders $m\leqslant n-1$. Unlike what happens in the exterior-zone method, the divergence-free condition is not verified by \smash{$\overline{h}{}\ab_{(n)}$} alone but, instead, by the truncated field \smash{$\overline{h}{}\ab_{(\leqslant n)}$}
which is such that \smash{$\partial{}_\beta \overline{h}{}\ab_{(\leqslant n)} = \mathcal{O}(G^{n+1})$}. This results from the fact that \smash{$\partial_\beta \overline{\tau}{}\ab_{(\leqslant n-1)} = \mathcal{O}(G^{n})$} vanishes only approximately.
Following Eq.~(2.18) of~\cite{BFT24} we then write
\begin{align}\label{eq:hndef}
	\mathop{\overline{h}}_{}\!{}\ab_{(n)} \equiv 
	 \overline{u}\ab_{(n)} + \overline{v}\ab_{(n)} \,,
\end{align}
where \smash{$\overline{u}{}\ab_{(n)} \equiv \Box^{-1}_\text{inst} \overline{\Sigma}{}\ab_{(n)} $} is a particular solution of the $n$-th order field equation~\eqref{eq:relaxed_EEn}, while \smash{$\overline{v}{}\ab_{(n)}$} represents some homogeneous solution specified below such that Eq.~\eqref{eq:harmonicity} holds. Note that the operator ${\Box}^{-1}_\text{inst}$ used here generalizes the symmetric integral operator ${\Box}^{-1}_\text{inst}$ of Eq.~\eqref{eq:h1def} to the case of non-compact supported sources, and involves the finite part regularization operator FP (see also the footnote~\ref{foot:FP}):
\begin{align} \label{eq:propsym}
	{\Box}^{-1}_{\text{inst}} \big[\overline{\Sigma}{}\ab_{(n)}\big] \equiv \FP\Box^{-1}_{\text{inst}} \big[\widetilde{r}^B \overline{\Sigma}\ab_{(n)}\big] = \sum_{k=0}^{+\infty}\left(\frac{\partial}{c\partial
		t}\right)^{\!\!2k}\FP\Delta^{-k-1} \big[\widetilde{r}^B \overline{\Sigma}\ab_{(n)}\big] \,.
\end{align}
The construction is similar to the MPM procedure in the exterior zone, with however three noticeable differences: (1) the quantities \smash{$\overline{h}{}\ab_{(n)}$} have the form of near-zone PN expansions instead of multipolar expansions; (2) \smash{$\overline{u}{}\ab_{(n)}$} is computed with the inverse operator $\Box^{-1}_\text{inst}$ instead of the retarded integral operator; (3) neither the divergence \smash{$\partial_\beta  \overline{h}{}\ab_{(n)}$},  nor even \smash{$\partial_\beta  \overline{h}{}\ab_{(\leqslant n)}$}, exactly vanish. Nevertheless let us show that \smash{$\partial_\beta  \overline{h}{}\ab_{(\leqslant n)}+\partial_\beta  \overline{u}{}\ab_{( n)}=\mathcal{O}(G^n)$}. Posing \smash{$\overline{u}{}\ab = \frac{16\pi G}{c^4}\, \Box^{-1}_{\text{inst}}\overline{T}\ab + \sum_{n \geqslant 2} G^n \overline{u}{}\ab_{(n)}$} and using the definition of \smash{$\overline{u}{}\ab_{(n)}$}, we get
\begin{align} \label{eq:divu}
\partial_\beta  \overline{u}\ab = \frac{16\pi G}{c^4} \partial_\beta {\Box}^{-1}_{\text{inst}} \big[  \overline{\tau}\ab \big] = \frac{16\pi G}{c^4} \bigl[\partial_\beta, {\Box}^{-1}_{\text{inst}}\bigr] \overline{\tau}\ab \,,
\end{align}
where the last equality comes from the fact that \smash{$\Box \partial{}_\beta \overline{h}{}\ab = \frac{16\pi G}{c^4} \partial{}_\beta \overline{\tau}{}\ab = 0$} in harmonic coordinates. We have introduced in~\eqref{eq:divu} the ``commutator'' operator $[\partial{}_\beta, {\Box}^{-1}_{\text{inst}}] =  \partial{}_\beta {\Box}^{-1}_{\text{inst}} - {\Box}^{-1}_{\text{inst}} \partial{}_\beta$, which is generically non-zero, due to the presence of the regularization factor $\widetilde{r}^B=(r/r_0)^B$ and the finite part FP in the definition of the operator~\eqref{eq:propsym}. For instance, the action of the commutator on regular functions $F$ for which $\Box^{-1}_\text{inst}$ is well-defined [see Eq.~\eqref{eq:F3d_infinity}] reads 
\begin{align} \label{eq:def_commutator}
\bigl[\partial{}_i, {\Box}^{-1}_{\text{inst}}\bigr] F \equiv  \partial{}_i {\Box}^{-1}_{\text{inst}} F - {\Box}^{-1}_{\text{inst}} \partial{}_i F = \FP \Box^{-1}_{\text{inst}}\Bigl[ B\, \widetilde{r}^B \frac{n^i}{r} F\Bigr]\,,
\end{align}
while we have obviously $[\partial{}_0, {\Box}^{-1}_{\text{inst}}\bigr]F=0$. Eq.~\eqref{eq:def_commutator} is a homogeneous solution of the vacuum wave equation, as can be checked by applying the d'Alembertian operator on the first equality. Since symmetric integrals or iterated Poisson operators commute with space (and time) derivatives when applied to compact-supported sources (since no FP regularization is needed in that case), only the non-compact piece of \smash{$\overline{\tau}{}\ab$} (\textit{i.e.}, \smash{$\Lambda{}\ab[\overline{h}]$}) contributes to the divergence \smash{$\overline{w}{}^\alpha \equiv \partial_\beta \overline{u}{}\ab$} in Eq.~\eqref{eq:divu}. Focusing on the terms in $G^n$, we thus conclude that
\begin{align} \label{eq:wn}
  \overline{w}^\alpha_{(n)} = \bigl[\partial_\beta, {\Box}^{-1}_{\text{inst}}\bigr] \Lambda\ab_{(n)}[\overline{h}]\,.
\end{align}
The tensor \smash{$\overline{v}{}\ab_{(n)}$} is then computed from \smash{$\overline{w}{}^\alpha_{(n)}$} in the near-zone with the help of the ``harmonicity'' algorithm described in the Appendix~A of Ref.~\cite{BFT24}. 

The extended BT metric and the harmonic metric are solutions of the same equations, with the same stress-energy tensor, but differ at linear order by a known, linear gauge transformation denoted \smash{$\overline{\varphi}{}^\alpha_{(1)}$}. Thus, modulo a possible homogeneous solution, regular over space and respecting the harmonic gauge condition, they must represent the same physical metric. This regular homogeneous solution can be written in the most general way as \smash{$\overline{\mathcal{H}}{}\ab_{\text{RR}\, {(1)}}[\delta \dI_L, \delta \dJ_L]$}, for some moments $\delta \dI_L$, $\delta \dJ_L$, plus a linear gauge transformation \smash{$\delta \overline{\zeta}{}^\alpha_{(1)} = \overline{\mathfrak{Z}}{}^\alpha_{(1)}[\delta \dW_L, \delta \dX_L, \delta \dY_L, \delta \dZ_L]$}, depending on other gauge-type moments $\delta \dW_L$, $\cdots$, $\delta \dZ_L$. At non-linear order the BT and harmonic metrics are thus related by the coordinate transformation~\eqref{Any_PM_metric_harmonic_to_BT}--\eqref{eq:gauge_transform}, up to a homogeneous solution of the previous type, and, as a corollary, the relation~\eqref{Relation_psi_omega_any_PM} linking the gauge vector \smash{$\overline{\varphi}^\alpha$} to the non-linear gauge transformation \smash{$\overline{\Omega}{}\ab=\Omega{}\ab[\overline{\varphi},\overline{h}]$}, must still hold.
Moreover, the moments $\delta \dI_L$, $\delta \dJ_L$ found above will imply that the source moments $\dI_L$, $\dJ_L$ in harmonic coordinates must be related to the corresponding moments defined in BT coordinates by $\dI_L = \dI^{\text{BT}}_L + \delta \dI_L$ and $\dJ_L = \dJ^{\text{BT}}_L + \delta \dJ_L$, and similarly for the four supplementary gauge type moments $\delta \dW_L$, $\delta \dX_L$, $\delta \dY_L$ and $\delta \dZ_L$. 

The construction of the gauge vector~\eqref{expphi} in the near zone is achieved by means of a method similar to the one sketched in section~\ref{sec:BTtoharmonic} (see \textit{e.g.},~\cite{BFL22}). In particular, we still proceed by recurrence on the PM order $n$, starting with the near zone expansion \smash{$\overline{\varphi}{}^\alpha_{(1)}$} of the linear gauge vector~\eqref{linearvarphi1} supplemented with Eqs.~\eqref{xi1} and~\eqref{gaugezetaasym2}.\footnote{The gauge vector used in~\cite{BFT24} to construct the extended BT coordinates was actually \smash{$\overline{\varphi}'{}^\alpha_{(1)} =  - \overline{\xi}{}^\alpha_{(1)}-\overline{\zeta}^\alpha_{(1)}+\overline{\psi}{}^\alpha_{(1)}$}, with two possible choices for \smash{$\overline{\psi}{}^\alpha_{(1)}$}, given in Eqs.~(A8a) or~(A8b) of~\cite{BFT24}, combined with~(A7). Both possible expressions depend on certain quantities $\mathcal{P}_L$, \smash{$\mathcal{Q}^{(+)}_L$}, \smash{$\mathcal{Q}^{(0)}_L$} and \smash{$\mathcal{Q}^{(-)}_L$} defined at the beginning of Appendix A in~\cite{BFT24}, and which are known to be at least of order 2PN. Moreover, $\mathcal{P}$ at 3PN order, as well as $\mathcal{P}_i$, $\mathcal{P}_{ij}$, \smash{$\mathcal{Q}{}^{(+)}_{i}$} and \smash{$\mathcal{Q}{}^{(+)}_{ij}$} at 2PN order, all vanish. Finally, we have checked, using techniques similar to those described around~\eqref{eq:P_expansion}--\eqref{eq:x3dP1dP2} below, that $\mathcal{P}_i$ and \smash{$\mathcal{Q}{}^{(+)}_{i}$} vanish at 3PN order, as so do  \smash{$\mathcal{Q}{}^{(0)}_i$}, \smash{$\mathcal{Q}{}^{(-)}$} and \smash{$\mathcal{Q}{}^{(-)}_i$} at 2PN order. This  is sufficient to prove that the gauge vector \smash{$\overline{\psi}{}^{\alpha}_{(1)}$} does not contribute to the contact transformation~\eqref{contact} at 4.5PN order, and can thus be ignored, so that \smash{$\overline{\varphi}{}^{\alpha}_{(1)}= - \overline{\varphi}'{}^{\alpha}_{(1)}$}. In conclusion, the gauge vector \smash{$\overline{\varphi}{}^{\alpha}_{(1)}$} is the required one to bring us back to the harmonic gauge at 4.5PN order.}
As recurrence hypothesis at order $n$, we assume that all the \smash{$\overline{\varphi}{}^\alpha_{(m)}$}'s in~\eqref{expphi} have been already determined up to order $n-1$ (with $n\geqslant 2$), and that the BT source moments $\dI_L^{\text{BT}}$, $\dJ_L^{\text{BT}}$ and gauge moments $\dW_L^{\text{BT}}$, $\dX_L^{\text{BT}}$, $\dY_L^{\text{BT}}$, $\dZ_L^{\text{BT}}$ are known in terms of the moments $\dI_L$, $\dJ_L$, $\cdots$, $\dZ_L$ of harmonic coordinates. At leading order both sets of moments agree. We thus write
\begin{align}
	\overline{\varphi}^\alpha_{(\leqslant n-1)} \equiv \sum_{m=1}^{n-1} G^m \overline{\varphi}^\alpha_{(m)}\,,
\end{align}
together with
\begin{subequations}\label{eq:iterated_quantities}
	\begin{align}\label{eq:iterated_moments}
	\dI^{\text{BT}}_{{(\leqslant n-1)}\,L} &\equiv \dI_L + \sum_{m=2}^{n-1} G^{m-1}
	\mathcal{I}_{{(m)}\,L}\left[\dI_K, \dJ_K, \cdots, \dZ_K\right]\,,\\ \dJ^{\text{BT}}_{{(\leqslant n-1)}\,L} &\equiv \dJ_L + \sum_{m=2}^{n-1} G^{m-1} \mathcal{J}_{{(m)}\,L}\left[\dI_K, \dJ_K, \cdots, \dZ_K\right]\,,\\
   & \makebox[0.2\textwidth]{\dotfill} \nn\\
    \dZ^{\text{BT}}_{{(\leqslant n-1)}\,L} &\equiv \dZ_L + \sum_{m=2}^{n-1} G^{m-1} \mathcal{Z}_{{(m)}\,L}\left[\dI_K, \dJ_K, \cdots, \dZ_K\right]\,,
\end{align}
\end{subequations}
where the functionals $\mathcal{I}_{{(m)}\, L}$, $\mathcal{J}_{{(m)}\, L}$, $\cdots$, $\mathcal{Z}_{{(m)}\, L}$ depend on the harmonic multipole moments. At linear order we have
\begin{align}
 \overline{h}\ab_{\text{BT}\,{(1)}} +\partial \overline{\varphi}\ab_{{(1)}} = \frac{16\pi}{c^4}\Box^{-1}_{\text{inst}} \overline{T}\ab + \overline{\mathcal{H}}\ab_{\text{RR}\,{(1)}}[\dI^{\text{BT}}_{(\leqslant n-1)}, \dJ^{\text{BT}}_{(\leqslant n-1)}] + \partial \overline{\mathfrak{Z}}\ab_{{(1)}}[\dW^{\text{BT}}_{(\leqslant n-1)}, \cdots, \dZ^{\text{BT}}_{(\leqslant n-1)}] \,. \label{eq:linear_order}
\end{align}
By construction, the relations (for $2\leqslant m\leqslant n-1$)
\begin{equation} \label{eq:hharmm}
\overline{h}\ab_{(m)} = \overline{h}\ab_{\text{BT}\,{(m)}} + \partial\overline{\varphi}\ab_{(m)} + \Omega\ab_{(m)}\bigl[\overline{\varphi}_{(1)}\cdots\overline{\varphi}_{(m-1)}; \overline{h}_{\text{BT}\,{(1)}}\cdots \overline{h}_{\text{BT}\,{(m-1)}}\bigr]\,,
\end{equation}
hold identically. Here, the $m$-th order piece of the metric in the BT gauge is given by a functional \smash{$\overline{\mathcal{H}}{}\ab_{\text{BT}\, {(m)}}$} of the matter stress-energy tensor \smash{$\overline{T}{}\ab$} and the so far determined BT moments: 
\begin{align} \label{eq:hBTm}
  \overline{h}\ab_{\text{BT}\,{(m)}} = \overline{\mathcal{H}}\ab_{\text{BT}\, {(m)}}[\overline{T},\dI^{\text{BT}}_{{\leqslant m-1}\,L}, \dJ^{\text{BT}}_{{(\leqslant m-1)}\,L}] \,,
\end{align}
which is obtained through the PN iteration scheme sketched around~\eqref{eq:hndef}--\eqref{eq:wn}. The non-linear functional \smash{$\overline{\Omega}\ab_{(m)}$} in~\eqref{eq:hharmm} can be calculated explicitly up to arbitrary high orders in terms of the \smash{$\overline{\varphi}{}^\alpha_{(p)}$}'s and the \smash{$\overline{h}{}\ab_{\text{BT}\,(p)}$}'s by means of Eqs.~\eqref{Any_PM_metric_harmonic_to_BT}--\eqref{eq:gauge_transform}. It is convenient to pose  \smash{$\overline{\Delta}{}^\alpha_{(n)} \equiv \partial_\beta \overline{\Omega}\ab_{(n)}$}. With this notation, \smash{$\overline{\varphi}^\alpha_{(n)}$} satisfies the equation
\begin{align} \label{eq:box_varphin}
  \Box \overline{\varphi}^\alpha_{(n)} =  \overline{\Delta}^\alpha_{(n)}\,,
\end{align}
which generalizes Eq.~\eqref{Relation_psi_omega_any_PM}. Finally, in order to obtain \smash{$\overline{\varphi}{}^\alpha_{(n)}$} at order $n$ and the corresponding modifications of the moments, we proceed as follows (see~\cite{BFL22}, \textit{mutatis mutandis}, for details). Relying first on harmonic analysis techniques described in Sec.~(5.2) of~\cite{MHLMFB20}, we compute \smash{$\overline{\varphi}{}^{\alpha}_{(n)}$} from Eq.~\eqref{eq:box_varphin} as 
\begin{align}
	\overline{\varphi}^{\alpha}_{(n)} = \Box^{-1}_{\text{inst}} \overline{\Delta}^\alpha_{(n)} \,.
\end{align}
This is the counterpart of the exterior-zone analysis in Sec.~\ref{sec:BTtoharmonic} where the gauge vector is found to be given by the same particular solution of the wave equation. Next we compute the corresponding homogeneous solution of the vacuum wave equation in terms of the commutators as (see Eqs.~(3.19)--(3.21) in~\cite{BFL22})
\begin{subequations}\label{exprcommutators}
	\begin{align}
		\overline{U}\ab_{{(n)}} &= \overline{X}\ab_{{(n)}} + \overline{Y}\ab_{{(n)}}\,,\\
		\overline{X}\ab_{{(n)}} &= \bigl[{\Box}^{-1}_{\text{inst}}, \Box\bigr] \overline{\Omega}\ab_{(n)} \,,\\
		\overline{Y}\ab_{{(n)}} &= \bigl[{\Box}^{-1}_{\text{inst}},2\partial^{(\alpha}\delta^{\beta)}_\lambda - \eta\ab \partial_\lambda \bigr] \overline{\Delta}{}^\lambda_{(n)} \,.
	\end{align}
\end{subequations}
The commutators are evaluated with the help of Eq.~\eqref{eq:def_commutator}. Eqs.~\eqref{eq:box_commutator}--\eqref{eq:der_commutator} below will give some even more explicit formulas from which it becomes apparent that \smash{$U{}\ab_{{(n)}}$} is a regular function in the near zone. Next we form the divergence \smash{$\overline{W}{}^\alpha_{(n)} = \partial_\beta \overline{U}{}\ab_{(n)}$} and construct a homogeneous solution \smash{$\overline{V}{}\ab_{(n)}$} of the wave equation such that \smash{$\partial_\beta \overline{V}{}\ab_{(n)} = - \overline{W}{}^\alpha_{(n)}$}, by means of the harmonicity algorithm (see~\cite{BFL22}). The quantity
\begin{align} \label{eq:Hab}
	\overline{H}\ab_{(n)} = \overline{U}\ab_{(n)} + \overline{V}\ab_{(n)}\,,
\end{align}
is a regular homogeneous solution of the vacuum wave equation satisfying \smash{$\partial_\beta \overline{H}{}\ab_{(n)} =0$}. As already mentioned, for such a solution there must exist a set of source and gauge-type moments $\delta \dI_{{(n)}\, L}$, $\cdots$, $\delta \dZ_{{(n)}\,L}$ such that
\begin{align} \label{eq:Hn}
	G^n \overline{H}\ab_{(n)} = G \overline{\mathcal{H}}\ab_{\mathrm{RR}\,{(1)}}[\delta \dI_{{(n)} \, L}, \delta \dJ_{{(n)} \, L}] + G \overline{\mathfrak{Z}}\ab_{(1)}[\delta \dW_{{(n)} \, L}, \cdots, \delta \dZ_{{(n)} \, L}]\,.
\end{align}
Finally, we add the expression~\eqref{eq:Hn} to the right-hand side of~\eqref{eq:linear_order} and obtain
\begin{align}
	\overline{\mathcal{H}}{}\ab_{\text{RR}\,{(1)}}[\dI^{\text{BT}}_{(\leqslant n-1)}, \dJ^{\text{BT}}_{(\leqslant n-1)}] + \overline{\mathcal{H}}{}\ab_{\text{RR}\,{(1)}}[\delta \dI^{\text{BT}}_{(n)}, \delta \dJ^{\text{BT}}_{(n)}] = \overline{\mathcal{H}}{}\ab_{\text{RR}\,{(1)}}[I^{\text{BT}}_{(\leqslant n-1)} + \delta \dI^{\text{BT}}_{(n)}, \dJ^{\text{BT}}_{(\leqslant n-1)} + \delta \dJ^{\text{BT}}_{(n)}] \,,
\end{align}
which defines the updated versions of the moments at order $n$, \text{i.e.},
\begin{align}
	 \dI^{\text{BT}}_{(\leqslant n)} = \mathcal{I}^{\text{BT}}_{(\leqslant n-1)} + \delta \dI^{\text{BT}}_{(n)}\,,\qquad\qquad \dJ^{\text{BT}}_{(\leqslant n)} = \mathcal{J}^{\text{BT}}_{(\leqslant n-1)} + \delta \dJ^{\text{BT}}_{(n)}\,.
\end{align}
The updated versions of the gauge moments are found similarly. Finally the algorithm permits us to reconstitute order by order the relevant moments in extended BT coordinates starting from those in harmonic coordinates.

\subsubsection{Application at 4.5PN order}

To quadratic order, for $n=2$, the source of Eq.~\eqref{box equation}, \textit{i.e.} the quantity $\overline{\Delta}\ab_{(2)}$, is given by~\eqref{Delta2}. At the requested order the metric can be replaced by the leading PN expressions
\begin{align}
  \overline{h}{}^{00} = -\frac{4 U}{c^2} + \mathcal{O}\left( \frac{1}{c^4} \right), \qquad \overline{h}{}^{0i} = - \frac{4U_i}{c^3} + \mathcal{O}\left( \frac{1}{c^5}\right) , \qquad \overline{h}{}^{ij} = \mathcal{O}\left( \frac{1}{c^4} \right) ,
\end{align}
where for compact binary sources $U = \frac{G m_1}{r_1} + \frac{G m_2}{r_2}$ and $U_i = \frac{G m_1}{r_1}v_1^i + \frac{G m_2}{r_2}v_2^i$. This yields
\begin{align}
  & \overline{\Delta}{}^0_{(2)} = \frac{16 \, G \, U \dW^{(3)}}{c^8} + \mathcal{O} \left(\frac{1}{c^{10}} \right) ,\nn \\
  & \overline{\Delta}{}^i_{(2)} = \frac{4G}{c^9} \left(4 U \dY^{(3)}_i - 2 U_j \dI^{(4)}_{ij} - x^j U \dI^{(5)}_{ij}\right) + \mathcal{O}\left( \frac{1}{c^9} \right) ,
\end{align}
where we have truncated consistently with the 4.5PN order required for $\overline{\varphi}{}^\alpha_{(2)}$ in the contact transformation~\eqref{contactacc}. The solution~\eqref{2PMcontributionShift} for \smash{$\overline{\varphi}{}^\alpha_{(2)}$} is recovered directly in the near zone from \smash{$\overline{\varphi}{}^\alpha_{(2)} = {\Box}^{-1}_{\text{inst}} \overline{\Delta}{}^\alpha_{(2)}$} (see Sec.~(4.2) of~\cite{BFT24}).

To compute the (possible) modifications of the multipole moments entering the extended BT metric we follow Eqs.~\eqref{exprcommutators}. We need the expression for the commutators $[{\Box}^{-1}_{\text{inst}},\Box]$ and $[{\Box}^{-1}_{\text{inst}},\partial_i]$ applied to a generic smooth function $F(\mathbf{x},t)$. The commutators come from the behaviour of the function for large $r$ (since the FP procedure we use is actually an infra-red regularization). We assume, in three dimensions, the asymptotic expansion when $r\to +\infty$:
\begin{align} \label{eq:F3d_infinity}
  F(\mathbf{x},t) = \sum_{N \leqslant p \leqslant p_1 \atop 0 \leqslant q \leqslant q_1} \!\!\!\! r^p \ln^q \widetilde{r} \, \sum_{\ell=0}^{+\infty} f^{L}_{p\,, q} \hat{n}_L\,,
\end{align}
where the coefficients $\mathop{f}{}^{L}_{p,\, q}$ are STF over the multi-index $L$, and $p_1$, $q_1$ are integers.
It is useful, following~\cite{BFHLT23b, BFT24}, to define, for each of the terms in~\eqref{eq:F3d_infinity}, its $\pi$-value as: $\pi(r^p \ln^q \widetilde{r} \, \hat{n}_L)= 0$ if $p+\ell$ is even, and $\pi(r^p \ln^q \widetilde{r} \,\hat{n}_L)= 1$ if $p+\ell$ is odd. If all terms in~\eqref{eq:F3d_infinity} have the same $\pi$-value $b$, then $F$ is stated to have a well-defined $\pi$-value, equal to $b$. Let us recall that the $\pi$-value of a function, whenever it exists, is preserved by space and time differentiation, as well as by the action of the regularized iterated inverse Poisson operator $\FP \Delta^{-k-1}\,\widetilde{r}^B$ (with $k\in \mathbb{N}$) or the ``instantaneous'' symmetric integral operator $\Box^{-1}_{\text{inst}}$. The $\pi$-value of a product is the sum, modulo 2, of the $\pi$-values of its factors, as long as they are meaningfully defined.

Adapting the method explained in Sec.~III.B of~\cite{BFHLT23b} we get (in three dimensions)
\begin{subequations} \label{eq:box_commutator}
\begin{align}
  \bigl[{\Box}^{-1}_{\text{inst}},\Box\bigr] F &= \sum_{\ell \geqslant 0\,, s \geqslant 0} \Delta^{-s} \hat{x}_L \frac{\hat{h}^{(2s)}_L}{c^{2s}} \, ,\\ \text{with} \quad \hat{h}_L &= - \sum_{k=0}^{+\infty} (-)^k \frac{(2\ell-2k-1)!!}{(2k)!!(2\ell+1)!!c^{2k}} \left[ (2\ell-4k+1) \hat{f}^{L\, (2k)}_{\ell-2k\,, 0} + 2 \hat{f}^{L\, (2k)}_{\ell-2k\,, 1} \right]\,.
\end{align}
\end{subequations}
The non-logarithmic part may be recovered straightforwardly by setting $d=3$ in the $d$-dimensional formula~(3.14) in~\cite{BFHLT23b}.\footnote{The normalization of the quantities $g_L$ in~\cite{BFHLT23b} differs from the one of our $\hat{h}^L$; they are related through \smash{$g_L = (-1)^\ell \ell! \,\hat{h}_L$}. Similarly, the coefficients \smash{$\hat{f}^L_{p,0}$} in~\cite{BFHLT23b} read \smash{$\ell!/(2\ell+1)!! \hat{f}^L_{p,0}$} with the present convention, which is more appropriate when gamma functions are expressed in terms of double factorials.} On the other hand, performing a similar calculation, the action of the commutator $[{\Box}^{-1}_{\text{inst}},\partial_i]$ is found to be
\begin{subequations} \label{eq:der_commutator}
\begin{align}
  \bigl[{\Box}^{-1}_{\text{inst}},\partial_i\bigr] F &= \sum_{\ell \geqslant 0\,, s \geqslant 0} \Delta^{-s} \hat{x}_L \frac{\hat{h}^{i\, (2s)}_L}{c^{2s}}\, , \\ \text{with} \quad \hat{h}^i_L &= - \sum_{k=0}^{+\infty} (-)^k \frac{(2\ell-2k-1)!!}{(2k)!!(2\ell+1)!!c^{2k}} \left[ \delta^{i\langle i_\ell} \hat{f}^{L-1\rangle\, (2k)}_{\ell-2k-1\,, 0} + \frac{(\ell+1)}{(2\ell+3)} \hat{f}^{iL\, (2k)}_{\ell-2k-1\,, 0} \right]\,.
\end{align}
\end{subequations}
Note that the $\pi$-value of the two commutators~\eqref{eq:box_commutator} and~\eqref{eq:der_commutator} is 0.

In our case, the commutators will apply either to the quadratic part of the non-linear gauge transformation \smash{$\overline{\Omega}{}\ab_{(2)}$} or to its divergence \smash{$\overline{\Delta}{}^\alpha_{(2)}$}. To control the metric at the 4.5PN order, its $00$, $0i$ and $ij$ components must be included up to $c^{-11}$, $c^{-10}$ and $c^{-9}$, respectively. Therefore the same truncation must hold for the non-linear gauge correction \smash{$\overline{\Omega}{}\ab_{(2)}$}, whereas \smash{$\overline{\Delta}{}^0_{(2)}$} (respectively \smash{$\overline{\Delta}{}^i_{(2)}$}) must be known at order $c^{-10}$ (respectively $c^{-9}$). Those quantities are made of elementary functions $F$, each of them taking the form of a regular factor $F_\text{reg} = r^{2p}\hat{x}^L$, with $\pi$-value zero, times products of (derivatives of) metric potentials. The potentials are for instance $V$ and $V_i$ defined by Eqs.~\eqref{dalV}--\eqref{dalVi} in Appendix~\ref{app:ΣandPotentials}, which generalize $U$ and $U_i$ above to higher order. The simplest potentials, say $\overline{P}{}^{(\text{C})}$, have a source $\overline{S}$ with compact support; for instance $V$ and $V_i$ are of this type. The corresponding multipole expansion, say $\mathcal{M}(\overline{P}{}^{(\text{C})})$, of a compact support potential, \textit{i.e.} the expansion when $r\to +\infty$, reads
\begin{align} \label{eq:MPC}
\mathcal{M}(\overline{P}{}^{(\text{C})}) = \sum_{k=0}^{+\infty} \frac{1}{(2k)!c^{2k}} \sum_{\ell=0}^{+\infty} \frac{(-)^\ell}{\ell!} \partial_L r^{2k-1} \mathcal{S}^{(2k)}_L(t) \,,
\end{align}
with non-STF multipole moments $\mathcal{S}_L(t) = \int \dd^3 \mathbf{x}'\, x'_L S(\mathbf{x}',t)$. All the terms in~\eqref{eq:MPC} have the same $\pi$-value 1. We conclude that the $\pi$-value of a source that is linear in the ``C''-type potentials $\overline{P}{}^{(\text{C})}$ is always 1, so that this kind of term can never contribute. Since it is the only kind that appears at leading PN order, non-zero commutators cannot arise below the relative 1PN order. Another consequence of this property is that a product $F_\text{reg}\, \partial \overline{P}{}^{(\text{C})}_1 \cdots \overline{P}{}^{(\text{C})}_s$, with $\pi$-value 0 for $s$ even and 1 for $s$ odd, can only contribute to the commutator when $s$ is even. In particular, commutators acting to such products that are cubic in C-potentials yield necessarily zero.

Apart from C-potentials, $F$, in our case of interest, may contain potentials with non-compact support. For instance potentials such as $\hat{W}_{ij}$ which are listed in Appendix~\ref{app:ΣandPotentials}. Notably, the quadratic potentials \smash{$\overline{P}{}^{(\text{quad})}$} are sourced by products of (derivatives of) two C-potentials, say $\overline{P}{}^{(\text{C})}_1$ and $\overline{P}{}^{(\text{C})}_2$. Now, it can be shown that
\begin{align}
  \mathcal{M}(\overline{P}{}^{(\text{quad})}) = \Box^{-1}_{\text{inst}}\left[ \mathcal{M}(\partial\overline{P}{}^{(\text{C})}_1) \mathcal{M}(\partial \overline{P}{}^{(\text{C})}_2) \right] + \mathcal{P}{}^{(\text{C})}\,,
\end{align}
see \textit{e.g.}, the Appendix B of~\cite{BBBFMa}, where $\mathcal{P}{}^{(\text{C})}$ is a homogeneous solution with asymptotic expansion at infinity of the form~\eqref{eq:MPC}. Therefore, when $F$ is linear in a (quad) potential, with a possible regular prefactor that does not affect the global $\pi$-value, the second term in the right-hand side of the above expansion, whose $\pi$-value is 1, cannot contribute to the commutator. By contrast, the first term, which will be referred to as the ``explicitly quadratic'' part \smash{$\mathcal{M}(\mathcal{P}{}^{(\text{exp quad})})$}, has the same $\pi$-value as the source under the operator $\Box^{-1}_{\text{inst}}$, namely zero, and may lead to non-zero contributions.

With this is mind, let us sketch the computation of the commutators $\overline{X}\ab_{{(2)}} = [{\Box}^{-1}_{\text{inst}},\Box] \overline{\Omega}{}\ab_{(2)}$. Ignoring the terms linear or cubic in C-potentials, we are left with terms that are quadratic in C-potentials or linear in the (exp quad) asymptotic potentials $\mathcal{P}{}^{(\text{exp quad})}$ associated with the quadratic parts of the non-compact potentials \smash{$\hat{W}_{ij}$}, \smash{$\hat{R}_i$}, \smash{$\hat{X}$} (see Appendix~\ref{app:ΣandPotentials}), as well as \smash{$\hat{Z}_{ij}$}, whose structure is similar to that of \smash{$\hat{W}_{ij}$}. At Newtonian order, C-potentials decay as $1/r$. This permits controlling the behaviour of the source of \smash{$\mathcal{P}{}^{(\text{exp quad})}$}, from its explicit expression, and, using the fact that $\FP \Delta^{-1} \widetilde{r}^B \mathcal{O}(r^p) = \mathcal{O}(r^{p+2})$ (or $\mathcal{O}(r^{p+2} \ln \widetilde{r})$ if logarithms arise), of \smash{$\mathcal{P}{}^{(\text{exp quad})}$} itself. We have near infinity
	\begin{align}\label{eq:P_expansion}
& \mathcal{M}(\mathcal{P}^{(\text{exp quad})}) =  \mathcal{O}\left(\frac{1}{r^2} \right)\quad\text{for}\quad \mathcal{P} \in\{\hat{\mathcal{W}}_{ij}, \hat{\mathcal{R}}_{i}, \hat{\mathcal{X}}, \hat{\mathcal{Z}}_{ij}\}\,.
\end{align}
It follows that the coefficients $\hat{f}^L_{0,q}$ corresponding to $F=x^K \partial_{L_1} P{}^{(\text{C})}_1 \partial_{L_2} P{}^{(\text{C})}_2$ (respectively $F=x^K \partial_{L} P{}^{(\text{exp quad})}$) are $\mathcal{O}(1/c^2)$ as soon as $\ell_1+\ell_2 \geqslant k-1$ (respectively $\ell \geqslant k-1$), hence the commutator is zero at leading order. Note that, to use this result, it may be convenient to transform the time derivative of $V$ into a space derivative by means of the gauge condition $\partial_t V + \partial_i V_i = \mathcal{O}(1/c^2)$. The terms that remain after taking those various considerations into account all belong to $\overline{\Omega}{}^{00}_{(2)}$. They define a potentially contributing piece at 4.5PN order given by
\begin{align}
 \delta \overline{\Omega}^{00}_{(2)} &= \frac{2G}{c^9} \left[ 4 Y^{(1)}_i (\partial_i \hat{W} + 8 V \partial_i V) - x^i \dM^{(3)}_{ij} (\partial_j \hat{W} + 8 V \partial_j V) + 4 W^{(1)} (\partial_t \hat{W} + 8 V\partial_t V) - 4 W^{(2)} (\hat{W} + 4 V^2) \right] \nn \\ &+ \frac{G}{3c^{11}} \left[ x^{ijk} \dM^{(5)}_{ij} (\partial_k \hat{W} + 8 V \partial_k V) - x^i r^2 \dM^{(5)}_{ij} (\partial_j \hat{W} + 8 V \partial_j V) + 4 x^{ij} \dM^{(5)}_{ij} (\hat{W} + 4 V^2) \right. \nn \\ & \left. \qquad \quad ~ - 4 r^2 W^{(4)} (\hat{W} + 4 V^2) - x^{ij} \dM^{(4)}_{ij} \partial_t \hat{W} + 4 r^2 W^{(3)} \partial_t \hat{W} \right]\,.
\end{align}
The analysis can be further simplified by noticing that $\hat{W}\equiv \hat{W}_{ii}$ may be written as $\hat{W} = -V^2/2- (\partial_t V)^2/c^2$, plus C-potentials and 2PN corrections, as explained around Eq.~(3.22) in~\cite{BFHLT23b}. This implies that $\hat{W}$ can be treated effectively as a quadratic term of the type \smash{$\partial P{}^{(\text{C})}_1 \partial P{}^{(\text{C})}_2$}. Assuming this rewriting has been performed, the terms of \smash{$\delta \overline{\Omega}{}^{00}_{(2)}$} fall into four categories, listed below:
\begin{enumerate}
\item Quadratic-type terms $\partial P{}^{(\text{C})}_1 \partial P{}^{(\text{C})}_2$ at relative 1PN order, with commutator
\begin{align}
  \bigl[\Box^{-1}_{\text{inst}},\Box\bigr] (P^{(\text{C})}_1 P^{(\text{C})}_2) = \frac{1}{c^2} \left(3 \mathcal{S}^{(1)}_1\mathcal{S}^{(1)}_2 + \mathcal{S}_1 \mathcal{S}^{(2)}_2 + \mathcal{S}^{(2)}_1 \mathcal{S}_2 \right) + \mathcal{O}\left(\frac{1}{c^4} \right)\,.
\end{align}
Since we always have $P{}^{(\text{C})}_1= P{}^{(\text{C})}_2 =V$ in our case, the monopoles $\mathcal{S}_A$ (with $A=1,2$) are both equal to the total mass $m$ of the binary, hence the commutator vanishes at the 1PN order.
\item Quadratic-type terms $x^i \partial_j  P{}^{(\text{C})}_1 \partial P{}^{(\text{C})}_2$ up to the relative 1PN order at most, with commutator
  \begin{align}
    \bigl[\Box^{-1}_{\text{inst}},\Box\bigr] (x^i \partial_j P{}^{(\text{C})}_1 P{}^{(\text{C})}_2) = - \frac{\delta_{ij}}{c^2} \left(\mathcal{S}^{(1)}\mathcal{S}^{(1)}_2 + \frac{2}{3}\mathcal{S}_1 \mathcal{S}^{(2)}_2 + \frac{1}{3}\mathcal{S}^{(2)}_1 \mathcal{S}_2  \right) + \mathcal{O} \left(\frac{1}{c^4} \right)\,.
  \end{align}
Again, $P{}^{(\text{C})}_1= P{}^{(\text{C})}_2 =V$ for us and the commutator reduces to zero, neglecting the 2PN remainder.
\item Quadratic-type terms $x^i x^j \partial P{}^{(\text{C})}_1 \partial P{}^{(\text{C})}_2$ at leading order, with commutator
\begin{align}
\bigl[\Box^{-1}_{\text{inst}},\Box\bigr] (x^i x^j P{}^{(\text{C})}_1 P{}^{(\text{C})}_2) = - \frac{1}{3} \delta_{ij}\mathcal{S}_1 \mathcal{S}_2 + \mathcal{O}\left(\frac{1}{c^2} \right)\,.
\end{align}
Again, the potentials $P{}^{(\text{C})}_1$ and $P{}^{(\text{C})}_2$ are both equal to $V$. In terms of the form $\partial P{}^{(\text{C})}_1 P{}^{(\text{C})}_2$ where the first potential is differentiated, we may directly apply the above equation provided we regard $\mathcal{S}_1$ as referring to the moment $\partial P{}^{(\text{C})}_1$ including the derivative. Now, at Newtonian order, $\partial U=\partial (m/r)+...$ has no monopole, since $\partial_t m = 0$ and $\partial_i (m/r) = m \, \partial_i (1/r)$ is a dipolar term. This entails the vanishing of the commutator. However, the two terms that are not differentiated may contribute. One of them is proportional to \smash{$x^i x^j \dM_{ij}^{(5)} V^2$}, so that the commutator is proportional to \smash{$\delta_{ij}\dM_{ij}^{(5)}=0$}. The other one, reducing to \smash{$-14/3\, G r^2 V^2 W{}^{(4)}$} after replacing $\hat{W}$ by $-V/2+...$, produces a net contribution.
\item Quadratic-type terms $x^i x^j x^k \partial_l P{}^{(\text{C})}_1 \partial P{}^{(\text{C})}_2$ at leading order, with commutator
\begin{align} \label{eq:x3dP1dP2}
  \bigl[\Box^{-1}_{\text{inst}},\Box\bigr] (x^i x^j x^k \partial_l P{}^{(\text{C})}_1 P{}^{(\text{C})}_2)  =   \frac{1}{5}\delta_{(ij} \delta_{kl)}  \mathcal{S}_1 \mathcal{S}_2 \,.
\end{align}
Those are contracted either with $\dM^{(5)}_{ij} \delta_{kl}$ or with $\dM^{(5)}_{il} \delta_{jk}$, yielding zero.
\end{enumerate}
Putting together the four categories of contributions, we obtain
\begin{align}
\bigl[\Box^{-1}_{\text{inst}},\Box\bigr] \overline{\Omega}^{00}_{(2)} = \frac{14 G^3 m^2}{3c^{11}} W^{(4)}\,,
\end{align}
while the commutator of the other components of \smash{$\overline{\Omega}{}\ab_{(2)}$} vanishes. A similar analysis, but involving only sub-cases of the previous one, may be applied to the commutator of \smash{$\overline{\Delta}{}^\alpha_{(2)}$}, which is found to be zero at this order. As a result, we obtain for the homogeneous solution coming from those commutators, at the relevant 4.5PN order,
\begin{subequations}
\begin{align}
  \overline{U}{}^{00}_{(2)} &=  A(t) + \mathcal{O}\left(\frac{1}{c^{13}}\right)\,,\\  
  \overline{U}{}^{0i}_{(2)} &= \mathcal{O}\left(\frac{1}{c^{10}}\right) \,,\\
  \overline{U}{}^{ij}_{(2)} &= \mathcal{O}\left(\frac{1}{c^{9}}\right)\,.
\end{align}
\end{subequations}
where $A(t)$ is a mere function of time given by
\begin{align}
A = \frac{14G^3 m^2}{3 c^{11}} W^{(4)}\,.
\end{align}
Applying the harmonicity algorithm to \smash{$\partial_\beta \overline{U}{}\ab_{(2)} = A^{(1)}(t) \delta^{0\alpha}/c$}, we get \smash{$\overline{V}{}^{00}_{(2)} = -A$}, all other components of $\overline{V}{}\ab_{(2)}$ being zero. We conclude that the full homogeneous solution \smash{$\overline{H}{}\ab_{(2)}=\overline{U}{}\ab_{(2)} + \overline{V}{}\ab_{(2)}$} vanishes, so that none of the gauge moments involved in the gauge vector \smash{$\overline{\zeta}{}^\alpha_{(1)}$} needs to be redefined. This confirms that the gauge vector to be used to construct the contact transformation at 4.5PN order is \smash{$\varphi{}^\alpha = G \varphi{}^\alpha_{(1)} + G^2 \varphi{}^\alpha_{(2)} + \mathcal{O}(G^3)$}, with $\varphi{}^\alpha_{(1)}$ given by~\eqref{linearvarphi1} and $\varphi{}^\alpha_{(2)}$ by~\eqref{2PMcontributionShift}, without redefinition of moments.

\section{Definitions of PN potentials and source terms}
\label{app:ΣandPotentials}

In this Appendix, we give the definition of the PN potentials and the explicit expressions of the source quantities $\Sigma$, $\Sigma_{i}$ and $\Sigma_{ij}$, defined by Eqs.~\eqref{eq:Sigma} in $d$ dimensions, which are necessary to compute the source and gauge moments $\dI_{L}$,  $\dJ_{L}$, $\dW_{L}$, $\dX_{L}$, $\dY_{L}$ and $\dZ_{L}$ at the required PN order. Posing
\begin{align}\label{eq:sigma}
	\sigma = \frac{2}{d-1}\,
	\frac{(d-2)T^{00}+T^{ii}}{c^2}\,,\qquad
	\sigma_i = \frac{T^{i0}}{c}\,,\qquad
	\sigma_{ij} = T^{ij}\,,
\end{align}

where $T^{\mu \nu}$ is the matter stress-energy tensor, we can rewrite these terms such that
\begin{align}
    \sigma = \tilde{\mu}_{1} \delta_{1} +( 1 \leftrightarrow 2 )\, ,\qquad \sigma^{i}= \mu_{1} v_{1}^{i}\delta_{1} +( 1 \leftrightarrow 2 )\, ,\qquad \sigma^{ij}= \mu_{1} v_{1}^{i}v_{1}^{j}\delta_{1} +( 1 \leftrightarrow 2 )\,,
\end{align}

where $\tilde{\mu_{1}}$ and $\mu_{1}$ are provided in the Supplemental Materials~\cite{SuppMaterial}.
The PN potentials are defined by the hierarchy of equations
\begin{subequations}\label{defpotentials}
\begin{align}
	\Box V &= - 4 \pi G\, \sigma \,,\label{dalV}\\
	\Box V_{i} &= - 4 \pi G\, \sigma_{i}\,,\label{dalVi}\\
	\Box K&=-4\pi G \sigma V\,,
	\label{dalK}\\
	\Box\hat W_{ij}&=-4\pi G\left(\sigma_{ij}
	-\delta_{ij}\,\frac{\sigma_{kk}}{d-2}\right)
	-\frac{1}{2}\left(\frac{d-1}{d-2}\right)\partial_i V \partial_j V\,,
	\label{dalWij}\\
	\Box\hat R_i&=-\frac{4\pi G}{d-2}\left(\frac{5-d}{2}\, V \sigma_i
	-\frac{d-1}{2}\, V_i\, \sigma\right)
	-\frac{d-1}{d-2}\,\partial_k V\partial_i V_k
	-\frac{d(d-1)}{4(d-2)^2}\,\partial_t V \partial_i V\,,
	\label{dalRi}\\
	\Box\hat X&=-4\pi G\left[\frac{V\sigma_{ii}}{d-2}
	+2\left(\frac{d-3}{d-1}\right)\sigma_i V_i
	+\left(\frac{d-3}{d-2}\right)^2
	\sigma\left(\frac{V^2}{2} +K\right)\right]\nn\\
	& +\hat W_{ij}\, \partial_{ij}V
	+2 V_i\,\partial_t\partial_i V
	+\frac{1}{2}\left(\frac{d-1}{d-2}\right) V \partial^2_t V
	+\frac{d(d-1)}{4(d-2)^2}\left(\partial_t V\right)^2
	-2 \partial_i V_j\, \partial_j V_i\,,
	\label{dalX}\\
    \Box\hat Z_{ij}&=-\frac{4\pi G}{d-2}\, V\left(\sigma_{ij}
	-\delta_{ij}\,\frac{\sigma_{kk}}{d-2}\right)
	-\frac{d-1}{d-2}\, \partial_t V_{(i}\, \partial_{j)}V
	+\partial_i V_k\, \partial_j V_k +\partial_k V_i\, \partial_k V_j -2 \partial_k V_{(i}\, \partial_{j)}V_k \nn \\
	&-\frac{\delta_{ij}}{d-2}\, \partial_k V_m
	\left(\partial_k V_m -\partial_m V_k\right) -\frac{d(d-1)}{8(d-2)^3}\, \delta_{ij}\left(\partial_t V\right)^2
	+\frac{(d-1)(d-3)}{2(d-2)^2}\, \partial_{(i} V\partial_{j)} K \,,
	\label{dalZij}\\
	\Box\hat Y_i&=-4\pi G
	\biggl[-\frac{1}{2}\left(\frac{d-1}{d-2}\right)\sigma\hat R_i
	-\frac{(5-d)(d-1)}{4(d-2)^2}\, \sigma V V_i
	+\frac{1}{2}\, \sigma_k\hat W_{ik}
	+\frac{1}{2}\sigma_{ik} V_k\nn\\
	& +\frac{1}{2(d-2)}\, \sigma_{kk}V_i
	-\frac{d-3}{(d-2)^2}\, \sigma_i \left(V^2 +\frac{5-d}{2}\,
	K\right)\biggr] +\hat W_{kl}\, \partial_{kl} V_i
	-\frac{1}{2}\left(\frac{d-1}{d-2}\right)
	\partial_t\hat W_{ik}\, \partial_k V
	+\partial_i\hat W_{kl}\, \partial_k V_l
	\nn\\
	& -\partial_k\hat W_{il}\, \partial_l V_k -\frac{d-1}{d-2}\, \partial_k V \partial_i \hat R_k
	-\frac{d(d-1)}{4 (d-2)^2}\, V_k\, \partial_i V \partial_k V
	-\frac{d(d-1)^2}{8 (d-2)^3}\, V\partial_t V\partial_i V
	\nn\\
	& -\frac{1}{2}\left(\frac{d-1}{d-2}\right)^2 V \partial_k V
	\partial_k V_i
	+\frac{1}{2}\left(\frac{d-1}{d-2}\right) V \partial^2_t V_i
	+2 V_k\, \partial_k\partial_t V_i+\frac{(d-1)(d-3)}{(d-2)^2}\, \partial_k K \partial_i V_k
	\nn\\
    & +\frac{d(d-1)(d-3)}{4(d-2)^3}
	\left(\partial_t V\partial_i K +\partial_i V\partial_t K\right)\,.
	\label{dalYi}
\end{align}
\end{subequations}
The source quantities in terms of the potentials at the 2PN order are
\begin{subequations}
\begin{align}
\Sigma ={}&\sigma
 + \frac{1}{c^{2}} \biggl [\frac{4 \sigma}{d -  2} V
 - \frac{1}{G \pi} \frac{ (d -  1)}{2 (d -  2)} \partial_{k}V \partial_{k}V\biggl]
 + \frac{1}{c^{4}} \biggl(\frac{1}{G \pi} \biggl [- \partial_{k}\hat{W} \partial_{k}V
 -  \frac{(d -  1) (3 d -  2)}{4 (d -  2)^2} V \partial_{k}V \partial_{k}V\nn\\
& + \frac{(d -  1) (d -  4)}{4 (d -  2)^2} (\partial_t V)^2
 -  2 V_{k} \partial_t \partial_{k}V
 -  \frac{d -  1}{2 (d -  2)} V \partial_t^{2} V
 + \frac{2 (d -  1) (d -  3)}{(d -  2)^2} \partial_{k}V \partial_{k}K
 -  \hat{W}_{kl} \partial_{l}\partial_{k}V\nn\\
& + 2 \partial_{k}V_{l} \partial_{l}V_{k}
 + \frac{2 (d -  3)}{d -  1} \partial_{l}V_{k} \partial_{l}V_{k}\biggl]
 + \biggl [- \frac{8 K (d -  3)}{(d -  2)^2}
 + \frac{8 V^2}{(d -  2)^2}
 + 2 \hat{W}\biggl] \sigma \biggl)\, , \\
\Sigma_{i}={}&\sigma_{i}
 + \frac{1}{c^{2}} \biggl(\frac{4 \sigma_{i}}{d -  2} V
 + \frac{1}{G \pi} \biggl [\frac{d (d -  1)}{8 (d -  2)^2} \partial_t V \partial_{i}V
 + \frac{d -  1}{2 (d -  2)} \partial_{i}V_{k} \partial_{k}V
 -  \frac{d -  1}{2 (d -  2)} \partial_{k}V_{i} \partial_{k}V\biggl]\biggl)\nn\\
& + \frac{1}{c^{4}} \Biggl[\frac{1}{G \pi} \biggl(\partial_{k}\hat{W}_{il} \partial_{l}V_{k}
 + \partial_{l}\hat{W}_{ik} \partial_{l}V_{k}
 + \frac{d -  1}{2 (d -  2)^2} V_{k} \partial_{i}V \partial_{k}V
 -  \frac{d (d -  1)}{4 (d -  2)^2} V_{i} \partial_{k}V \partial_{k}V
 + \biggl [\frac{d (d -  1)^2}{4 (d -  2)^3} \partial_t V \partial_{i}V\nn\\
& + \frac{(d -  1)^2}{2 (d -  2)^2} \partial_{i}V_{k} \partial_{k}V
 -  \frac{(d -  1)^2}{2 (d -  2)^2} \partial_{k}V_{i} \partial_{k}V\biggl] V
 + \frac{(d -  1) \partial_{i}\hat{W}}{4 (d -  2)} \partial_t V
 + \frac{1}{2} \partial_{i}\hat{W} \partial_{k}V_{k}
 -  \partial_{k}\hat{W} \partial_{k}V_{i}\nn\\
& + \frac{d -  1}{2 (d -  2)} \partial_t V \partial_t V_{i}
 -  2 V_{k} \partial_t \partial_{k}V_{i}
 -  \frac{d -  1}{2 (d -  2)} V \partial_t^{2} V_{i}
 -  \frac{d (d -  1) (d -  3)}{4 (d -  2)^3} \partial_t V \partial_{i}K
 -  \frac{d (d -  1) (d -  3)}{4 (d -  2)^3} \partial_t K \partial_{i}V\nn\\
& + \frac{d -  1}{d -  2} \partial_{i}\hat{R}_{k} \partial_{k}V
 -  \frac{(d -  1) (d -  3)}{(d -  2)^2} \partial_{i}V_{k} \partial_{k}K
 + \frac{(d -  1) (d -  3)}{(d -  2)^2} \partial_{k}V_{i} \partial_{k}K
 -  \frac{d -  1}{d -  2} \partial_{k}V \partial_{k}\hat{R}_{i}\nn\\
& + \frac{d -  1}{2 (d -  2)} \partial_t \hat{W}_{ik} \partial_{k}V
 -  \hat{W}_{kl} \partial_{l}\partial_{k}V_{i}
 -  \partial_{i}\hat{W}_{kl} \partial_{l}V_{k}\biggl)
 + \biggl [- \frac{8 K (d -  3)}{(d -  2)^2}
 + \frac{8 V^2}{(d -  2)^2}
 + 2 \hat{W}\biggl] \sigma_{i}\Biggl]\, , \\
\Sigma_{ij}={}& \mathop{\text{Sym}}_{i,j}~ \sigma_{ij}
 + \frac{1}{G \pi} \biggl [- \frac{d -  1}{16 (d -  2)} \delta_{ij} \partial_{k}V \partial_{k}V
 + \frac{d -  1}{8 (d -  2)} \partial_{i}V \partial_{j}V\biggl]
 + \frac{1}{c^{2}} \Biggl[\frac{1}{G \pi} \biggl(- \frac{(d -  1) (d -  3)}{4 (d -  2)^2} \partial_{i}V \partial_{j}K\nn\\
& + \frac{d -  1}{d -  2} \partial_t V_{i} \partial_{j}V
 -  \frac{(d -  1) (d -  3)}{4 (d -  2)^2} \partial_{i}K \partial_{j}V
 -  \partial_{i}V_{k} \partial_{j}V_{k}
 + 2 \partial_{j}V_{k} \partial_{k}V_{i}
 -  \partial_{k}V_{j} \partial_{k}V_{i}\nn\\
& + \delta_{ij} \biggl [- \frac{d (d -  1)}{16 (d -  2)^2} (\partial_t V)^2
 + \frac{(d -  1) (d -  3)}{4 (d -  2)^2} \partial_{k}V \partial_{k}K
 -  \frac{d -  1}{2 (d -  2)} \partial_t V_{k} \partial_{k}V
 -  \frac{1}{2} \partial_{k}V_{l} \partial_{l}V_{k}
 + \frac{1}{2} \partial_{l}V_{k} \partial_{l}V_{k}\biggl]\biggl)\nn\\
& + \frac{4 \sigma_{ij}}{d -  2} V\Biggl]
 + \frac{1}{c^{4}} \Biggl \{\frac{1}{G \pi} \Biggl[- \frac{1}{2} \partial_{i}\hat{W} \partial_{j}\hat{W}
 + \partial_{k}\hat{W}_{jl} \partial_{l}\hat{W}_{ik}
 + \partial_{l}\hat{W}_{jk} \partial_{l}\hat{W}_{ik}
 + \frac{(d -  1)^2}{(d -  2)^2} V \partial_t V_{i} \partial_{j}V\nn\\
& + \frac{2 (d -  1)}{d -  2} V_{k} \partial_{j}V \partial_{k}V_{i}
 -  \frac{d -  1}{d -  2} \hat{W}_{ik} \partial_{j}V \partial_{k}V
 + \frac{d -  1}{4 (d -  2)} \hat{W}_{ij} \partial_{k}V \partial_{k}V
 + \biggl [\frac{d -  1}{(d -  2)^2} \partial_t V \partial_{j}V\nn\\
& + \frac{2 (d -  1)}{d -  2} \partial_{j}V_{k} \partial_{k}V
 -  \frac{2 (d -  1)}{d -  2} \partial_{k}V_{j} \partial_{k}V\biggl] V_{i}
 + \partial_{j}\hat{W} \partial_t V_{i}
 + \partial_{j}\hat{W} \partial_{k}\hat{W}_{ik}
 -  \partial_{k}\hat{W} \partial_{k}\hat{W}_{ij}
 + 2 \partial_t V_{i} \partial_t V_{j}\nn\\
& -  2 V_{k} \partial_t \partial_{k}\hat{W}_{ij}
 -  \frac{d -  1}{2 (d -  2)} V \partial_t^{2} \hat{W}_{ij}
 -  \frac{2 (d -  1) (d -  3)}{(d -  2)^2} \partial_t V_{i} \partial_{j}K
 + \frac{(d -  1) (d -  3)^2}{2 (d -  2)^3} \partial_{i}K \partial_{j}K
 -  2 \partial_{i}V_{k} \partial_{j}\hat{R}_{k}\nn\\
& + \frac{2 (d -  1)}{d -  2} \partial_t \hat{R}_{i} \partial_{j}V
 + \frac{(d -  1) \hat{W}}{4 (d -  2)} \partial_{i}V \partial_{j}V
 + \frac{d -  1}{2 (d -  2)} \partial_{i}\hat{X} \partial_{j}V
 -  2 \partial_{i}\hat{R}_{k} \partial_{j}V_{k}
 -  2 \partial_t \hat{W}_{ik} \partial_{j}V_{k}\nn\\
& + \frac{1}{2} \partial_{i}\hat{W}_{kl} \partial_{j}\hat{W}_{kl}
 + \frac{d -  1}{2 (d -  2)} \partial_{i}V \partial_{j}\hat{X}
 + 4 \partial_{j}\hat{R}_{k} \partial_{k}V_{i}
 + 4 \partial_{j}V_{k} \partial_{k}\hat{R}_{i}
 -  4 \partial_{k}V_{i} \partial_{k}\hat{R}_{j}
 + 2 \partial_t \hat{W}_{ik} \partial_{k}V_{j}\nn\\
& -  \hat{W}_{kl} \partial_{l}\partial_{k}\hat{W}_{ij}
 -  2 \partial_{j}\hat{W}_{kl} \partial_{l}\hat{W}_{ik}
 + \delta_{ij} \biggl(\partial_t \hat{W}_{kl} \partial_{l}V_{k}
 + \frac{1}{2} \partial_{l}\hat{W} \partial_{l}\hat{W}
 + \biggl [- \frac{d (d -  1)^2}{8 (d -  2)^3} (\partial_t V)^2\nn\\
& -  \frac{(d -  1)^2}{2 (d -  2)^2} \partial_t V_{k} \partial_{k}V\biggl] V
 + \frac{d -  1}{4 (d -  2)} \hat{W}_{kl} \partial_{k}V \partial_{l}V
 + \biggl [- \frac{d -  1}{2 (d -  2)^2} \partial_t V \partial_{k}V
 -  \frac{d -  1}{d -  2} \partial_{k}V_{l} \partial_{l}V\biggl] V_{k}\nn\\
& + \frac{(d -  1) \partial_t^{2} \hat{W}}{4 (d -  2)} V
 + \partial_t \partial_{k}\hat{W} V_{k}
 + \frac{1}{2} \partial_{l}\partial_{k}\hat{W} \hat{W}_{kl}
 -  \frac{(d -  1) \partial_t \hat{W}}{4 (d -  2)} \partial_t V
 -  \frac{1}{2} \partial_{k}\hat{W} \partial_t V_{k}
 -  \frac{1}{2} \partial_t \hat{W} \partial_{k}V_{k}\nn\\
& -  \frac{1}{2} \partial_{l}\hat{W} \partial_{m}\hat{W}_{lm}
 + \frac{d (d -  1) (d -  3)}{4 (d -  2)^3} \partial_t K \partial_t V
 + \frac{(d -  1) (d -  3)}{(d -  2)^2} \partial_t V_{k} \partial_{k}K
 -  \frac{(d -  1) (d -  3)^2}{4 (d -  2)^3} \partial_{k}K \partial_{k}K\nn\\
& -  \frac{d -  1}{d -  2} \partial_t \hat{R}_{k} \partial_{k}V
 -  \frac{(d -  1) \hat{W}}{8 (d -  2)} \partial_{k}V \partial_{k}V
 -  \frac{d -  1}{2 (d -  2)} \partial_{k}\hat{X} \partial_{k}V
 -  2 \partial_{k}V_{l} \partial_{l}\hat{R}_{k}
 + 2 \partial_{l}V_{k} \partial_{l}\hat{R}_{k}\nn\\
& + \frac{1}{2} \partial_{l}\hat{W}_{km} \partial_{m}\hat{W}_{kl}
 -  \frac{1}{4} \partial_{m}\hat{W}_{kl} \partial_{m}\hat{W}_{kl}\biggl)\Biggl]
 + \biggl [- \frac{8 K (d -  3)}{(d -  2)^2}
 + \frac{8 V^2}{(d -  2)^2}
 + 2 \hat{W}\biggl] \sigma_{ij}\Biggl \}\, .
\end{align}
\end{subequations}

\section{Radiation-reaction acceleration as a function of moments}
\label{app:accMoments}

In this Appendix, we provide the expressions of the RR acceleration as a function of the multipole moments up to 3.5PN. The expression at 4.5PN is long and has been included in the Supplemental Materials~\cite{SuppMaterial}. Note that, even if in the BT coordinates the RR force description in terms of multipole moments is much simpler than the explicit formulation~\cite{BFT24}, this is not the case in harmonic coordinates where the multipole moment formulation is more complicated than the explicitly expanded one. At 2.5PN order the result agrees with Eq.~\eqref{eq:aharm} in the Introduction.
\begin{subequations}
\begin{align}
a^{i}_{2.5\text{PN}} &= \frac{G}{c^5} \Bigl(-3 n_{12}^{iab} \frac{G m_{2} \dI^{(3)}_{ab}}{r_{12}^2}
 -  \frac{8 G m_{2} n_{12}^{i} \dW^{(2)}_{}}{r_{12}^2}
 + 4 v_{1}^{i} \dW^{(3)}_{}
 - 4 \dY^{(3)}_{i\,\text{3D}}
 + 2 v_{1}^{a} \dI^{(4)}_{ia}
 + \frac{3}{5} y_{1}^{a} \dI^{(5)}_{ia}\Bigr)\, ,\label{a25pn} \\
a^{i}_{3.5\text{PN}} &= \frac{G}{c^7} \biggl\{\Bigl(20 n_{12}^{abi} \frac{m_{1} m_{2}}{r_{12}^3}
 + 16 n_{12}^{abi} \frac{m_{2}^2}{r_{12}^3}\Bigr) G^2 \dI^{(3)}_{ab}
 - 2 n_{12}^{ia} \frac{G m_{2} \dY^{(3)}_{a}}{r_{12}}
 - 2 n_{12}^{iab} \frac{G m_{2} \dY^{(3)}_{ab}}{r_{12}^2}
 -  \frac{4}{3} n_{12}^{ijk} \frac{G m_{2} \varepsilon_{abk} y_{1}^{a} \dJ^{(4)}_{bj}}{r_{12}^2}\nn\\
& -  \frac{1}{6} n_{12}^{iab} G m_{2} \dI^{(5)}_{ab}
 + \frac{1}{3} n_{12}^{ibj} \frac{G m_{2} y_{1}^{a} \dI^{(5)}_{abj}}{r_{12}^2}
 -  \frac{1}{6} n_{12}^{iabj} \frac{G m_{2} \dI^{(5)}_{abj}}{r_{12}}
 + G^2 \dW^{(2)}_{} \Bigl(\frac{40 m_{1} m_{2} n_{12}^{i}}{r_{12}^3}
 + \frac{32 m_{2}^2 n_{12}^{i}}{r_{12}^3}\Bigr)\nn\\
& + 12 n_{12}^{ib} \frac{G m_{2} \dY^{(2)}_{b}}{r_{12}^2} (n_{12}{} v_{2}{})
 + 4 n_{12}^{ib} \frac{G m_{2} \dW^{(3)}_{b}}{r_{12}^2} (n_{12}{} v_{2}{})
 - 2 n_{12}^{ij} \frac{G m_{2} y_{1}^{b} \dI^{(4)}_{bj}}{r_{12}^2} (n_{12}{} v_{2}{})\nn\\
& + \frac{5}{2} n_{12}^{ibj} \frac{G m_{2} \dI^{(4)}_{bj}}{r_{12}} (n_{12}{} v_{2}{})
 + \frac{15}{2} n_{12}^{ijk} \frac{G m_{2} \dI^{(3)}_{jk}}{r_{12}^2} (n_{12}{} v_{2}{})^2
 + G m_{2} \dI^{(4)}_{ib} \Bigl(-4 \frac{n_{12}^{b}}{r_{12}} (n_{12}{} v_{1}{})
 + 3 \frac{n_{12}^{b}}{r_{12}} (n_{12}{} v_{2}{})\Bigr)\nn\\
& + G m_{2} \dI^{(3)}_{bj} \Bigl(12 \frac{v_{1}^{i}}{r_{12}^2} n_{12}^{bj} (n_{12}{} v_{1}{})
 - 12 \frac{v_{2}^{i}}{r_{12}^2} n_{12}^{bj} (n_{12}{} v_{1}{})
 - 6 n_{12}^{ij} \frac{v_{2}^{b}}{r_{12}^2} (n_{12}{} v_{2}{})
 - 9 \frac{v_{1}^{i}}{r_{12}^2} n_{12}^{bj} (n_{12}{} v_{2}{})\nn\\
& + 9 \frac{v_{2}^{i}}{r_{12}^2} n_{12}^{bj} (n_{12}{} v_{2}{})\Bigr)
 + 4 n_{12}^{ib} \frac{G m_{2} \dY^{(3)}_{b}}{r_{12}^2} (n_{12}{} y_{1}{})
 -  \frac{8}{3} \dY^{(4)}_{i} (v_{1}{} y_{1}{})
 + \frac{2}{15} y_{1}^{b} \dI^{(6)}_{ib} (v_{1}{} y_{1}{})
 + \dY^{(3)}_{i} \Bigl(\frac{2 G m_{2}}{3 r_{12}}
 -  \frac{4}{3} v_{1}^{2}\Bigr)\nn\\
& + G m_{2} \dI^{(3)}_{bj} \Bigl(12 n_{12}^{bij} \frac{1}{r_{12}^{2}} (v_{1}{} v_{2}{})
 - 3 n_{12}^{bij} \frac{1}{r_{12}^{2}} v_{1}^{2}
 - 6 n_{12}^{bij} \frac{1}{r_{12}^{2}} v_{2}^{2}\Bigr)
 + \frac{2}{3} v_{1}^{i} \dW^{(5)}_{} y_{1}^{2}
 -  \frac{2}{3} \dY^{(5)}_{i} y_{1}^{2}
 + \frac{1}{3} v_{1}^{a} \dI^{(6)}_{ia} y_{1}^{2}\nn\\
& + \frac{13}{210} y_{1}^{b} \dI^{(7)}_{ib} y_{1}^{2}
 + G m_{2} \dI^{(5)}_{bj} \Bigl(\frac{1}{2} n_{12}^{bij} \frac{1}{r_{12}} (n_{12}{} y_{1}{})
 -  \frac{1}{2} n_{12}^{bij} \frac{1}{r_{12}^{2}} y_{1}^{2}\Bigr)
 + \dI^{(5)}_{iab} \Bigl(\frac{1}{18} \frac{G m_{2}}{r_{12}} n_{12}^{ab}
 -  \frac{1}{9} v_{1}^{ab}\Bigr)\nn\\
& + G m_{2} \dY^{(2)}_{a} \Bigl(\frac{8 n_{12}^{i} v_{1}^{a}}{r_{12}^2}
 + \frac{4 n_{12}^{a} v_{1}^{i}}{r_{12}^2}
 -  \frac{4 n_{12}^{a} v_{2}^{i}}{r_{12}^2}\Bigr)
 + G m_{2} \dW^{(3)}_{a} \Bigl(\frac{8 n_{12}^{i} v_{1}^{a}}{3 r_{12}^2}
 + \frac{4 n_{12}^{a} v_{1}^{i}}{3 r_{12}^2}
 -  \frac{4 n_{12}^{a} v_{2}^{i}}{3 r_{12}^2}\Bigr)\nn\\
& + G m_{2} \dI^{(3)}_{ab} \Bigl(-8 \frac{n_{12}^{a}}{r_{12}^2} v_{1}^{bi}
 - 6 \frac{n_{12}^{a}}{r_{12}^2} v_{2}^{bi}
 + 2 \frac{n_{12}^{i}}{r_{12}^2} v_{1}^{ab}
 -  \frac{8 n_{12}^{i} v_{1}^{a} v_{2}^{b}}{r_{12}^2}
 + \frac{6 n_{12}^{a} v_{1}^{i} v_{2}^{b}}{r_{12}^2}
 + \frac{8 n_{12}^{a} v_{1}^{b} v_{2}^{i}}{r_{12}^2}
 + 4 \frac{n_{12}^{i}}{r_{12}^2} v_{2}^{ab}\Bigr)\nn\\
& + G m_{2} \dI^{(5)}_{ia} \Bigl(\frac{47}{30} n_{12}^{a}
 + \frac{y_{1}^{a}}{30 r_{12}}\Bigr)
 + \dI^{(5)}_{ib} \Bigl(- \frac{1}{3} \frac{G m_{2} n_{12}^{b}}{r_{12}} (n_{12}{} y_{1}{})
 + \frac{2}{3} (v_{1}{} y_{1}{}) v_{1}^{b}
 -  \frac{1}{15} v_{1}^{2} y_{1}^{b}\Bigr)\nn\\
& + G m_{2} \dI^{(4)}_{ab} \Bigl(4 n_{12}^{bi} \frac{v_{1}^{a}}{r_{12}}
 - 4 n_{12}^{bi} \frac{v_{2}^{a}}{r_{12}}
 - 3 \frac{v_{1}^{i}}{r_{12}} n_{12}^{ab}
 + \frac{19}{6} \frac{v_{2}^{i}}{r_{12}} n_{12}^{ab}
 -  \frac{4 n_{12}^{i} v_{1}^{a} y_{1}^{b}}{3 r_{12}^2}
 -  \frac{2 n_{12}^{a} v_{1}^{i} y_{1}^{b}}{3 r_{12}^2}
 + \frac{2 n_{12}^{a} v_{2}^{i} y_{1}^{b}}{3 r_{12}^2}\Bigr)\nn\\
& + \frac{4}{45} \varepsilon_{ibj} \dJ^{(6)}_{aj} y_{1}^{ab}
 -  \frac{5}{126} \dI^{(7)}_{iab} y_{1}^{ab}
 + \biggl [\frac{4}{3} v_{1}^{i} v_{1}^{2}
 + G m_{2} \Bigl(2 \frac{n_{12}^{i}}{r_{12}} (n_{12}{} v_{2}{})
 -  \frac{4}{3} \frac{v_{1}^{i}}{r_{12}^2} (n_{12}{} y_{1}{})
 + \frac{4}{3} \frac{v_{2}^{i}}{r_{12}^2} (n_{12}{} y_{1}{})\nn\\
& - 4 \frac{n_{12}^{i}}{r_{12}^2} (n_{12}{} v_{2}{}) (n_{12}{} y_{1}{})
 -  \frac{8}{3} \frac{n_{12}^{i}}{r_{12}^2} (v_{1}{} y_{1}{})
 -  \frac{2 v_{2}^{i}}{3 r_{12}}\Bigr)\biggl] \dW^{(3)}_{}
 -  \frac{4 G m_{2} n_{12}^{i} \dX^{(3)}_{}}{r_{12}^2}
 -  \frac{1}{3} v_{1}^{iab} \dI^{(4)}_{ab}
 + \Bigl(\frac{4}{9} \frac{G m_{2} \varepsilon_{ijb}}{r_{12}} n_{12}^{ab}\nn\\
& -  \frac{8}{9} \varepsilon_{ijb} v_{1}^{ab}\Bigr) \dJ^{(4)}_{aj}
 + \Bigl(\frac{8}{3} v_{1}^{i} (v_{1}{} y_{1}{})
 -  \frac{4}{3} \frac{G m_{2} n_{12}^{i}}{r_{12}^2} y_{1}^{2}\Bigr) \dW^{(4)}_{}
 + \frac{8 G m_{2} n_{12}^{i} y_{1}^{a} \dW^{(4)}_{a}}{3 r_{12}^2}
 -  \frac{8}{3} v_{1}^{ia} \dW^{(4)}_{a}
 + \frac{8}{3} v_{1}^{i} \dX^{(4)}_{}\nn\\
& + \frac{4}{3} v_{1}^{a} \dY^{(4)}_{ia}
 -  \frac{4}{3} \varepsilon_{iab} v_{1}^{a} \dZ^{(4)}_{b}
 + \biggl [- \frac{1}{3} y_{1}^{i} v_{1}^{ab}
 + \frac{4}{15} v_{1}^{bi} y_{1}^{a}
 + G m_{2} \Bigl(\frac{1}{30} n_{12}^{bi} \frac{y_{1}^{a}}{r_{12}}
 + \frac{1}{6} \frac{y_{1}^{i}}{r_{12}} n_{12}^{ab}
 -  \frac{1}{6} \frac{n_{12}^{i}}{r_{12}^2} y_{1}^{ab}\Bigr)\biggl] \dI^{(5)}_{ab}\nn\\
& + \frac{8}{15} \varepsilon_{ibj} v_{1}^{a} y_{1}^{b} \dJ^{(5)}_{aj}
 + \frac{8}{45} \varepsilon_{iaj} v_{1}^{a} y_{1}^{b} \dJ^{(5)}_{bj}
 + \frac{16}{45} \varepsilon_{abj} v_{1}^{a} y_{1}^{b} \dJ^{(5)}_{ij}
 -  \frac{4}{3} v_{1}^{i} y_{1}^{a} \dW^{(5)}_{a}
 + \frac{4}{3} y_{1}^{i} \dX^{(5)}_{}
 -  \frac{4}{3} \dX^{(5)}_{i}
 + \frac{2}{3} y_{1}^{a} \dY^{(5)}_{ia}\nn\\
& -  \frac{2}{3} \varepsilon_{iab} y_{1}^{a} \dZ^{(5)}_{b}
 + \Bigl(- \frac{2}{15} y_{1}^{bi} v_{1}^{a}
 + \frac{1}{10} v_{1}^{i} y_{1}^{ab}\Bigr) \dI^{(6)}_{ab}
 -  \frac{2}{9} v_{1}^{a} y_{1}^{b} \dI^{(6)}_{iab}
 -  \frac{1}{210} y_{1}^{iab} \dI^{(7)}_{ab}\biggl\}.
\end{align}
\end{subequations}

\section{Explicit expressions for the moments}
\label{app:jauge}

In this Appendix we provide the explicit expressions of the multipoles moments in a general frame obtained with the procedure described in Sec.~\ref{sec:MultipoleMoments}. This includes the explicit expression for $\dI^{ij}$, $\dI^{ijk}$, $\dJ^{ij}$, $\dW$, $\dW^i$, $\dX$, $\dY^i$ and $\dZ^i$ at 1PN order and $\dI^{ijkl}$, $\dJ^{ijk}$, $\dW^{ij}$, $\dX^i$, $\dX^{ij}$, $\dY^{ij}$, $\dY^{ijk}$ and $\dZ^{ij}$  at Newtonian order. The other useful moments to derive the acceleration in harmonic coordinates at 4.5PN are $\dI^{ij}$, $\dW$ and $\dY^i$ at 2PN order and $\dX^i$ and $\dY^{ij}$ at 1PN order. These moments are too complicated to be included in their expanded form in the text and can be found in the Supplemental Material~\cite{SuppMaterial}.
\begin{subequations}
\begin{align}
    \dI^{ij}={}& \mathop{\text{STF}}_{i,j} m_{1} \Bigl(- \frac{1}{2} n_{12}^{j} r_{12} y_{1}^{i}
+ \frac{1}{2} n_{12}^{i} r_{12} y_{1}^{j}
+ y_{1}^{ij}\Bigr)+ \frac{1}{c^{2}} \biggl [m_{1} \Bigl(\frac{2}{7} n_{12}^{j} r_{12} v_{1}^{i} (v_{1}{} y_{1}{})
 -  \frac{2}{7} n_{12}^{i} r_{12} v_{1}^{j} (v_{1}{} y_{1}{})
 -  \frac{4}{7} v_{1}^{i} y_{1}^{j} (v_{1}{} y_{1}{})
 \nn\\
& -  \frac{29}{84} n_{12}^{j} r_{12} y_{1}^{i} v_{1}^{2} + \frac{29}{84} n_{12}^{i} r_{12} y_{1}^{j} v_{1}^{2}
+ \frac{29}{42} y_{1}^{ij} v_{1}^{2} + \frac{11}{21} v_{1}^{ij} y_{1}^{2}\Bigr)
 + G m_{1} m_{2} \Bigl(- \frac{5}{14} n_{12}^{ij} r_{12}
 -  \frac{25}{42} \frac{1}{r_{12}} y_{1}^{ij}
 + \frac{5}{42} n_{12}^{ij} (n_{12}{} y_{1}{})
 \nn\\
&+ \frac{2}{7} \frac{n_{12}^{i} y_{1}^{j}}{r_{12}} (n_{12}{} y_{1}{}) -  \frac{11}{42} n_{12}^{ij} \frac{1}{r_{12}} y_{1}^{2} + \frac{19}{42} n_{12}^{i} y_{1}^{j}\Bigr)\biggl]
 + \big( 1 \leftrightarrow 2 \big)+ \dI^{ij}_{2\text{PN}}\, , \\
 \dI^{ijk}={}& \mathop{\text{STF}}_{i,j,k} m_{1} \Bigl(- n_{12}^{ik} r_{12}^2 y_{1}^{j}
+ \frac{1}{2} n_{12}^{ij} r_{12}^2 y_{1}^{k}
-  \frac{1}{2} n_{12}^{k} r_{12} y_{1}^{ij}
-  \frac{1}{2} n_{12}^{j} r_{12} y_{1}^{ik}
+ \frac{1}{2} r_{12}^2 y_{1}^{i} n_{12}^{jk}
+ n_{12}^{i} r_{12} y_{1}^{jk}
+ y_{1}^{ijk}\Bigr)\nn\\
&+ \frac{1}{c^{2}} \biggl [G m_{1} m_{2} \Bigl(- n_{12}^{ij} r_{12} y_{1}^{k}
 -  \frac{2}{3} \frac{1}{r_{12}} y_{1}^{ijk}
 -  \frac{1}{6} n_{12}^{ijk} r_{12} (n_{12}{} y_{1}{})
 + \frac{1}{4} n_{12}^{ijk} y_{1}^{2}
 -  \frac{1}{2} n_{12}^{ij} \frac{y_{1}^{k}}{r_{12}} y_{1}^{2}
 + \frac{3}{4} n_{12}^{i} y_{1}^{jk}\nn\\
& + \frac{1}{2} \frac{n_{12}^{i}}{r_{12}} (n_{12}{} y_{1}{}) y_{1}^{jk}\Bigr)
 + m_{1} \Bigl(n_{12}^{ik} r_{12}^2 v_{1}^{j} (v_{1}{} y_{1}{})
 -  \frac{1}{2} n_{12}^{ij} r_{12}^2 v_{1}^{k} (v_{1}{} y_{1}{})
 + \frac{1}{2} n_{12}^{k} r_{12} v_{1}^{i} y_{1}^{j} (v_{1}{} y_{1}{})\nn\\
& + \frac{1}{2} n_{12}^{j} r_{12} v_{1}^{i} y_{1}^{k} (v_{1}{} y_{1}{})
 -  n_{12}^{i} r_{12} v_{1}^{j} y_{1}^{k} (v_{1}{} y_{1}{})
 -  \frac{1}{2} r_{12}^2 v_{1}^{i} n_{12}^{jk} (v_{1}{} y_{1}{})
 -  \frac{5}{6} n_{12}^{ik} r_{12}^2 y_{1}^{j} v_{1}^{2}
 + \frac{5}{12} n_{12}^{ij} r_{12}^2 y_{1}^{k} v_{1}^{2}\nn\\
& -  \frac{5}{12} n_{12}^{k} r_{12} y_{1}^{ij} v_{1}^{2}
 -  \frac{5}{12} n_{12}^{j} r_{12} y_{1}^{ik} v_{1}^{2}
 + \frac{5}{6} y_{1}^{ijk} v_{1}^{2}
 + \frac{5}{12} r_{12}^2 y_{1}^{i} n_{12}^{jk} v_{1}^{2}
 -  \frac{1}{2} n_{12}^{k} r_{12} v_{1}^{ij} y_{1}^{2}
 + \frac{1}{2} n_{12}^{i} r_{12} y_{1}^{2} v_{1}^{jk}\nn\\
& + v_{1}^{ij} y_{1}^{2} y_{1}^{k}
 -  v_{1}^{i} (v_{1}{} y_{1}{}) y_{1}^{jk}
 + \frac{5}{6} n_{12}^{i} r_{12} v_{1}^{2} y_{1}^{jk}\Bigr)\biggl]
+\big( 1 \leftrightarrow 2 \big)\, , \\
\dI^{ijkl}={}&\mathop{\text{STF}}_{i,j,k,l} m_{1} \Bigl(\frac{3}{2} n_{12}^{ikl} r_{12}^3 y_{1}^{j}
 -  \frac{3}{2} n_{12}^{ijl} r_{12}^3 y_{1}^{k}
 + \frac{1}{2} n_{12}^{ijk} r_{12}^3 y_{1}^{l}
 -  \frac{1}{2} n_{12}^{l} r_{12} y_{1}^{ijk}
 -  \frac{1}{2} n_{12}^{k} r_{12} y_{1}^{ijl}
 -  \frac{1}{2} n_{12}^{j} r_{12} y_{1}^{ikl}
 + \frac{1}{2} r_{12}^2 y_{1}^{il} n_{12}^{jk}\nn\\
& + \frac{1}{2} r_{12}^2 y_{1}^{ik} n_{12}^{jl}
 + \frac{1}{2} r_{12}^2 y_{1}^{ij} n_{12}^{kl}
 -  \frac{1}{2} r_{12}^3 y_{1}^{i} n_{12}^{jkl}
 -  \frac{3}{2} n_{12}^{il} r_{12}^2 y_{1}^{jk}
 -  \frac{3}{2} n_{12}^{ik} r_{12}^2 y_{1}^{jl}
 + \frac{3}{2} n_{12}^{ij} r_{12}^2 y_{1}^{kl}
 + \frac{3}{2} n_{12}^{i} r_{12} y_{1}^{jkl}\nn\\
& + y_{1}^{ijkl}\Bigr)+\big( 1 \leftrightarrow 2 \big)\, ,\\
\dJ^{ij}={}& \mathop{\text{STF}}_{i,j} -m_{1} \varepsilon_{iab} v_{1}^{a} y_{1}^{jb} + \frac{1}{c^{2}} \biggl [G m_{1} m_{2} \varepsilon_{iab} \Bigl(\frac{5}{14} r_{12} v_{1}^{a} n_{12}^{bj}
 -  \frac{5}{112} \frac{n_{12}^{j} v_{1}^{a} y_{1}^{b}}{r_{12}} (n_{12}{} y_{1}{})
 -  \frac{1}{16} \frac{n_{12}^{a} v_{1}^{j} y_{1}^{b}}{r_{12}} (n_{12}{} y_{1}{})
 \nn\\
&+ \frac{15}{112} \frac{n_{12}^{j} v_{2}^{a} y_{1}^{b}}{r_{12}} (n_{12}{} y_{1}{}) + \frac{17}{112} \frac{n_{12}^{a} v_{2}^{j} y_{1}^{b}}{r_{12}} (n_{12}{} y_{1}{}) -  \frac{1}{56} \frac{n_{12}^{a} v_{1}^{b} y_{1}^{j}}{r_{12}} (n_{12}{} y_{1}{})
 + \frac{1}{56} \frac{n_{12}^{a} v_{2}^{b} y_{1}^{j}}{r_{12}} (n_{12}{} y_{1}{})
 \nn\\
&-  \frac{15}{112} \frac{y_{1}^{a}}{r_{12}} n_{12}^{bj} (n_{12}{} v_{1}{}) (n_{12}{} y_{1}{}) + \frac{15}{112} \frac{y_{1}^{a}}{r_{12}} n_{12}^{bj} (n_{12}{} v_{2}{}) (n_{12}{} y_{1}{})
 -  \frac{1}{16} \frac{y_{1}^{a}}{r_{12}} n_{12}^{bj} (v_{1}{} y_{1}{})
 -  \frac{3}{112} \frac{y_{1}^{a}}{r_{12}} n_{12}^{bj} (v_{2}{} y_{1}{})
 + \frac{3}{28} \frac{n_{12}^{j} v_{1}^{a} y_{1}^{b}}{r_{12}^2} y_{1}^{2}
 \nn\\
&+ \frac{3}{28} \frac{n_{12}^{a} v_{1}^{j} y_{1}^{b}}{r_{12}^2} y_{1}^{2} -  \frac{3}{28} \frac{n_{12}^{j} v_{2}^{a} y_{1}^{b}}{r_{12}^2} y_{1}^{2}
 -  \frac{3}{28} \frac{n_{12}^{a} v_{2}^{j} y_{1}^{b}}{r_{12}^2} y_{1}^{2}
 -  \frac{3}{56} \frac{v_{1}^{a}}{r_{12}} n_{12}^{bj} y_{1}^{2}
 + \frac{3}{56} \frac{v_{2}^{a}}{r_{12}} n_{12}^{bj} y_{1}^{2}
 + \frac{9}{56} \frac{y_{1}^{a}}{r_{12}^2} n_{12}^{bj} (n_{12}{} v_{1}{}) y_{1}^{2}\nn\\
& -  \frac{9}{56} \frac{y_{1}^{a}}{r_{12}^2} n_{12}^{bj} (n_{12}{} v_{2}{}) y_{1}^{2}
 + \frac{5}{112} n_{12}^{bj} (n_{12}{} y_{1}{}) v_{1}^{a}
 -  \frac{5}{56} n_{12}^{bj} (n_{12}{} y_{1}{}) v_{2}^{a}
 + \frac{25}{112} n_{12}^{bj} (n_{12}{} v_{1}{}) y_{1}^{a}
 + \frac{3}{56} n_{12}^{bj} (n_{12}{} v_{2}{}) y_{1}^{a}\nn\\
& -  \frac{23}{28} n_{12}^{j} v_{1}^{a} y_{1}^{b}
 + \frac{15}{16} n_{12}^{a} v_{1}^{j} y_{1}^{b}
 + \frac{15}{28} n_{12}^{j} v_{2}^{a} y_{1}^{b}
 -  \frac{55}{56} n_{12}^{a} v_{2}^{j} y_{1}^{b}
 + \frac{197}{112} n_{12}^{a} v_{1}^{b} y_{1}^{j}
 -  \frac{85}{56} n_{12}^{a} v_{2}^{b} y_{1}^{j}
 + \frac{29}{112} \frac{v_{1}^{a}}{r_{12}} y_{1}^{bj}\nn\\
& + \frac{25}{112} \frac{v_{2}^{a}}{r_{12}} y_{1}^{bj}
 + \frac{25}{112} \frac{n_{12}^{a}}{r_{12}} (n_{12}{} v_{1}{}) y_{1}^{bj}
 + \frac{27}{112} \frac{n_{12}^{a}}{r_{12}} (n_{12}{} v_{2}{}) y_{1}^{bj}
 -  \frac{1}{28} \frac{v_{1}^{a}}{r_{12}^2} (n_{12}{} y_{1}{}) y_{1}^{bj}
 + \frac{1}{28} \frac{v_{2}^{a}}{r_{12}^2} (n_{12}{} y_{1}{}) y_{1}^{bj}\nn\\
& + \frac{3}{56} \frac{n_{12}^{a}}{r_{12}^2} (n_{12}{} v_{1}{}) (n_{12}{} y_{1}{}) y_{1}^{bj}
 -  \frac{3}{56} \frac{n_{12}^{a}}{r_{12}^2} (n_{12}{} v_{2}{}) (n_{12}{} y_{1}{}) y_{1}^{bj}
 -  \frac{1}{28} \frac{n_{12}^{a}}{r_{12}^2} (v_{1}{} y_{1}{}) y_{1}^{bj}
 + \frac{1}{28} \frac{n_{12}^{a}}{r_{12}^2} (v_{2}{} y_{1}{}) y_{1}^{bj}\Bigr)\nn\\
& + m_{1} \varepsilon_{iab} \Bigl(\frac{5}{28} (v_{1}{} y_{1}{}) v_{1}^{bj} y_{1}^{a}
 -  \frac{13}{28} v_{1}^{2} v_{1}^{a} y_{1}^{bj}\Bigr)\biggl]
 +\big( 1 \leftrightarrow 2 \big)\, ,\\
 \dJ^{ijk}={}& \mathop{\text{STF}}_{i,j,k} m_{1} v_{1}^{a} \varepsilon_{iab} \Bigl(\frac{1}{2} n_{12}^{k} r_{12}  y_{1}^{bj}
 -  \frac{1}{2} n_{12}^{j} r_{12} y_{1}^{bk} - y_{1}^{jkb}\Bigr) +\big( 1 \leftrightarrow 2 \big)\, ,
\end{align}
\end{subequations}
\begin{subequations}
\begin{align}
\dW={}&\frac{1}{3} m_{1} (v_{1}{} y_{1}{})
 + \frac{1}{c^{2}} \biggl [\frac{3}{10} m_{1} (v_{1}{} y_{1}{}) v_{1}^{2}
 + G m_{1} m_{2} \Bigl(\frac{5}{3} (n_{12}{} v_{1}{})
 -  \frac{2}{15} \frac{1}{r_{12}} (n_{12}{} v_{1}{}) (n_{12}{} y_{1}{})
 -  \frac{1}{6} \frac{1}{r_{12}} (n_{12}{} v_{2}{}) (n_{12}{} y_{1}{})\nn\\
& + \frac{1}{40} \frac{1}{r_{12}^{2}} (n_{12}{} v_{1}{}) (n_{12}{} y_{1}{})^2
 -  \frac{1}{40} \frac{1}{r_{12}^{2}} (n_{12}{} v_{2}{}) (n_{12}{} y_{1}{})^2
 -  \frac{1}{10} \frac{1}{r_{12}} (v_{1}{} y_{1}{})
 -  \frac{1}{30} \frac{1}{r_{12}^{2}} (n_{12}{} y_{1}{}) (v_{1}{} y_{1}{})
 -  \frac{2}{15} \frac{1}{r_{12}} (v_{2}{} y_{1}{})\nn\\
& + \frac{1}{30} \frac{1}{r_{12}^{2}} (n_{12}{} y_{1}{}) (v_{2}{} y_{1}{})
 -  \frac{1}{40} \frac{1}{r_{12}^{2}} (n_{12}{} v_{1}{}) y_{1}^{2}
 + \frac{1}{40} \frac{1}{r_{12}^{2}} (n_{12}{} v_{2}{}) y_{1}^{2}\Bigr)\biggl]  +\big( 1 \leftrightarrow 2 \big)+ \dW_{2\text{PN}}\, ,\\
\dW^{i}={}&m_{1} \Bigl(\frac{3}{10} y_{1}^{i} (v_{1}{} y_{1}{})
 -  \frac{1}{10} v_{1}^{i} y_{1}^{2}\Bigr)
 + \frac{1}{c^{2}} \biggl [m_{1} \Bigl(\frac{3}{140} v_{1}^{i} (v_{1}{} y_{1}{})^2
 + \frac{31}{140} y_{1}^{i} (v_{1}{} y_{1}{}) v_{1}^{2}
 -  \frac{1}{70} v_{1}^{i} v_{1}^{2} y_{1}^{2}\Bigr)\nn\\
& + G m_{1} m_{2} \Bigl(\frac{19}{40} y_{1}^{i} (n_{12}{} v_{1}{})
 -  \frac{3}{140} v_{1}^{i} (n_{12}{} y_{1}{})
 -  \frac{1}{56} \frac{y_{1}^{i}}{r_{12}} (n_{12}{} v_{1}{}) (n_{12}{} y_{1}{})
 -  \frac{3}{280} \frac{v_{1}^{i}}{r_{12}} (n_{12}{} y_{1}{})^2\nn\\
& + \frac{3}{280} \frac{y_{1}^{i}}{r_{12}^2} (n_{12}{} v_{1}{}) (n_{12}{} y_{1}{})^2
 -  \frac{129}{280} n_{12}^{i} (v_{1}{} y_{1}{})
 + \frac{677}{560} \frac{y_{1}^{i}}{r_{12}} (v_{1}{} y_{1}{})
 -  \frac{3}{280} \frac{n_{12}^{i}}{r_{12}} (n_{12}{} y_{1}{}) (v_{1}{} y_{1}{})\nn\\
& -  \frac{3}{35} \frac{y_{1}^{i}}{r_{12}^2} (n_{12}{} y_{1}{}) (v_{1}{} y_{1}{})
 -  \frac{3}{280} \frac{n_{12}^{i}}{r_{12}^2} (n_{12}{} y_{1}{})^2 (v_{1}{} y_{1}{})
 -  \frac{9}{280} \frac{y_{1}^{i}}{r_{12}^3} (n_{12}{} y_{1}{})^2 (v_{1}{} y_{1}{})
 -  \frac{93}{70} \frac{y_{1}^{i}}{r_{12}} (v_{2}{} y_{1}{})\nn\\
& -  \frac{1}{40} \frac{y_{1}^{i}}{r_{12}^2} (n_{12}{} y_{1}{}) (v_{2}{} y_{1}{})
 -  \frac{3}{280} \frac{y_{1}^{i}}{r_{12}^3} (n_{12}{} y_{1}{})^2 (v_{2}{} y_{1}{})
 -  \frac{223}{560} \frac{v_{1}^{i}}{r_{12}} y_{1}^{2}
 + \frac{241}{560} \frac{v_{2}^{i}}{r_{12}} y_{1}^{2}
 + \frac{31}{560} \frac{n_{12}^{i}}{r_{12}} (n_{12}{} v_{1}{}) y_{1}^{2}\nn\\
& -  \frac{27}{280} \frac{y_{1}^{i}}{r_{12}^2} (n_{12}{} v_{1}{}) y_{1}^{2}
 -  \frac{23}{560} \frac{y_{1}^{i}}{r_{12}^2} (n_{12}{} v_{2}{}) y_{1}^{2}
 + \frac{1}{56} \frac{v_{1}^{i}}{r_{12}^2} (n_{12}{} y_{1}{}) y_{1}^{2}
 -  \frac{1}{56} \frac{v_{2}^{i}}{r_{12}^2} (n_{12}{} y_{1}{}) y_{1}^{2}\nn\\
& -  \frac{3}{140} \frac{n_{12}^{i}}{r_{12}^2} (n_{12}{} v_{1}{}) (n_{12}{} y_{1}{}) y_{1}^{2}
 + \frac{1}{140} \frac{n_{12}^{i}}{r_{12}^2} (v_{1}{} y_{1}{}) y_{1}^{2}
 + \frac{3}{560} \frac{y_{1}^{i}}{r_{12}^3} (v_{1}{} y_{1}{}) y_{1}^{2}
 + \frac{3}{140} \frac{n_{12}^{i}}{r_{12}^3} (n_{12}{} y_{1}{}) (v_{1}{} y_{1}{}) y_{1}^{2}\nn\\
& + \frac{3}{70} \frac{y_{1}^{i}}{r_{12}^4} (n_{12}{} y_{1}{}) (v_{1}{} y_{1}{}) y_{1}^{2}
 + \frac{27}{560} \frac{n_{12}^{i}}{r_{12}^2} (v_{2}{} y_{1}{}) y_{1}^{2}
 -  \frac{3}{112} \frac{y_{1}^{i}}{r_{12}^3} (v_{2}{} y_{1}{}) y_{1}^{2}
 + \frac{3}{140} \frac{n_{12}^{i}}{r_{12}^3} (n_{12}{} y_{1}{}) (v_{2}{} y_{1}{}) y_{1}^{2}
 -  \frac{1}{80} \frac{v_{1}^{i}}{r_{12}^3} y_{1}^{4}\nn\\
& + \frac{1}{80} \frac{v_{2}^{i}}{r_{12}^3} y_{1}^{4}
 + \frac{9}{560} \frac{n_{12}^{i}}{r_{12}^3} (n_{12}{} v_{1}{}) y_{1}^{4}
 + \frac{3}{560} \frac{n_{12}^{i}}{r_{12}^3} (n_{12}{} v_{2}{}) y_{1}^{4}
 -  \frac{3}{280} \frac{y_{1}^{i}}{r_{12}^4} (n_{12}{} v_{2}{}) y_{1}^{4}
 -  \frac{3}{280} \frac{n_{12}^{i}}{r_{12}^4} (v_{1}{} y_{1}{}) y_{1}^{4}\nn\\
& -  \frac{3}{280} \frac{y_{1}^{i}}{r_{12}^5} (v_{1}{} y_{1}{}) y_{1}^{4}
 -  \frac{3}{140} \frac{n_{12}^{i}}{r_{12}^4} (v_{2}{} y_{1}{}) y_{1}^{4}
 + \frac{3}{280} \frac{y_{1}^{i}}{r_{12}^5} (v_{2}{} y_{1}{}) y_{1}^{4}
 -  \frac{1}{56} r_{12} v_{1}^{i}\Bigr)\biggl]+\big( 1 \leftrightarrow 2 \big)\, , \\
 \dW^{ij}={}& \mathop{\text{STF}}_{i,j} ~\frac{1}{21} m_{1} \Bigl(5 y_{1}^{ij} (v_{1}{} y_{1}{})
 - 2 v_{1}^{i} y_{1}^{j} y_{1}^{2}\Bigr)+\big( 1 \leftrightarrow 2 \big)\, , \\
\dX={}&G m_{1} m_{2} \Bigl(- \frac{1}{30} r_{12}
 + \frac{1}{60} (n_{12}{} y_{1}{})
 -  \frac{1}{40} \frac{1}{r_{12}} (n_{12}{} y_{1}{})^2
 + \frac{1}{120} \frac{1}{r_{12}} y_{1}^{2}\Bigr)
 + m_{1} \Bigl(\frac{1}{20} (v_{1}{} y_{1}{})^2
 -  \frac{1}{60} v_{1}^{2} y_{1}^{2}\Bigr)\nn\\
& + \frac{1}{c^{2}} \biggl [m_{1} \Bigl(\frac{43}{840} (v_{1}{} y_{1}{})^2 v_{1}^{2}
 -  \frac{1}{168} v_{1}^{4} y_{1}^{2}\Bigr)
 + G^2 m_{1}^2 m_{2} \Bigl(\frac{11}{420}
 + \frac{23}{210} \frac{1}{r_{12}} (n_{12}{} y_{1}{})
 + \frac{4}{15} \frac{1}{r_{12}^{2}} (n_{12}{} y_{1}{})^2\nn\\
& -  \frac{1}{70} \frac{1}{r_{12}^{3}} (n_{12}{} y_{1}{})^3
 -  \frac{1}{15} \frac{1}{r_{12}^{2}} y_{1}^{2}
 -  \frac{1}{210} \frac{1}{r_{12}^{3}} (n_{12}{} y_{1}{}) y_{1}^{2}
 + \frac{1}{280} \frac{1}{r_{12}^{4}} (n_{12}{} y_{1}{})^2 y_{1}^{2}
 -  \frac{1}{840} \frac{1}{r_{12}^{4}} y_{1}^{4}\Bigr)\nn\\
& + G m_{1} m_{2} \Bigl(- \frac{1}{40} r_{12} (n_{12}{} v_{1}{})^2
 -  \frac{11}{420} r_{12} (n_{12}{} v_{1}{}) (n_{12}{} v_{2}{})
 + \frac{3}{70} r_{12} (v_{1}{} v_{2}{})
 -  \frac{1}{56} (n_{12}{} v_{1}{})^2 (n_{12}{} y_{1}{})\nn\\
& + \frac{41}{840} (n_{12}{} v_{1}{}) (n_{12}{} v_{2}{}) (n_{12}{} y_{1}{})
 -  \frac{1}{60} (n_{12}{} v_{2}{})^2 (n_{12}{} y_{1}{})
 -  \frac{1}{15} (v_{1}{} v_{2}{}) (n_{12}{} y_{1}{})
 -  \frac{1}{56} \frac{1}{r_{12}} (n_{12}{} v_{1}{})^2 (n_{12}{} y_{1}{})^2\nn\\
& + \frac{19}{560} \frac{1}{r_{12}} (n_{12}{} v_{1}{}) (n_{12}{} v_{2}{}) (n_{12}{} y_{1}{})^2
 + \frac{3}{140} \frac{1}{r_{12}} (n_{12}{} v_{2}{})^2 (n_{12}{} y_{1}{})^2
 + \frac{137}{1680} \frac{1}{r_{12}} (v_{1}{} v_{2}{}) (n_{12}{} y_{1}{})^2\nn\\
& + \frac{3}{112} \frac{1}{r_{12}^{2}} (n_{12}{} v_{1}{})^2 (n_{12}{} y_{1}{})^3
 -  \frac{3}{56} \frac{1}{r_{12}^{2}} (n_{12}{} v_{1}{}) (n_{12}{} v_{2}{}) (n_{12}{} y_{1}{})^3
 + \frac{3}{112} \frac{1}{r_{12}^{2}} (n_{12}{} v_{2}{})^2 (n_{12}{} y_{1}{})^3\nn\\
& + \frac{3}{280} \frac{1}{r_{12}^{2}} (v_{1}{} v_{2}{}) (n_{12}{} y_{1}{})^3
 + \frac{5}{28} (n_{12}{} v_{1}{}) (v_{1}{} y_{1}{})
 -  \frac{23}{840} (n_{12}{} v_{2}{}) (v_{1}{} y_{1}{})
 + \frac{3}{35} \frac{1}{r_{12}} (n_{12}{} v_{1}{}) (n_{12}{} y_{1}{}) (v_{1}{} y_{1}{})\nn\\
& -  \frac{27}{280} \frac{1}{r_{12}} (n_{12}{} v_{2}{}) (n_{12}{} y_{1}{}) (v_{1}{} y_{1}{})
 -  \frac{3}{280} \frac{1}{r_{12}^{2}} (n_{12}{} v_{1}{}) (n_{12}{} y_{1}{})^2 (v_{1}{} y_{1}{})
 + \frac{3}{280} \frac{1}{r_{12}^{2}} (n_{12}{} v_{2}{}) (n_{12}{} y_{1}{})^2 (v_{1}{} y_{1}{})\nn\\
& -  \frac{1}{120} \frac{1}{r_{12}} (v_{1}{} y_{1}{})^2
 -  \frac{3}{140} \frac{1}{r_{12}^{2}} (n_{12}{} y_{1}{}) (v_{1}{} y_{1}{})^2
 + \frac{19}{140} (n_{12}{} v_{1}{}) (v_{2}{} y_{1}{})
 -  \frac{83}{420} (n_{12}{} v_{2}{}) (v_{2}{} y_{1}{})\nn\\
& -  \frac{16}{105} \frac{1}{r_{12}} (n_{12}{} v_{1}{}) (n_{12}{} y_{1}{}) (v_{2}{} y_{1}{})
 + \frac{17}{280} \frac{1}{r_{12}} (n_{12}{} v_{2}{}) (n_{12}{} y_{1}{}) (v_{2}{} y_{1}{})
 + \frac{3}{56} \frac{1}{r_{12}^{2}} (n_{12}{} v_{1}{}) (n_{12}{} y_{1}{})^2 (v_{2}{} y_{1}{})\nn\\
& -  \frac{3}{56} \frac{1}{r_{12}^{2}} (n_{12}{} v_{2}{}) (n_{12}{} y_{1}{})^2 (v_{2}{} y_{1}{})
 + \frac{1}{140} \frac{1}{r_{12}} (v_{1}{} y_{1}{}) (v_{2}{} y_{1}{})
 -  \frac{1}{70} \frac{1}{r_{12}^{2}} (n_{12}{} y_{1}{}) (v_{1}{} y_{1}{}) (v_{2}{} y_{1}{})
 -  \frac{31}{840} \frac{1}{r_{12}} (v_{2}{} y_{1}{})^2\nn\\
& + \frac{1}{28} \frac{1}{r_{12}^{2}} (n_{12}{} y_{1}{}) (v_{2}{} y_{1}{})^2
 -  \frac{31}{840} r_{12} v_{1}^{2}
 -  \frac{19}{420} (n_{12}{} y_{1}{}) v_{1}^{2}
 -  \frac{59}{1680} \frac{1}{r_{12}} (n_{12}{} y_{1}{})^2 v_{1}^{2}
 -  \frac{3}{560} \frac{1}{r_{12}^{2}} (n_{12}{} y_{1}{})^3 v_{1}^{2}\nn\\
& + \frac{31}{280} (n_{12}{} y_{1}{}) v_{2}^{2}
 -  \frac{79}{1680} \frac{1}{r_{12}} (n_{12}{} y_{1}{})^2 v_{2}^{2}
 -  \frac{3}{560} \frac{1}{r_{12}^{2}} (n_{12}{} y_{1}{})^3 v_{2}^{2}
 -  \frac{13}{420} \frac{1}{r_{12}} (n_{12}{} v_{1}{})^2 y_{1}^{2}\nn\\
& + \frac{97}{1680} \frac{1}{r_{12}} (n_{12}{} v_{1}{}) (n_{12}{} v_{2}{}) y_{1}^{2}
 -  \frac{23}{840} \frac{1}{r_{12}} (n_{12}{} v_{2}{})^2 y_{1}^{2}
 -  \frac{3}{112} \frac{1}{r_{12}} (v_{1}{} v_{2}{}) y_{1}^{2}
 + \frac{13}{560} \frac{1}{r_{12}^{2}} (n_{12}{} v_{1}{})^2 (n_{12}{} y_{1}{}) y_{1}^{2}\nn\\
& -  \frac{1}{35} \frac{1}{r_{12}^{2}} (n_{12}{} v_{1}{}) (n_{12}{} v_{2}{}) (n_{12}{} y_{1}{}) y_{1}^{2}
 + \frac{3}{560} \frac{1}{r_{12}^{2}} (n_{12}{} v_{2}{})^2 (n_{12}{} y_{1}{}) y_{1}^{2}
 + \frac{11}{840} \frac{1}{r_{12}^{2}} (v_{1}{} v_{2}{}) (n_{12}{} y_{1}{}) y_{1}^{2}\nn\\
& -  \frac{3}{112} \frac{1}{r_{12}^{3}} (n_{12}{} v_{1}{})^2 (n_{12}{} y_{1}{})^2 y_{1}^{2}
 + \frac{3}{56} \frac{1}{r_{12}^{3}} (n_{12}{} v_{1}{}) (n_{12}{} v_{2}{}) (n_{12}{} y_{1}{})^2 y_{1}^{2}
 -  \frac{3}{112} \frac{1}{r_{12}^{3}} (n_{12}{} v_{2}{})^2 (n_{12}{} y_{1}{})^2 y_{1}^{2}\nn\\
& -  \frac{3}{280} \frac{1}{r_{12}^{3}} (v_{1}{} v_{2}{}) (n_{12}{} y_{1}{})^2 y_{1}^{2}
 -  \frac{1}{70} \frac{1}{r_{12}^{2}} (n_{12}{} v_{1}{}) (v_{1}{} y_{1}{}) y_{1}^{2}
 + \frac{1}{420} \frac{1}{r_{12}^{2}} (n_{12}{} v_{2}{}) (v_{1}{} y_{1}{}) y_{1}^{2}\nn\\
& + \frac{9}{280} \frac{1}{r_{12}^{3}} (n_{12}{} v_{1}{}) (n_{12}{} y_{1}{}) (v_{1}{} y_{1}{}) y_{1}^{2}
 -  \frac{9}{280} \frac{1}{r_{12}^{3}} (n_{12}{} v_{2}{}) (n_{12}{} y_{1}{}) (v_{1}{} y_{1}{}) y_{1}^{2}
 -  \frac{1}{140} \frac{1}{r_{12}^{3}} (v_{1}{} y_{1}{})^2 y_{1}^{2}\nn\\
& + \frac{13}{840} \frac{1}{r_{12}^{2}} (n_{12}{} v_{1}{}) (v_{2}{} y_{1}{}) y_{1}^{2}
 -  \frac{1}{280} \frac{1}{r_{12}^{2}} (n_{12}{} v_{2}{}) (v_{2}{} y_{1}{}) y_{1}^{2}
 -  \frac{9}{280} \frac{1}{r_{12}^{3}} (n_{12}{} v_{1}{}) (n_{12}{} y_{1}{}) (v_{2}{} y_{1}{}) y_{1}^{2}\nn\\
& + \frac{9}{280} \frac{1}{r_{12}^{3}} (n_{12}{} v_{2}{}) (n_{12}{} y_{1}{}) (v_{2}{} y_{1}{}) y_{1}^{2}
 + \frac{1}{70} \frac{1}{r_{12}^{3}} (v_{1}{} y_{1}{}) (v_{2}{} y_{1}{}) y_{1}^{2}
 -  \frac{1}{140} \frac{1}{r_{12}^{3}} (v_{2}{} y_{1}{})^2 y_{1}^{2}
 + \frac{29}{1680} \frac{1}{r_{12}} v_{1}^{2} y_{1}^{2}\nn\\
& -  \frac{1}{80} \frac{1}{r_{12}^{2}} (n_{12}{} y_{1}{}) v_{1}^{2} y_{1}^{2}
 + \frac{3}{560} \frac{1}{r_{12}^{3}} (n_{12}{} y_{1}{})^2 v_{1}^{2} y_{1}^{2}
 + \frac{19}{1680} \frac{1}{r_{12}} v_{2}^{2} y_{1}^{2}
 -  \frac{1}{1680} \frac{1}{r_{12}^{2}} (n_{12}{} y_{1}{}) v_{2}^{2} y_{1}^{2}\nn\\
& + \frac{3}{560} \frac{1}{r_{12}^{3}} (n_{12}{} y_{1}{})^2 v_{2}^{2} y_{1}^{2}
 -  \frac{1}{560} \frac{1}{r_{12}^{3}} (n_{12}{} v_{1}{})^2 y_{1}^{4}
 + \frac{1}{280} \frac{1}{r_{12}^{3}} (n_{12}{} v_{1}{}) (n_{12}{} v_{2}{}) y_{1}^{4}
 -  \frac{1}{560} \frac{1}{r_{12}^{3}} (n_{12}{} v_{2}{})^2 y_{1}^{4}\nn\\
& -  \frac{1}{840} \frac{1}{r_{12}^{3}} (v_{1}{} v_{2}{}) y_{1}^{4}
 + \frac{1}{1680} \frac{1}{r_{12}^{3}} v_{1}^{2} y_{1}^{4}
 + \frac{1}{1680} \frac{1}{r_{12}^{3}} v_{2}^{2} y_{1}^{4}\Bigr)\biggl]+\big( 1 \leftrightarrow 2 \big)\, ,\\
\dX^{i}={}&m_{1} \Bigl(\frac{1}{28} y_{1}^{i} (v_{1}{} y_{1}{})^2
 -  \frac{1}{70} v_{1}^{i} (v_{1}{} y_{1}{}) y_{1}^{2}
 -  \frac{1}{140} y_{1}^{i} v_{1}^{2} y_{1}^{2}\Bigr)
 + G m_{1} m_{2} \Bigl(\frac{1}{560} n_{12}^{i} r_{12} (n_{12}{} y_{1}{})
 + \frac{1}{70} y_{1}^{i} (n_{12}{} y_{1}{})\nn\\
& + \frac{1}{560} n_{12}^{i} (n_{12}{} y_{1}{})^2
 -  \frac{1}{56} \frac{y_{1}^{i}}{r_{12}} (n_{12}{} y_{1}{})^2
 -  \frac{3}{560} n_{12}^{i} y_{1}^{2}
 + \frac{1}{280} \frac{y_{1}^{i}}{r_{12}} y_{1}^{2}
 + \frac{1}{140} \frac{n_{12}^{i}}{r_{12}} (n_{12}{} y_{1}{}) y_{1}^{2}
 -  \frac{19}{560} r_{12} y_{1}^{i}\Bigr)\nn \\
 &+\big( 1 \leftrightarrow 2 \big)+\dX^{i}_{1\text{PN}}\, ,\\
\dX^{ij}={}& \mathop{\text{STF}}_{i,j} ~ m_{1} \Bigl(\frac{5}{216} y_{1}^{ij} (v_{1}{} y_{1}{})^2
 -  \frac{5}{378} v_{1}^{i} y_{1}^{j} (v_{1}{} y_{1}{}) y_{1}^{2}
 -  \frac{5}{1512} y_{1}^{ij} v_{1}^{2} y_{1}^{2}
 + \frac{1}{756} v_{1}^{ij} y_{1}^{4}\Bigr)
 + G m_{1} m_{2} \Bigl(- \frac{115}{3024} r_{12} y_{1}^{ij}\nn\\
& + \frac{1}{378} n_{12}^{j} r_{12} y_{1}^{i} (n_{12}{} y_{1}{})
 + \frac{5}{378} y_{1}^{ij} (n_{12}{} y_{1}{})
 + \frac{13}{3024} n_{12}^{j} y_{1}^{i} (n_{12}{} y_{1}{})^2
 -  \frac{19}{3024} \frac{1}{r_{12}} y_{1}^{ij} (n_{12}{} y_{1}{})^2
 -  \frac{25}{3024} n_{12}^{j} y_{1}^{i} y_{1}^{2}\nn\\
& + \frac{13}{3024} \frac{1}{r_{12}} y_{1}^{ij} y_{1}^{2}
 + \frac{1}{378} \frac{n_{12}^{j} y_{1}^{i}}{r_{12}} (n_{12}{} y_{1}{}) y_{1}^{2}
 -  \frac{1}{189} \frac{1}{r_{12}^{2}} y_{1}^{ij} (n_{12}{} y_{1}{}) y_{1}^{2}
 -  \frac{1}{336} n_{12}^{ij} \frac{1}{r_{12}} y_{1}^{4}
 + \frac{5}{3024} \frac{n_{12}^{j} y_{1}^{i}}{r_{12}^2} y_{1}^{4}\nn\\
& + \frac{1}{336} \frac{n_{12}^{i} y_{1}^{j}}{r_{12}^2} y_{1}^{4}
 + \frac{5}{378} n_{12}^{j} r_{12}^2 y_{1}^{i}\Bigr)+\big( 1 \leftrightarrow 2 \big)\, ,\\
\dY^{i}={}&m_{1} \Bigl(\frac{3}{10} v_{1}^{i} (v_{1}{} y_{1}{})
 -  \frac{1}{10} y_{1}^{i} v_{1}^{2}\Bigr)
 + G m_{1} m_{2} \Bigl(- \frac{3}{20} \frac{n_{12}^{i}}{r_{12}} (n_{12}{} y_{1}{})
 + \frac{y_{1}^{i}}{20 r_{12}}\Bigr)
 + \frac{1}{c^{2}} \biggl [m_{1} \Bigl(\frac{39}{140} v_{1}^{i} (v_{1}{} y_{1}{}) v_{1}^{2}
 -  \frac{13}{140} y_{1}^{i} v_{1}^{4}\Bigr)\nn\\
& + G m_{1} m_{2} \Bigl(\frac{163}{140} v_{1}^{i} (n_{12}{} v_{1}{})
 + \frac{1}{28} n_{12}^{i} (n_{12}{} v_{1}{})^2
 -  \frac{87}{560} \frac{y_{1}^{i}}{r_{12}} (n_{12}{} v_{1}{})^2
 -  \frac{163}{280} v_{1}^{i} (n_{12}{} v_{2}{})
 + \frac{61}{140} \frac{y_{1}^{i}}{r_{12}} (n_{12}{} v_{1}{}) (n_{12}{} v_{2}{})\nn\\
& -  \frac{19}{112} \frac{y_{1}^{i}}{r_{12}} (n_{12}{} v_{2}{})^2
 -  \frac{27}{140} \frac{y_{1}^{i}}{r_{12}} (v_{1}{} v_{2}{})
 + \frac{99}{280} \frac{v_{1}^{i}}{r_{12}} (n_{12}{} v_{1}{}) (n_{12}{} y_{1}{})
 -  \frac{141}{280} \frac{v_{2}^{i}}{r_{12}} (n_{12}{} v_{1}{}) (n_{12}{} y_{1}{})\nn\\
& -  \frac{51}{560} \frac{n_{12}^{i}}{r_{12}} (n_{12}{} v_{1}{})^2 (n_{12}{} y_{1}{})
 -  \frac{33}{280} \frac{y_{1}^{i}}{r_{12}^2} (n_{12}{} v_{1}{})^2 (n_{12}{} y_{1}{})
 -  \frac{43}{140} \frac{v_{1}^{i}}{r_{12}} (n_{12}{} v_{2}{}) (n_{12}{} y_{1}{})
 + \frac{5}{28} \frac{v_{2}^{i}}{r_{12}} (n_{12}{} v_{2}{}) (n_{12}{} y_{1}{})\nn\\
& + \frac{3}{14} \frac{n_{12}^{i}}{r_{12}} (n_{12}{} v_{1}{}) (n_{12}{} v_{2}{}) (n_{12}{} y_{1}{})
 + \frac{9}{140} \frac{y_{1}^{i}}{r_{12}^2} (n_{12}{} v_{1}{}) (n_{12}{} v_{2}{}) (n_{12}{} y_{1}{})
 + \frac{57}{560} \frac{n_{12}^{i}}{r_{12}} (n_{12}{} v_{2}{})^2 (n_{12}{} y_{1}{})\nn\\
& + \frac{3}{56} \frac{y_{1}^{i}}{r_{12}^2} (n_{12}{} v_{2}{})^2 (n_{12}{} y_{1}{})
 + \frac{71}{140} \frac{n_{12}^{i}}{r_{12}} (v_{1}{} v_{2}{}) (n_{12}{} y_{1}{})
 -  \frac{1}{28} \frac{y_{1}^{i}}{r_{12}^2} (v_{1}{} v_{2}{}) (n_{12}{} y_{1}{})
 -  \frac{9}{70} \frac{v_{1}^{i}}{r_{12}^2} (n_{12}{} v_{1}{}) (n_{12}{} y_{1}{})^2\nn\\
& + \frac{27}{140} \frac{v_{2}^{i}}{r_{12}^2} (n_{12}{} v_{1}{}) (n_{12}{} y_{1}{})^2
 + \frac{3}{14} \frac{n_{12}^{i}}{r_{12}^2} (n_{12}{} v_{1}{})^2 (n_{12}{} y_{1}{})^2
 + \frac{3}{28} \frac{y_{1}^{i}}{r_{12}^3} (n_{12}{} v_{1}{})^2 (n_{12}{} y_{1}{})^2
 + \frac{9}{70} \frac{v_{1}^{i}}{r_{12}^2} (n_{12}{} v_{2}{}) (n_{12}{} y_{1}{})^2\nn\\
& -  \frac{27}{140} \frac{v_{2}^{i}}{r_{12}^2} (n_{12}{} v_{2}{}) (n_{12}{} y_{1}{})^2
 -  \frac{3}{7} \frac{n_{12}^{i}}{r_{12}^2} (n_{12}{} v_{1}{}) (n_{12}{} v_{2}{}) (n_{12}{} y_{1}{})^2
 -  \frac{3}{14} \frac{y_{1}^{i}}{r_{12}^3} (n_{12}{} v_{1}{}) (n_{12}{} v_{2}{}) (n_{12}{} y_{1}{})^2\nn\\
& + \frac{3}{14} \frac{n_{12}^{i}}{r_{12}^2} (n_{12}{} v_{2}{})^2 (n_{12}{} y_{1}{})^2
 + \frac{3}{28} \frac{y_{1}^{i}}{r_{12}^3} (n_{12}{} v_{2}{})^2 (n_{12}{} y_{1}{})^2
 + \frac{3}{35} \frac{n_{12}^{i}}{r_{12}^2} (v_{1}{} v_{2}{}) (n_{12}{} y_{1}{})^2
 + \frac{3}{70} \frac{y_{1}^{i}}{r_{12}^3} (v_{1}{} v_{2}{}) (n_{12}{} y_{1}{})^2\nn\\
& -  \frac{3}{40} \frac{v_{1}^{i}}{r_{12}} (v_{1}{} y_{1}{})
 + \frac{3}{280} \frac{v_{2}^{i}}{r_{12}} (v_{1}{} y_{1}{})
 + \frac{57}{280} \frac{n_{12}^{i}}{r_{12}} (n_{12}{} v_{1}{}) (v_{1}{} y_{1}{})
 + \frac{3}{28} \frac{y_{1}^{i}}{r_{12}^2} (n_{12}{} v_{1}{}) (v_{1}{} y_{1}{})
 -  \frac{43}{140} \frac{n_{12}^{i}}{r_{12}} (n_{12}{} v_{2}{}) (v_{1}{} y_{1}{})\nn\\
& + \frac{1}{140} \frac{y_{1}^{i}}{r_{12}^2} (n_{12}{} v_{2}{}) (v_{1}{} y_{1}{})
 -  \frac{9}{140} \frac{v_{1}^{i}}{r_{12}^2} (n_{12}{} y_{1}{}) (v_{1}{} y_{1}{})
 -  \frac{3}{140} \frac{v_{2}^{i}}{r_{12}^2} (n_{12}{} y_{1}{}) (v_{1}{} y_{1}{})
 + \frac{3}{70} \frac{n_{12}^{i}}{r_{12}^2} (n_{12}{} v_{1}{}) (n_{12}{} y_{1}{}) (v_{1}{} y_{1}{})\nn\\
& -  \frac{9}{70} \frac{y_{1}^{i}}{r_{12}^3} (n_{12}{} v_{1}{}) (n_{12}{} y_{1}{}) (v_{1}{} y_{1}{})
 -  \frac{3}{70} \frac{n_{12}^{i}}{r_{12}^2} (n_{12}{} v_{2}{}) (n_{12}{} y_{1}{}) (v_{1}{} y_{1}{})
 + \frac{9}{70} \frac{y_{1}^{i}}{r_{12}^3} (n_{12}{} v_{2}{}) (n_{12}{} y_{1}{}) (v_{1}{} y_{1}{})\nn\\
& -  \frac{3}{28} \frac{n_{12}^{i}}{r_{12}^2} (v_{1}{} y_{1}{})^2
 + \frac{1}{35} \frac{y_{1}^{i}}{r_{12}^3} (v_{1}{} y_{1}{})^2
 + \frac{17}{280} \frac{v_{1}^{i}}{r_{12}} (v_{2}{} y_{1}{})
 -  \frac{59}{280} \frac{v_{2}^{i}}{r_{12}} (v_{2}{} y_{1}{})
 -  \frac{113}{280} \frac{n_{12}^{i}}{r_{12}} (n_{12}{} v_{1}{}) (v_{2}{} y_{1}{})\nn\\
& -  \frac{11}{140} \frac{y_{1}^{i}}{r_{12}^2} (n_{12}{} v_{1}{}) (v_{2}{} y_{1}{})
 + \frac{8}{35} \frac{n_{12}^{i}}{r_{12}} (n_{12}{} v_{2}{}) (v_{2}{} y_{1}{})
 -  \frac{1}{28} \frac{y_{1}^{i}}{r_{12}^2} (n_{12}{} v_{2}{}) (v_{2}{} y_{1}{})
 -  \frac{17}{140} \frac{v_{1}^{i}}{r_{12}^2} (n_{12}{} y_{1}{}) (v_{2}{} y_{1}{})\nn\\
& + \frac{29}{140} \frac{v_{2}^{i}}{r_{12}^2} (n_{12}{} y_{1}{}) (v_{2}{} y_{1}{})
 + \frac{33}{140} \frac{n_{12}^{i}}{r_{12}^2} (n_{12}{} v_{1}{}) (n_{12}{} y_{1}{}) (v_{2}{} y_{1}{})
 + \frac{9}{70} \frac{y_{1}^{i}}{r_{12}^3} (n_{12}{} v_{1}{}) (n_{12}{} y_{1}{}) (v_{2}{} y_{1}{})\nn\\
& -  \frac{33}{140} \frac{n_{12}^{i}}{r_{12}^2} (n_{12}{} v_{2}{}) (n_{12}{} y_{1}{}) (v_{2}{} y_{1}{})
 -  \frac{9}{70} \frac{y_{1}^{i}}{r_{12}^3} (n_{12}{} v_{2}{}) (n_{12}{} y_{1}{}) (v_{2}{} y_{1}{})
 + \frac{1}{35} \frac{n_{12}^{i}}{r_{12}^2} (v_{1}{} y_{1}{}) (v_{2}{} y_{1}{})\nn\\
& -  \frac{2}{35} \frac{y_{1}^{i}}{r_{12}^3} (v_{1}{} y_{1}{}) (v_{2}{} y_{1}{})
 + \frac{11}{140} \frac{n_{12}^{i}}{r_{12}^2} (v_{2}{} y_{1}{})^2
 + \frac{1}{35} \frac{y_{1}^{i}}{r_{12}^3} (v_{2}{} y_{1}{})^2
 -  \frac{59}{140} n_{12}^{i} v_{1}^{2}
 + \frac{57}{560} \frac{y_{1}^{i}}{r_{12}} v_{1}^{2}
 -  \frac{129}{560} \frac{n_{12}^{i}}{r_{12}} (n_{12}{} y_{1}{}) v_{1}^{2}\nn\\
& + \frac{3}{40} \frac{y_{1}^{i}}{r_{12}^2} (n_{12}{} y_{1}{}) v_{1}^{2}
 -  \frac{3}{70} \frac{n_{12}^{i}}{r_{12}^2} (n_{12}{} y_{1}{})^2 v_{1}^{2}
 -  \frac{3}{140} \frac{y_{1}^{i}}{r_{12}^3} (n_{12}{} y_{1}{})^2 v_{1}^{2}
 + \frac{89}{560} \frac{y_{1}^{i}}{r_{12}} v_{2}^{2}
 -  \frac{149}{560} \frac{n_{12}^{i}}{r_{12}} (n_{12}{} y_{1}{}) v_{2}^{2}\nn\\
& -  \frac{11}{280} \frac{y_{1}^{i}}{r_{12}^2} (n_{12}{} y_{1}{}) v_{2}^{2}
 -  \frac{3}{70} \frac{n_{12}^{i}}{r_{12}^2} (n_{12}{} y_{1}{})^2 v_{2}^{2}
 -  \frac{3}{140} \frac{y_{1}^{i}}{r_{12}^3} (n_{12}{} y_{1}{})^2 v_{2}^{2}
 -  \frac{27}{280} \frac{v_{1}^{i}}{r_{12}^2} (n_{12}{} v_{1}{}) y_{1}^{2}
 + \frac{3}{35} \frac{v_{2}^{i}}{r_{12}^2} (n_{12}{} v_{1}{}) y_{1}^{2}\nn\\
& + \frac{9}{70} \frac{n_{12}^{i}}{r_{12}^2} (n_{12}{} v_{1}{})^2 y_{1}^{2}
 -  \frac{3}{280} \frac{y_{1}^{i}}{r_{12}^3} (n_{12}{} v_{1}{})^2 y_{1}^{2}
 + \frac{1}{280} \frac{v_{1}^{i}}{r_{12}^2} (n_{12}{} v_{2}{}) y_{1}^{2}
 + \frac{1}{140} \frac{v_{2}^{i}}{r_{12}^2} (n_{12}{} v_{2}{}) y_{1}^{2}\nn\\
& -  \frac{33}{280} \frac{n_{12}^{i}}{r_{12}^2} (n_{12}{} v_{1}{}) (n_{12}{} v_{2}{}) y_{1}^{2}
 + \frac{3}{140} \frac{y_{1}^{i}}{r_{12}^3} (n_{12}{} v_{1}{}) (n_{12}{} v_{2}{}) y_{1}^{2}
 -  \frac{3}{280} \frac{n_{12}^{i}}{r_{12}^2} (n_{12}{} v_{2}{})^2 y_{1}^{2}
 -  \frac{3}{280} \frac{y_{1}^{i}}{r_{12}^3} (n_{12}{} v_{2}{})^2 y_{1}^{2}\nn\\
& + \frac{2}{35} \frac{n_{12}^{i}}{r_{12}^2} (v_{1}{} v_{2}{}) y_{1}^{2}
 -  \frac{1}{140} \frac{y_{1}^{i}}{r_{12}^3} (v_{1}{} v_{2}{}) y_{1}^{2}
 + \frac{9}{56} \frac{v_{1}^{i}}{r_{12}^3} (n_{12}{} v_{1}{}) (n_{12}{} y_{1}{}) y_{1}^{2}
 -  \frac{9}{56} \frac{v_{2}^{i}}{r_{12}^3} (n_{12}{} v_{1}{}) (n_{12}{} y_{1}{}) y_{1}^{2}\nn\\
& -  \frac{15}{56} \frac{n_{12}^{i}}{r_{12}^3} (n_{12}{} v_{1}{})^2 (n_{12}{} y_{1}{}) y_{1}^{2}
 -  \frac{9}{56} \frac{v_{1}^{i}}{r_{12}^3} (n_{12}{} v_{2}{}) (n_{12}{} y_{1}{}) y_{1}^{2}
 + \frac{9}{56} \frac{v_{2}^{i}}{r_{12}^3} (n_{12}{} v_{2}{}) (n_{12}{} y_{1}{}) y_{1}^{2}\nn\\
& + \frac{15}{28} \frac{n_{12}^{i}}{r_{12}^3} (n_{12}{} v_{1}{}) (n_{12}{} v_{2}{}) (n_{12}{} y_{1}{}) y_{1}^{2}
 -  \frac{15}{56} \frac{n_{12}^{i}}{r_{12}^3} (n_{12}{} v_{2}{})^2 (n_{12}{} y_{1}{}) y_{1}^{2}
 -  \frac{3}{28} \frac{n_{12}^{i}}{r_{12}^3} (v_{1}{} v_{2}{}) (n_{12}{} y_{1}{}) y_{1}^{2}\nn\\
& -  \frac{1}{14} \frac{v_{1}^{i}}{r_{12}^3} (v_{1}{} y_{1}{}) y_{1}^{2}
 + \frac{1}{14} \frac{v_{2}^{i}}{r_{12}^3} (v_{1}{} y_{1}{}) y_{1}^{2}
 + \frac{9}{56} \frac{n_{12}^{i}}{r_{12}^3} (n_{12}{} v_{1}{}) (v_{1}{} y_{1}{}) y_{1}^{2}
 -  \frac{9}{56} \frac{n_{12}^{i}}{r_{12}^3} (n_{12}{} v_{2}{}) (v_{1}{} y_{1}{}) y_{1}^{2}
 + \frac{1}{14} \frac{v_{1}^{i}}{r_{12}^3} (v_{2}{} y_{1}{}) y_{1}^{2}\nn\\
& -  \frac{1}{14} \frac{v_{2}^{i}}{r_{12}^3} (v_{2}{} y_{1}{}) y_{1}^{2}
 -  \frac{9}{56} \frac{n_{12}^{i}}{r_{12}^3} (n_{12}{} v_{1}{}) (v_{2}{} y_{1}{}) y_{1}^{2}
 + \frac{9}{56} \frac{n_{12}^{i}}{r_{12}^3} (n_{12}{} v_{2}{}) (v_{2}{} y_{1}{}) y_{1}^{2}
 -  \frac{3}{40} \frac{n_{12}^{i}}{r_{12}^2} v_{1}^{2} y_{1}^{2}
 + \frac{1}{280} \frac{y_{1}^{i}}{r_{12}^3} v_{1}^{2} y_{1}^{2}\nn\\
& + \frac{3}{56} \frac{n_{12}^{i}}{r_{12}^3} (n_{12}{} y_{1}{}) v_{1}^{2} y_{1}^{2}
 + \frac{1}{56} \frac{n_{12}^{i}}{r_{12}^2} v_{2}^{2} y_{1}^{2}
 + \frac{1}{280} \frac{y_{1}^{i}}{r_{12}^3} v_{2}^{2} y_{1}^{2}
 + \frac{3}{56} \frac{n_{12}^{i}}{r_{12}^3} (n_{12}{} y_{1}{}) v_{2}^{2} y_{1}^{2}\Bigr)
 + G^2 m_{1}^2 m_{2} \Bigl(\frac{29 n_{12}^{i}}{70 r_{12}}\nn\\
& + \frac{63}{40} \frac{n_{12}^{i}}{r_{12}^2} (n_{12}{} y_{1}{})
 + \frac{3}{35} \frac{y_{1}^{i}}{r_{12}^3} (n_{12}{} y_{1}{})
 -  \frac{4}{35} \frac{n_{12}^{i}}{r_{12}^3} (n_{12}{} y_{1}{})^2
 -  \frac{1}{70} \frac{y_{1}^{i}}{r_{12}^4} (n_{12}{} y_{1}{})^2
 -  \frac{2}{35} \frac{n_{12}^{i}}{r_{12}^3} y_{1}^{2}
 -  \frac{1}{140} \frac{y_{1}^{i}}{r_{12}^4} y_{1}^{2}\nn\\
& + \frac{1}{28} \frac{n_{12}^{i}}{r_{12}^4} (n_{12}{} y_{1}{}) y_{1}^{2}
 -  \frac{21 y_{1}^{i}}{40 r_{12}^2}\Bigr)\biggl]+\big( 1 \leftrightarrow 2 \big)+ \dY^{i}_{2\text{PN}}\, ,\label{Yi3D}\\
\dY^{ij}={}&\mathop{\text{STF}}_{i,j} ~ m_{1} \Bigl(- \frac{3}{7} n_{12}^{j} r_{12} v_{1}^{i} (v_{1}{} y_{1}{})
 + \frac{3}{7} n_{12}^{i} r_{12} v_{1}^{j} (v_{1}{} y_{1}{})
 + \frac{6}{7} v_{1}^{i} y_{1}^{j} (v_{1}{} y_{1}{})
 + \frac{1}{7} n_{12}^{j} r_{12} y_{1}^{i} v_{1}^{2}
 -  \frac{1}{7} n_{12}^{i} r_{12} y_{1}^{j} v_{1}^{2}
 -  \frac{2}{7} y_{1}^{ij} v_{1}^{2}\nn\\
& -  \frac{2}{7} v_{1}^{ij} y_{1}^{2}\Bigr)
 + G m_{1} m_{2} \Bigl(\frac{1}{28} n_{12}^{ij} r_{12}
 + \frac{1}{7} \frac{1}{r_{12}} y_{1}^{ij}
 + \frac{1}{14} n_{12}^{ij} (n_{12}{} y_{1}{})
 -  \frac{3}{7} \frac{n_{12}^{i} y_{1}^{j}}{r_{12}} (n_{12}{} y_{1}{})
 + \frac{1}{7} n_{12}^{ij} \frac{1}{r_{12}} y_{1}^{2}\nn\\
& + \frac{1}{14} n_{12}^{i} y_{1}^{j}\Bigr) +\big( 1 \leftrightarrow 2 \big)+\dY^{ij}_{1\text{PN}}\, ,\\
\dY^{ijk}={}& \mathop{\text{STF}}_{i,j,k} ~ G m_{1} m_{2} \Bigl(\frac{3}{8} \frac{1}{r_{12}} y_{1}^{ijk}
 + \frac{1}{8} y_{1}^{ij} n_{12}^{k}
 -  \frac{1}{4} \frac{n_{12}^{k}}{r_{12}} y_{1}^{ij} (n_{12}{} y_{1}{})
 -  \frac{1}{3} \frac{1}{r_{12}^{2}} y_{1}^{ijk} (n_{12}{} y_{1}{})
 + \frac{1}{4} \frac{n_{12}^{k}}{r_{12}^2} y_{1}^{ij} y_{1}^{2}
 -  \frac{1}{8} \frac{n_{12}^{j}}{r_{12}^2} y_{1}^{ik} y_{1}^{2}\nn\\
& + \frac{1}{8} \frac{y_{1}^{i}}{r_{12}} n_{12}^{jk} y_{1}^{2}\Bigr)
 + m_{1} \Bigl(- \frac{5}{12} y_{1}^{ijk} v_{1}^{2}
 -  \frac{1}{2} v_{1}^{ij} y_{1}^{2} y_{1}^{k}
 + \frac{5}{4} v_{1}^{i} (v_{1}{} y_{1}{}) y_{1}^{jk}\Bigr)+\big( 1 \leftrightarrow 2 \big)\, ,\\
\dZ^{i}={}&- \frac{1}{5} m_{1} \varepsilon_{ibj} v_{1}^{b} y_{1}^{j} (v_{1}{} y_{1}{})
 + G m_{1} m_{2} \Bigl(\frac{1}{10} \frac{\varepsilon_{ibj} n_{12}^{b} y_{1}^{j}}{r_{12}} (n_{12}{} y_{1}{})
 -  \frac{1}{20} \varepsilon_{iab} n_{12}^{a} y_{1}^{b}\Bigr)
 + \frac{1}{c^{2}} \biggl(- \frac{13}{70} m_{1} \varepsilon_{ijk} v_{1}^{j} y_{1}^{k} (v_{1}{} y_{1}{}) v_{1}^{2}\nn\\
& + G^2 m_{1}^2 m_{2} \Bigl(- \frac{21}{20} \frac{\varepsilon_{ibj} n_{12}^{b} y_{1}^{j}}{r_{12}^2} (n_{12}{} y_{1}{})
 + \frac{2}{35} \frac{\varepsilon_{ijk} n_{12}^{j} y_{1}^{k}}{r_{12}^3} (n_{12}{} y_{1}{})^2
 + \frac{1}{35} \frac{\varepsilon_{iaj} n_{12}^{a} y_{1}^{j}}{r_{12}^3} y_{1}^{2}
 -  \frac{1}{70} \frac{\varepsilon_{ibk} n_{12}^{b} y_{1}^{k}}{r_{12}^4} (n_{12}{} y_{1}{}) y_{1}^{2}\nn\\
& -  \frac{29 \varepsilon_{iab} n_{12}^{a} y_{1}^{b}}{70 r_{12}}\Bigr)
 + G m_{1} m_{2} \biggl [\varepsilon_{ikl} \Bigl(\frac{9}{280} \frac{n_{12}^{k} y_{1}^{l}}{r_{12}} (n_{12}{} v_{1}{})^2 (n_{12}{} y_{1}{})
 -  \frac{3}{35} \frac{n_{12}^{k} y_{1}^{l}}{r_{12}} (n_{12}{} v_{1}{}) (n_{12}{} v_{2}{}) (n_{12}{} y_{1}{})\nn\\
& -  \frac{27}{280} \frac{n_{12}^{k} y_{1}^{l}}{r_{12}} (n_{12}{} v_{2}{})^2 (n_{12}{} y_{1}{})
 + \frac{3}{70} \frac{n_{12}^{k} v_{1}^{l}}{r_{12}} (n_{12}{} v_{1}{}) (n_{12}{} y_{1}{})^2
 -  \frac{3}{70} \frac{n_{12}^{k} v_{2}^{l}}{r_{12}} (n_{12}{} v_{1}{}) (n_{12}{} y_{1}{})^2\nn\\
& + \frac{3}{70} \frac{v_{1}^{k} y_{1}^{l}}{r_{12}^2} (n_{12}{} v_{1}{}) (n_{12}{} y_{1}{})^2
 -  \frac{3}{35} \frac{v_{2}^{k} y_{1}^{l}}{r_{12}^2} (n_{12}{} v_{1}{}) (n_{12}{} y_{1}{})^2
 -  \frac{3}{70} \frac{n_{12}^{k} v_{1}^{l}}{r_{12}} (n_{12}{} v_{2}{}) (n_{12}{} y_{1}{})^2\nn\\
& + \frac{3}{70} \frac{n_{12}^{k} v_{2}^{l}}{r_{12}} (n_{12}{} v_{2}{}) (n_{12}{} y_{1}{})^2
 -  \frac{3}{70} \frac{v_{1}^{k} y_{1}^{l}}{r_{12}^2} (n_{12}{} v_{2}{}) (n_{12}{} y_{1}{})^2
 + \frac{3}{35} \frac{v_{2}^{k} y_{1}^{l}}{r_{12}^2} (n_{12}{} v_{2}{}) (n_{12}{} y_{1}{})^2\Bigr)\nn\\
& + \varepsilon_{ilm} \Bigl(- \frac{3}{28} \frac{n_{12}^{l} y_{1}^{m}}{r_{12}^2} (n_{12}{} v_{1}{})^2 (n_{12}{} y_{1}{})^2
 + \frac{3}{14} \frac{n_{12}^{l} y_{1}^{m}}{r_{12}^2} (n_{12}{} v_{1}{}) (n_{12}{} v_{2}{}) (n_{12}{} y_{1}{})^2
 -  \frac{3}{28} \frac{n_{12}^{l} y_{1}^{m}}{r_{12}^2} (n_{12}{} v_{2}{})^2 (n_{12}{} y_{1}{})^2\Bigr)\nn\\
& + \varepsilon_{ijl} \Bigl(- \frac{3}{70} \frac{n_{12}^{j} y_{1}^{l}}{r_{12}^2} (v_{1}{} v_{2}{}) (n_{12}{} y_{1}{})^2
 -  \frac{9}{70} \frac{n_{12}^{j} y_{1}^{l}}{r_{12}^2} (n_{12}{} v_{1}{}) (n_{12}{} y_{1}{}) (v_{2}{} y_{1}{})
 + \frac{9}{70} \frac{n_{12}^{j} y_{1}^{l}}{r_{12}^2} (n_{12}{} v_{2}{}) (n_{12}{} y_{1}{}) (v_{2}{} y_{1}{})\nn\\
& + \frac{3}{140} \frac{n_{12}^{j} y_{1}^{l}}{r_{12}^2} (n_{12}{} y_{1}{})^2 v_{1}^{2}
 + \frac{3}{140} \frac{n_{12}^{j} y_{1}^{l}}{r_{12}^2} (n_{12}{} y_{1}{})^2 v_{2}^{2}
 -  \frac{3}{56} \frac{n_{12}^{j} y_{1}^{l}}{r_{12}^2} (n_{12}{} v_{1}{})^2 y_{1}^{2}
 + \frac{3}{70} \frac{n_{12}^{j} y_{1}^{l}}{r_{12}^2} (n_{12}{} v_{1}{}) (n_{12}{} v_{2}{}) y_{1}^{2}\nn\\
& + \frac{3}{280} \frac{n_{12}^{j} y_{1}^{l}}{r_{12}^2} (n_{12}{} v_{2}{})^2 y_{1}^{2}
 -  \frac{9}{140} \frac{v_{1}^{j} y_{1}^{l}}{r_{12}^3} (n_{12}{} v_{1}{}) (n_{12}{} y_{1}{}) y_{1}^{2}
 + \frac{9}{140} \frac{v_{2}^{j} y_{1}^{l}}{r_{12}^3} (n_{12}{} v_{1}{}) (n_{12}{} y_{1}{}) y_{1}^{2}\nn\\
& + \frac{9}{140} \frac{v_{1}^{j} y_{1}^{l}}{r_{12}^3} (n_{12}{} v_{2}{}) (n_{12}{} y_{1}{}) y_{1}^{2}
 -  \frac{9}{140} \frac{v_{2}^{j} y_{1}^{l}}{r_{12}^3} (n_{12}{} v_{2}{}) (n_{12}{} y_{1}{}) y_{1}^{2}\Bigr)
 + \varepsilon_{ikm} \Bigl(\frac{3}{28} \frac{n_{12}^{k} y_{1}^{m}}{r_{12}^3} (n_{12}{} v_{1}{})^2 (n_{12}{} y_{1}{}) y_{1}^{2}\nn\\
& -  \frac{3}{14} \frac{n_{12}^{k} y_{1}^{m}}{r_{12}^3} (n_{12}{} v_{1}{}) (n_{12}{} v_{2}{}) (n_{12}{} y_{1}{}) y_{1}^{2}
 + \frac{3}{28} \frac{n_{12}^{k} y_{1}^{m}}{r_{12}^3} (n_{12}{} v_{2}{})^2 (n_{12}{} y_{1}{}) y_{1}^{2}\Bigr)
 + \varepsilon_{ibk} \Bigl(- \frac{5}{14} \frac{n_{12}^{b} y_{1}^{k}}{r_{12}} (v_{1}{} v_{2}{}) (n_{12}{} y_{1}{})\nn\\
& -  \frac{3}{20} \frac{n_{12}^{b} y_{1}^{k}}{r_{12}} (n_{12}{} v_{1}{}) (v_{1}{} y_{1}{})
 + \frac{1}{5} \frac{n_{12}^{b} y_{1}^{k}}{r_{12}} (n_{12}{} v_{2}{}) (v_{1}{} y_{1}{})
 + \frac{1}{70} \frac{n_{12}^{b} v_{2}^{k}}{r_{12}} (n_{12}{} y_{1}{}) (v_{1}{} y_{1}{})
 + \frac{37}{140} \frac{n_{12}^{b} y_{1}^{k}}{r_{12}} (n_{12}{} v_{1}{}) (v_{2}{} y_{1}{})\nn\\
& -  \frac{9}{70} \frac{n_{12}^{b} y_{1}^{k}}{r_{12}} (n_{12}{} v_{2}{}) (v_{2}{} y_{1}{})
 -  \frac{2}{35} \frac{n_{12}^{b} v_{2}^{k}}{r_{12}} (n_{12}{} y_{1}{}) (v_{2}{} y_{1}{})
 + \frac{3}{70} \frac{v_{1}^{b} y_{1}^{k}}{r_{12}^2} (n_{12}{} y_{1}{}) (v_{2}{} y_{1}{})
 + \frac{43}{280} \frac{n_{12}^{b} y_{1}^{k}}{r_{12}} (n_{12}{} y_{1}{}) v_{1}^{2}\nn\\
& + \frac{11}{56} \frac{n_{12}^{b} y_{1}^{k}}{r_{12}} (n_{12}{} y_{1}{}) v_{2}^{2}
 + \frac{3}{70} \frac{v_{1}^{b} y_{1}^{k}}{r_{12}^2} (n_{12}{} v_{1}{}) y_{1}^{2}
 -  \frac{1}{28} \frac{v_{2}^{b} y_{1}^{k}}{r_{12}^2} (n_{12}{} v_{1}{}) y_{1}^{2}
 -  \frac{1}{140} \frac{v_{2}^{b} y_{1}^{k}}{r_{12}^2} (n_{12}{} v_{2}{}) y_{1}^{2}
 + \frac{1}{35} \frac{v_{1}^{b} y_{1}^{k}}{r_{12}^3} (v_{1}{} y_{1}{}) y_{1}^{2}\nn\\
& -  \frac{1}{35} \frac{v_{2}^{b} y_{1}^{k}}{r_{12}^3} (v_{1}{} y_{1}{}) y_{1}^{2}
 + \frac{1}{35} \frac{v_{2}^{b} y_{1}^{k}}{r_{12}^3} (v_{2}{} y_{1}{}) y_{1}^{2}\Bigr)
 + \varepsilon_{iak} \Bigl(\frac{3}{70} \frac{n_{12}^{a} y_{1}^{k}}{r_{12}^2} (v_{1}{} y_{1}{})^2
 -  \frac{3}{70} \frac{n_{12}^{a} y_{1}^{k}}{r_{12}^2} (v_{2}{} y_{1}{})^2\nn\\
& -  \frac{3}{140} \frac{n_{12}^{a} y_{1}^{k}}{r_{12}^2} (v_{1}{} v_{2}{}) y_{1}^{2}
 -  \frac{1}{35} \frac{v_{1}^{a} y_{1}^{k}}{r_{12}^3} (v_{2}{} y_{1}{}) y_{1}^{2}
 + \frac{9}{280} \frac{n_{12}^{a} y_{1}^{k}}{r_{12}^2} v_{1}^{2} y_{1}^{2}
 -  \frac{3}{280} \frac{n_{12}^{a} y_{1}^{k}}{r_{12}^2} v_{2}^{2} y_{1}^{2}\Bigr)\nn\\
& + \varepsilon_{ibl} \Bigl(\frac{3}{70} \frac{n_{12}^{b} y_{1}^{l}}{r_{12}^3} (v_{1}{} v_{2}{}) (n_{12}{} y_{1}{}) y_{1}^{2}
 -  \frac{9}{140} \frac{n_{12}^{b} y_{1}^{l}}{r_{12}^3} (n_{12}{} v_{1}{}) (v_{1}{} y_{1}{}) y_{1}^{2}
 + \frac{9}{140} \frac{n_{12}^{b} y_{1}^{l}}{r_{12}^3} (n_{12}{} v_{2}{}) (v_{1}{} y_{1}{}) y_{1}^{2}\nn\\
& + \frac{9}{140} \frac{n_{12}^{b} y_{1}^{l}}{r_{12}^3} (n_{12}{} v_{1}{}) (v_{2}{} y_{1}{}) y_{1}^{2}
 -  \frac{9}{140} \frac{n_{12}^{b} y_{1}^{l}}{r_{12}^3} (n_{12}{} v_{2}{}) (v_{2}{} y_{1}{}) y_{1}^{2}
 -  \frac{3}{140} \frac{n_{12}^{b} y_{1}^{l}}{r_{12}^3} (n_{12}{} y_{1}{}) v_{1}^{2} y_{1}^{2}\nn\\
& -  \frac{3}{140} \frac{n_{12}^{b} y_{1}^{l}}{r_{12}^3} (n_{12}{} y_{1}{}) v_{2}^{2} y_{1}^{2}\Bigr)
 + \varepsilon_{iab} \Bigl(- \frac{1}{140} \frac{v_{1}^{a} v_{2}^{b}}{r_{12}} y_{1}^{2}
 -  \frac{3}{140} \frac{n_{12}^{a} v_{1}^{b}}{r_{12}^2} (v_{2}{} y_{1}{}) y_{1}^{2}
 + \frac{11}{140} n_{12}^{a} (v_{2}{} y_{1}{}) v_{1}^{b}\Bigr)\nn\\
& + \varepsilon_{iaj} \Bigl(- \frac{3}{140} \frac{v_{1}^{a} y_{1}^{j}}{r_{12}} (v_{2}{} y_{1}{})
 + \frac{3}{140} \frac{n_{12}^{a} v_{1}^{j}}{r_{12}^2} (v_{1}{} y_{1}{}) y_{1}^{2}
 -  \frac{1}{140} \frac{n_{12}^{a} v_{2}^{j}}{r_{12}^2} (v_{1}{} y_{1}{}) y_{1}^{2}
 + \frac{1}{140} \frac{n_{12}^{a} v_{2}^{j}}{r_{12}^2} (v_{2}{} y_{1}{}) y_{1}^{2}\nn\\
& + \frac{1}{4} n_{12}^{a} (v_{1}{} y_{1}{}) v_{1}^{j}
 -  \frac{3}{28} n_{12}^{a} (v_{1}{} y_{1}{}) v_{2}^{j}
 -  \frac{3}{20} n_{12}^{a} (v_{2}{} y_{1}{}) v_{2}^{j}
 + \frac{53}{280} n_{12}^{a} (v_{1}{} v_{2}{}) y_{1}^{j}
 + \frac{1}{56} n_{12}^{a} v_{1}^{2} y_{1}^{j}
 -  \frac{57}{280} n_{12}^{a} v_{2}^{2} y_{1}^{j}\Bigr)\nn\\
& + \varepsilon_{ibj} \Bigl(- \frac{1}{70} n_{12}^{b} r_{12} v_{1}^{j} (n_{12}{} v_{1}{})
 -  \frac{29}{280} n_{12}^{b} r_{12} v_{1}^{j} (n_{12}{} v_{2}{})
 + \frac{1}{20} \frac{v_{1}^{b} y_{1}^{j}}{r_{12}} (v_{1}{} y_{1}{})
 -  \frac{1}{140} \frac{v_{2}^{b} y_{1}^{j}}{r_{12}} (v_{1}{} y_{1}{})
 + \frac{17}{140} \frac{v_{2}^{b} y_{1}^{j}}{r_{12}} (v_{2}{} y_{1}{})\nn\\
& + \frac{3}{70} \frac{n_{12}^{b} v_{1}^{j}}{r_{12}} (n_{12}{} y_{1}{}) (v_{2}{} y_{1}{})
 + \frac{3}{140} \frac{n_{12}^{b} v_{1}^{j}}{r_{12}} (n_{12}{} v_{1}{}) y_{1}^{2}
 -  \frac{1}{70} \frac{n_{12}^{b} v_{2}^{j}}{r_{12}} (n_{12}{} v_{1}{}) y_{1}^{2}
 -  \frac{1}{140} \frac{n_{12}^{b} v_{2}^{j}}{r_{12}} (n_{12}{} v_{2}{}) y_{1}^{2}\nn\\
& + \frac{1}{70} \frac{v_{1}^{b} v_{2}^{j}}{r_{12}^2} (n_{12}{} y_{1}{}) y_{1}^{2}
 -  \frac{13}{70} (n_{12}{} y_{1}{}) v_{1}^{b} v_{2}^{j}
 -  \frac{13}{28} (n_{12}{} v_{1}{}) v_{1}^{b} y_{1}^{j}
 + \frac{87}{280} (n_{12}{} v_{2}{}) v_{1}^{b} y_{1}^{j}
 -  \frac{21}{40} (n_{12}{} v_{1}{}) v_{2}^{b} y_{1}^{j}\nn\\
& + \frac{18}{35} (n_{12}{} v_{2}{}) v_{2}^{b} y_{1}^{j}\Bigr)
 + \varepsilon_{ijk} \Bigl(- \frac{27}{140} \frac{v_{1}^{j} y_{1}^{k}}{r_{12}} (n_{12}{} v_{1}{}) (n_{12}{} y_{1}{})
 + \frac{41}{140} \frac{v_{2}^{j} y_{1}^{k}}{r_{12}} (n_{12}{} v_{1}{}) (n_{12}{} y_{1}{})
 + \frac{1}{5} \frac{v_{1}^{j} y_{1}^{k}}{r_{12}} (n_{12}{} v_{2}{}) (n_{12}{} y_{1}{})\nn\\
& -  \frac{4}{35} \frac{v_{2}^{j} y_{1}^{k}}{r_{12}} (n_{12}{} v_{2}{}) (n_{12}{} y_{1}{})
 -  \frac{1}{70} \frac{v_{1}^{j} v_{2}^{k}}{r_{12}} (n_{12}{} y_{1}{})^2
 + \frac{3}{70} \frac{v_{1}^{j} y_{1}^{k}}{r_{12}^2} (n_{12}{} y_{1}{}) (v_{1}{} y_{1}{})
 + \frac{1}{70} \frac{v_{2}^{j} y_{1}^{k}}{r_{12}^2} (n_{12}{} y_{1}{}) (v_{1}{} y_{1}{})\nn\\
& -  \frac{1}{10} \frac{v_{2}^{j} y_{1}^{k}}{r_{12}^2} (n_{12}{} y_{1}{}) (v_{2}{} y_{1}{})
 -  \frac{3}{70} \frac{n_{12}^{j} v_{1}^{k}}{r_{12}^2} (n_{12}{} v_{1}{}) (n_{12}{} y_{1}{}) y_{1}^{2}
 + \frac{3}{140} \frac{n_{12}^{j} v_{2}^{k}}{r_{12}^2} (n_{12}{} v_{1}{}) (n_{12}{} y_{1}{}) y_{1}^{2}\nn\\
& + \frac{3}{70} \frac{n_{12}^{j} v_{1}^{k}}{r_{12}^2} (n_{12}{} v_{2}{}) (n_{12}{} y_{1}{}) y_{1}^{2}
 -  \frac{3}{140} \frac{n_{12}^{j} v_{2}^{k}}{r_{12}^2} (n_{12}{} v_{2}{}) (n_{12}{} y_{1}{}) y_{1}^{2}
 -  \frac{13}{140} n_{12}^{j} (n_{12}{} v_{1}{}) (n_{12}{} y_{1}{}) v_{1}^{k}\nn\\
& + \frac{1}{10} n_{12}^{j} (n_{12}{} v_{2}{}) (n_{12}{} y_{1}{}) v_{1}^{k}
 + \frac{3}{20} n_{12}^{j} (n_{12}{} v_{1}{}) (n_{12}{} y_{1}{}) v_{2}^{k}
 -  \frac{9}{140} n_{12}^{j} (n_{12}{} v_{2}{}) (n_{12}{} y_{1}{}) v_{2}^{k}
 + \frac{3}{35} n_{12}^{j} (n_{12}{} v_{1}{})^2 y_{1}^{k}\nn\\
& -  \frac{59}{280} n_{12}^{j} (n_{12}{} v_{1}{}) (n_{12}{} v_{2}{}) y_{1}^{k}
 + \frac{3}{28} n_{12}^{j} (n_{12}{} v_{2}{})^2 y_{1}^{k}\Bigr)\biggl]\biggl)+\big( 1 \leftrightarrow 2 \big)\, ,\\
\dZ^{ij}={}&\mathop{\text{STF}}_{i,j} ~ m_{1} \varepsilon_{ibk} \Bigl(- \frac{1}{28} y_{1}^{2} v_{1}^{jk} y_{1}^{b}
 -  \frac{5}{28} (v_{1}{} y_{1}{}) v_{1}^{b} y_{1}^{jk}\Bigr)
 + G m_{1} m_{2} \biggl [- \frac{9}{112} \varepsilon_{iab} n_{12}^{a} y_{1}^{jb}
 + \frac{3}{56} \frac{\varepsilon_{iak} n_{12}^{a}}{r_{12}^2} y_{1}^{2} y_{1}^{jk}\nn\\
& + \varepsilon_{ibk} \Bigl(\frac{5}{112} \frac{y_{1}^{b}}{r_{12}} n_{12}^{jk} y_{1}^{2}
 + \frac{1}{28} \frac{n_{12}^{b}}{r_{12}} (n_{12}{} y_{1}{}) y_{1}^{jk}\Bigr)\biggl]+\big( 1 \leftrightarrow 2 \big)\, .
\end{align}
\end{subequations}

For completeness, we also provide the expressions of these multipole moments in the center-of-mass frame which are simpler and sufficient for most computations. With the notations $\hat{n}^L=\text{STF}_L \,n^L$, $\hat{v}^L=\text{STF}_L \,v^L$ and where the angled brackets $\langle \,\cdots\, \rangle$ applied on indices also indicate STF projection, they read

\begin{subequations}
\begin{align}
    \dI_{ij} &= m \nu r^2   \Bigg\{\hat{n}^{ij}    +\frac{1}{c^2}\Bigg[\hat{n}^{ij} \Bigg(v^2 \left(\frac{29}{42}-\frac{29 \nu }{14}\right)+\frac{G m}{r} \left(-\frac{5}{7}+\frac{8 \nu }{7}\right)\Bigg) +\dot{r}
   \left(-\frac{4}{7}+\frac{12 \nu }{7}\right) n^{\langle i}v^{j \rangle}+\left(\frac{11}{21}-\frac{11 \nu }{7}\right) \hat{v}^{ij} \Bigg] \nn\\
    &\quad +\frac{1}{c^4}\Bigg[\hat{n}^{ij}\Bigg(\frac{G m}{r} v^2 \left(\frac{2021}{756}-\frac{5947 \nu }{756}-\frac{4883
   \nu ^2}{756}\right) +\frac{G m}{r} \dot{r}^2 \left(-\frac{131}{756}+\frac{907 \nu }{756}-\frac{1273 \nu ^2}{756}\right) \nn\\
    &\qquad\quad\qquad +\frac{G^2 m^2}{r^2} \left(-\frac{355}{252}-\frac{953 \nu }{126}+\frac{337 \nu
   ^2}{252}\right) +v^4 \left(\frac{253}{504}-\frac{1835 \nu }{504}+\frac{3545 \nu ^2}{504}\right)\Bigg)\nn\\
   &\qquad +\dot{r} n^{\langle i}v^{j \rangle} \Bigg(v^2 \left(-\frac{26}{63}+\frac{202 \nu }{63}-\frac{418 \nu
   ^2}{63}\right)+\frac{G m}{r} \left(-\frac{155}{54}+\frac{4057 \nu }{378}+\frac{209 \nu ^2}{54}\right)\Bigg)\nn\\
   &\qquad  + \hat{v}^{ij} \Bigg(\frac{G m}{r} \left(\frac{106}{27}-\frac{335 \nu }{189}-\frac{985 \nu
   ^2}{189}\right)+\dot{r}^2 \left(\frac{5}{63}-\frac{25 \nu }{63}+\frac{25 \nu ^2}{63}\right)+v^2 \left(\frac{41}{126}-\frac{337 \nu }{126}+\frac{733 \nu ^2}{126}\right)\Bigg) \Bigg]\Bigg\}\, , \\
   \dI_{ijk} &= -m \nu \Delta  r^3    \Bigg\{\hat{n}^{ijk}+\frac{1}{c^2}\bigg[\hat{n}^{ijk} \left(v^2 \left(\frac{5}{6}-\frac{19 \nu }{6}\right)+\frac{G m }{r} \left(-\frac{5}{6}+\frac{13 \nu }{6}\right) \right)\ +\dot{r} (-1+2 \nu
   ) n^{\langle ij}v^{k \rangle} +(1-2 \nu ) n^{\langle i}v^{j k \rangle} \bigg]\Bigg\}\, , \\
   \dI_{ijkl} &= m \nu (1-3 \nu) r^4 \hat{n}^{ijkl}, \\
   \dJ_{ij} &= -m \nu \Delta r^2 n^a v^b \varepsilon^{ab\langle i} \Bigg\{n^{j \rangle}+\frac{1 }{c^2}\Bigg[n^{j\rangle} \left(v^2 \left(\frac{13}{28}-\frac{17 \nu }{7}\right)+\frac{G m }{r}\left(\frac{27}{14}+\frac{15 \nu }{7}\right)\right) +\dot{r} v^{j\rangle} \left(\frac{5}{28}-\frac{5
   \nu }{14}\right)  \Bigg]\Bigg\}\, ,  \\
   \dJ_{ijk} &= m \nu  (1-3 \nu ) r^3   n^a v^b \varepsilon^{ab\langle  i} n^{jk\rangle }\, ,
   \end{align}
    \end{subequations}
    \begin{subequations}
    \begin{align}
    \dW &= m \nu  r \dot{r} \Bigg\{\frac{1}{3}+\frac{1}{c^2} \left[v^2 \left(\frac{3}{10}-\frac{9\nu}{10}  \right)+\frac{G m}{r}\left(\frac{7}{5}+\frac{17\nu}{15}  \right)\right]\nn \\
    & \qquad +\frac{1}{c^4}\Bigg[\frac{G m}{r} \dot{r}^2 \left(-\frac{7}{120}-\frac{7 \nu}{40} -\frac{89\nu^2}{120} \right) +  \frac{G m}{r}   v^2 \left(\frac{2071}{840}-\frac{1171 \nu }{840}-\frac{1537 \nu
   ^2}{280}\right) \nn\\
   &\qquad\qquad +\frac{G^2 m^2}{r^2}  \left(\frac{323}{105}+\frac{53 \nu }{210}+\frac{661 \nu ^2}{210}\right) +v^4   \left(\frac{233}{840}-\frac{533 \nu }{280}+\frac{967 \nu
   ^2}{280}\right) \Bigg]\Bigg\}\, , \\
   \dW_i &= m  \nu \Delta  r^2   \Bigg\{-\frac{3}{10} \dot{r} n^i+\frac{1}{10}v^i  +\frac{1}{c^2}\Bigg[  n^i\left(\frac{G m }{r}\dot{r} \left(-\frac{38}{35}-\frac{137 \nu }{140}\right)+v^2 \dot{r} \left(-\frac{31}{140}+\frac{25\nu}{28} \right)\right) \nn\\
   &\qquad\qquad\qquad\qquad\qquad\qquad\qquad +v^i\left(v^2 \left(\frac{1}{70}-\frac{5\nu}{28}  \right)+\dot{r}^2 \left(-\frac{3}{140}+\frac{3\nu}{70} \right)+\frac{G m}{r} \left(\frac{31}{70}+\frac{23\nu}{140}\right) \right) \Bigg]\Bigg\}\, , \\
   \dW_{ij} &= m \nu  r^3  \left\{\hat{n}^{ij} \dot{r} \left(\frac{5}{21}-\frac{5\nu}{7}  \right)  +n^{\langle i}v^{j \rangle} \left(-\frac{2}{21}+\frac{2\nu}{7}  \right) \right\}\, , \\
   \dX &= m \nu  r^2  \Bigg\{\dot{r}^2 \left(\frac{1}{20}-\frac{3 \nu }{20}\right)+\frac{G m}{r} \left(-\frac{1}{15}+\frac{\nu }{30}\right)+v^2 \left(-\frac{1}{60}+\frac{\nu }{20}\right) \nn \\
   &\qquad +\frac{1}{c^2}\Bigg[v^4
   \left(-\frac{1}{168}+\frac{53 \nu }{840}-\frac{137 \nu ^2}{840}\right)+\frac{G m v^2}{r} \left(-\frac{47}{420}+\frac{12 \nu }{35}+\frac{2 \nu ^2}{35}\right) +\frac{G^2 m^2}{r^2} \left(\frac{11}{420}+\frac{27 \nu }{70}+\frac{41 \nu ^2}{420}\right) \nn \\
   &\qquad \qquad  + \frac{G m \dot{r}^2}{r}\left(\frac{23}{140}-\frac{4 \nu }{7}-\frac{16 \nu ^2}{21}\right)  +v^2 \dot{r}^2 
   \left(\frac{43}{840}-\frac{299 \nu }{840}+\frac{551 \nu ^2}{840}\right) \Bigg]\Bigg\}\, , \\
   \dX_i &= m  \nu \Delta  r^3  \Bigg\{n^i \Bigg(v^2 \left(\frac{1}{140}-\frac{\nu }{70}\right)+\frac{G m}{r}\left(\frac{1}{35}-\frac{\nu }{140}\right)+\dot{r}^2 \left(-\frac{1}{28}+\frac{\nu }{14}\right)\Bigg)
   + v^i  \dot{r} \left(\frac{1}{70}-\frac{\nu }{35}\right)\nn  \\
   &\qquad +\frac{1}{c^2}\Bigg[n^i\Bigg(\frac{G^2 m^2}{r^2}\left(-\frac{37}{1890}-\frac{191 \nu }{1890}-\frac{\nu ^2}{36}\right) +\frac{G m}{r} v^2
   \left(\frac{23}{504}-\frac{289 \nu }{2520}-\frac{23 \nu ^2}{840}\right)  +v^4 \left(\frac{1}{840}-\frac{19 \nu }{840}+\frac{2 \nu ^2}{35}\right)  \nn\\
   &\qquad\qquad\qquad +  \dot{r}^2 v^2 
   \left(-\frac{29}{840}+\frac{191 \nu }{840}-\frac{13 \nu ^2}{35}\right) +\frac{G m }{r}\dot{r}^2 \left(-\frac{281}{2520}+\frac{115 \nu }{504}+\frac{349 \nu ^2}{840}\right)\Bigg) \nn \\
   &\qquad\qquad  + v^i \dot{r}\Bigg(\dot{r}^2
   \left(-\frac{1}{630}+\frac{2 \nu }{315}-\frac{\nu ^2}{210}\right)+  \frac{G m}{r} \left(\frac{16}{315}-\frac{157 \nu }{1260}-\frac{13 \nu ^2}{84}\right) +v^2
   \left(\frac{1}{84}-\frac{\nu }{12}+\frac{\nu ^2}{7}\right)  \Bigg)\Bigg]\Bigg\}\, , \\
  \dX_{ij} &= m\nu   r^4  \Bigg\{ \hat{n}^{ij}\Bigg[v^2 \left(-\frac{5}{1512}+\frac{25 \nu }{1512}-\frac{25 \nu ^2}{1512}\right)+\frac{G m}{r} \left(-\frac{1}{63}+\frac{\nu }{18}-\frac{\nu ^2}{126}\right) +\dot{r}^2
   \left(\frac{5}{216}-\frac{25 \nu }{216}+\frac{25 \nu ^2}{216}\right)\Bigg] \nn \\
   &\qquad\qquad  + n^{\langle i}v^{j \rangle}  \dot{r} \left(-\frac{5}{378}+\frac{25 \nu }{378}-\frac{25 \nu ^2}{378}\right) +  \hat{v}^{ij}\left(\frac{1}{756}-\frac{5 \nu }{756}+\frac{5 \nu ^2}{756}\right)\Bigg\}\, , \\
  \dY_{i}^\text{3D} &= m \nu  \Delta   r  \Bigg\{\frac{1}{10}n^i \left(\frac{G m}{r}+v^2\right) -\frac{3}{10} \dot{r} v^i \nn\\
  &\qquad +\frac{1}{c^2}\Bigg[n^i \Bigg(v^4 \left(\frac{13}{140}-\frac{47 \nu }{140}\right)+\frac{G m}{r}\dot{r}^2
   \left(-\frac{3}{20}+\frac{\nu }{10}\right)   +\frac{G^2 m^2}{r^2} \left(\frac{29}{70}+\frac{6 \nu }{35}\right)+\frac{G m }{r}v^2 \left(\frac{39}{70}+\frac{13 \nu }{70}\right)\Bigg)\nn \\
   &\qquad\qquad\qquad  +\dot{r} v^i \Bigg(\frac{G m}{r} \left(-\frac{87}{70}-\frac{46 \nu }{35}\right) +v^2 \left(-\frac{39}{140}+\frac{141 \nu }{140}\right)\Bigg) \Bigg] \nn\\
   &\qquad  +\frac{1}{c^4}\Bigg[n^i \Bigg(\frac{G m}{r}v^4
   \left(\frac{443}{560}-\frac{907 \nu }{560}-\frac{809 \nu ^2}{560}\right) +\frac{G m }{r}\dot{r}^4 \left(-\frac{3}{40}-\frac{99 \nu }{560}+\frac{9 \nu ^2}{112}\right) \nn \\
   &\qquad\qquad\quad +\frac{G^2 m^2}{r^2}v^2
   \left(\frac{69}{28}-\frac{41 \nu }{40}+\frac{37 \nu ^2}{140}\right) +\frac{G^3 m^3}{r^3} \left(-\frac{5261}{945}-\frac{8 \nu }{27}+\frac{88 \nu ^2}{315}\right)\nn  \\
  &\qquad\qquad\quad +v^6 \left(\frac{139}{1680}-\frac{227 \nu }{336}+\frac{781 \nu ^2}{560}\right)+ \frac{G m}{r} \dot{r}^2 v^2 \left(\frac{2}{35}-\frac{8 \nu }{35}-\frac{107 \nu ^2}{140}\right) +\frac{G^2 m^2
   }{r^2}\dot{r}^2 \left(-\frac{47}{35}-\frac{97 \nu }{140}+\frac{\nu ^2}{5}\right) \Bigg) \nn \\
  &\qquad\qquad +  v^i \dot{r} \Bigg(\frac{G m }{r}\dot{r}^2 \left(\frac{27}{140}+\frac{201 \nu }{140}+\frac{131 \nu
   ^2}{140}\right) +  v^4 \left(-\frac{139}{560}+\frac{227 \nu }{112}-\frac{2343 \nu ^2}{560}\right)  \nn \\
  &\qquad\qquad\qquad +\frac{G^2 m^2}{r^2} \left(-\frac{75}{28}+\frac{193 \nu }{280}-\frac{281 \nu
   ^2}{70}\right) +\frac{G m}{r} v^2 \left(-\frac{381}{140}+\frac{93 \nu }{35}+\frac{34 \nu ^2}{5}\right) \Bigg) \Bigg]\Bigg\}\, , \\
   \dY_i^\text{pole} &=  -\frac{G^3 m^4  \nu  \Delta}{c^4 r^2} n^i \left(\frac{1}{\varepsilon }-3 \ln\left[\frac{\sqrt{{\bar{q}}} r}{\ell_0}\right]\right)\, , \\
      Y_{ij} &= m \nu  r^2  \Bigg\{\hat{n}^{ij} \Bigg(\frac{G m }{r}\left(\frac{1}{14}+\frac{2 \nu }{7}\right) +v^2 \left(-\frac{2}{7}+\frac{6 \nu }{7}\right)\Bigg) +\dot{r}  n^{\langle i}v^{j \rangle} \left(\frac{6}{7}-\frac{18 \nu }{7}\right)
   + \hat{v}^{ij} \left(-\frac{2}{7}+\frac{6 \nu }{7}\right) \nn \\
   &\qquad +\frac{1}{c^2}\Bigg[\hat{n}^{ij} \Bigg(v^4 \left(-\frac{13}{63}+\frac{101 \nu }{63}-\frac{209 \nu ^2}{63}\right)+\frac{G m}{r} \dot{r}^2
   \left(\frac{157}{756}-\frac{113 \nu }{189}-\frac{58 \nu ^2}{189}\right) \nn \\
   &\qquad\qquad +\frac{G^2 m^2}{r^2} \left(-\frac{85}{252}+\frac{515 \nu }{252}+\frac{7 \nu ^2}{9}\right) +\frac{G m}{r} v^2
   \left(-\frac{172}{189}+\frac{1081 \nu }{378}+\frac{61 \nu ^2}{27}\right)\Bigg) \nn \\
   &\qquad\qquad +\dot{r} n^{\langle i}v^{j \rangle}  \Bigg(\frac{G m}{r} \left(\frac{452}{189}-\frac{2189 \nu }{378}-\frac{1907 \nu
   ^2}{189}\right) +v^2 \left(\frac{37}{63}-\frac{293 \nu }{63}+\frac{617 \nu ^2}{63}\right)\Bigg) \nn \\
   &\qquad + \hat{v}^{ij} \Bigg(v^2 \left(-\frac{8}{63}+\frac{76 \nu }{63}-\frac{184 \nu
   ^2}{63}\right)+\dot{r}^2 \left(\frac{5}{63}-\frac{25 \nu }{63}+\frac{25 \nu ^2}{63}\right)+\frac{G m}{r} \left(-\frac{145}{378}+\frac{851 \nu }{378}+\frac{470 \nu ^2}{189}\right) \Bigg) \Bigg]\Bigg\}\, , \\
   \dY_{ijk} &= m  \nu \Delta   r^3  \Bigg\{\hat{n}^{ijk}\Bigg[v^2 \left(\frac{5}{12}-\frac{5 \nu }{6}\right)+\frac{G m}{r}\left(-\frac{1}{6}-\frac{\nu }{6}\right)\Bigg] +\dot{r} n^{\langle ij}v^{k \rangle}\left(-\frac{5}{4}+\frac{5 \nu
   }{2}\right)  +n^{\langle i}v^{jk \rangle} \left(\frac{1}{2}-\nu \right)  \Bigg\}\, , \\
   \dZ_{i} &= m \nu r^2 \dot{r}  \, n^a v^b \varepsilon^{ab   i }   \Bigg\{ \frac{1}{5}-\frac{3 \nu }{5}+\frac{1}{c^2} \Bigg[\frac{G m }{r}\left(\frac{7}{10}-\frac{43 \nu }{20}-\frac{29 \nu ^2}{10}\right)+v^2
   \left(\frac{13}{70}-\frac{93 \nu }{70}+\frac{177 \nu ^2}{70}\right)\Bigg]\Bigg\}\, ,  \\
   \dZ_{ij} &= m \nu \Delta r^3 \, n^a v^b \varepsilon^{ab \langle i }  \Bigg\{\dot{r} n^{j \rangle}\left(-\frac{5}{28}+\frac{5 \nu }{14}\right) +v^{j \rangle} \left(\frac{1}{28}-\frac{\nu }{14}\right) \Bigg\}\, .  
\end{align}
\end{subequations}

\bibliography{RefList_BFST25.bib}

\end{document}